\documentclass[11pt]{article}

\usepackage[T1]{fontenc}
\usepackage[margin=1in]{geometry}
\usepackage[font=small,labelfont=bf]{caption}

\newcommand{\parencite}{\cite}
\newcommand{\textcite}{\cite}

\usepackage{color}
\usepackage[x11names]{xcolor} 
\usepackage{hyperref}
\hypersetup{
    colorlinks,
    citecolor=gray,
    linkcolor=Blue4,
}
\usepackage{amsmath}
\usepackage{amsfonts}
\usepackage{floatpag}
\floatpagestyle{plain}

\usepackage{authblk}          
\usepackage{rotating}          

\title{Human Inference in Changing Environments With Temporal Structure}
\date{}

\author[1,$\dagger$]{Arthur Prat-Carrabin}
\author[3,$\ddagger$]{Robert C. Wilson}
\author[3]{Jonathan D. Cohen}
\author[1,2,3,4*]{Rava Azeredo da Silveira}

\affil[1]{Laboratoire de Physique de l’École Normale Supérieure, ENS, Université PSL, CNRS, Sorbonne Université, Université de Paris, 75005 Paris, France}
\affil[2]{IOB, Faculty of Science, University of Basel, Basel, Switzerland}
\affil[3]{Princeton Neuroscience Institute, Princeton University, Princeton, USA}
\affil[4]{Department of Neurobiology, Weizmann Institute of Science, Rehovot, Israel}
\affil[$\dagger$]{\textit{Present address:} Department of Economics, Columbia University, USA}
\affil[$\ddagger$]{\textit{Present address:} Department of Psychology and Cognitive Science Program, University of Arizona, Tucson, USA}
\affil[*]{\textit{For correspondence:} rava@ens.fr}

\graphicspath{ {./materials/figures/} }
\usepackage{bbm} 
\newcommand{\D}{\mathrm{d}} 
\newcommand{\E}{\mathbb{E}} 

\begin{document}
\maketitle

\begin{abstract}
To make informed decisions in natural environments that change over time, humans must update their beliefs as new observations are gathered. Studies exploring human inference as a dynamical process that unfolds in time have focused on situations in which the statistics of observations are history-independent. Yet temporal structure is everywhere in nature, and yields history-dependent observations. Do humans modify their inference processes depending on the latent temporal statistics of their observations? We investigate this question experimentally and theoretically using a change-point inference task. We show that humans adapt their inference process to fine aspects of the temporal structure in the statistics of stimuli. As such, humans behave qualitatively in a Bayesian fashion, but, quantitatively, deviate away from optimality. Perhaps more importantly, humans behave suboptimally in that their responses are not deterministic, but variable. We show that this variability itself is modulated by the temporal statistics of stimuli. To elucidate the cognitive algorithm that yields this behavior, we investigate a broad array of existing and new models that characterize different sources of suboptimal deviations away from Bayesian inference. While models with `output noise' that corrupts the response-selection process are natural candidates, human behavior is best described by sampling-based inference models, in which the main ingredient is a compressed approximation of the posterior, represented through a modest set of random samples and updated over time. This result comes to complement a growing literature on sample-based representation and learning in humans.
\end{abstract}

\section*{}
{\small
\textcopyright 2021, American Psychological Association. This paper is not the copy of record and may not exactly replicate the final, authoritative version of the article. Please do not copy or cite without authors' permission. The final article will be available, upon publication, via its DOI: 10.1037/rev0000276}

\section*{}
In a variety of inference tasks, human subjects use sensory cues as well as prior information in a manner consistent with Bayesian models. In tasks requiring the combination of a visual cue (such as the shape, position, texture, or motion of an object) with a haptic \parencite{Ernst2002, Battaglia2011}, auditory \parencite{Battaglia2003}, proprioceptive \parencite{vanBeers1999}, or a secondary visual cue \parencite{Jacobs1999, Hillis2004, Knill2007}, human subjects weigh information coming from each cue according to its uncertainty, in agreement with an optimal, probabilistic approach. Moreover, subjects appear also to integrate optimally prior knowledge on spatial \parencite{Kording2004,Kording2006} and temporal \parencite{Miyazaki2005, Jazayeri2010} variables relevant to inference, in line with Bayes' rule.

The Bayesian paradigm hence offers an elegant and mathematically principled account of the way in which humans carry inference in the presence of uncertainty. In most experimental designs, however, successive trials are unrelated to each other. Yet, in many natural situations, the brain receives a stream of evidence from the environment: inference, then, unfolds in time. Moreover, natural mechanisms introduce sophisticated temporal statistics in the course of events (e.g., rhythmicity in locomotion, day-night cycles, and various structures found in speech). Are these temporal dynamics used by the brain to refine its \textit{online} inference of the state of the environment?

Furthermore, most studies that support a Bayesian account of human inference discuss average behaviors of subjects, and, thereby, side-step the issue of the variability in human responses. While an optimal Bayesian model yields a unique, deterministic action in response to a given set of observations, human subjects exhibit noisy, and thus suboptimal, responses. Methods commonly used to model response variability, such as `softmax' and probability-matching response-selection strategies, or, more recently, stochastic inference processes, correspond to different forms of departure from Bayesian optimality. One would like to identify the nature of the deviations from Bayesian models that can account for the observed discrepancies from optimality in human behavior.

To explore these questions, we use an online inference task based on a `change-point' paradigm, i.e., with random stimuli originating from a hidden state that is subject to abrupt, occasional variations, which are referred to as `change points'. A growing theoretical and experimental literature examines inference problems for this class of signals \parencite{Adams2007, Fearnhead2007, Brown2009, Wilson2010, Nassar2010, Wilson2013, Nassar2012, Veliz-Cuba2016, Radillo2017, Glaze2015, Glaze2018, Piet2018, Radillo2019, Gallistel2014, Khaw2017}. All these studies, with the exception of the work of \textcite{Fearnhead2007}, focus on the history-independent case of random change points that obey Poisson temporal statistics. Such problems are characterized by the absence of temporal structure: the probability of occurrence of a change point does not depend on the realization of past change points. \textcite{Wilson2010, Piet2018} and \textcite{Glaze2018} extend their studies beyond this simple framework by considering `hierarchical-Poisson' models in which the change probability is itself subject to random variations; but, here also, the occurence of a change point does not depend on the timing of earlier change points. The experimental studies among the ones cited above have investigated the way in which human subjects and rodents infer hidden states, and whether they learn history-independent change probabilities.

Because of the pervasiveness of temporal structure in natural environments, we decided to study human inference in the presence of `history-dependent' statistics in which the occurrence of a change point depends on the timing of earlier change points. This introduces considerable complexity in the optimal inference model (as the hidden state is no longer Markovian), and serves as a first step toward a more ecological approach to human inference. For the purpose of comparison, we consider two different statistics of change points: the first one is the Poisson statistics commonly used in earlier studies; the second is the simplest non-Markovian statistics, in which the probability of a change point is a function of the timing of the preceding change point. This setup allows us to examine the effect of the latent temporal structure on both human behavior and model responses.

In these two contrasting conditions, the behavior of the Bayesian model and that of human subjects exhibit both similarities and discrepancies. A salient departure from optimality exhibited by subjects is the variability in their responses. What is more, the shape of the distribution of responses is not constant, but, rather, subject to modulations during the course of inference. The standard deviation and skewness of the empirical response distribution are correlated with that of the optimal, Bayesian posterior; this suggests that the randomness in subjects' responses does not reflect some `passive' source of noise but is in fact related to the uncertainty of the Bayesian observer.

To account for this non-trivial variability in human responses and other deviations from optimality, we investigate in what ways approximations of the Bayesian model alter behavior, in our task. The optimal estimation of a hidden state can be split into two steps: Bayesian posterior inference (computing optimally the belief distribution over the state space) and optimal response selection (using the belief distribution to choose the response that maximizes the expected reward). Suboptimal models introduce systematic errors or stochastic errors in the inference step or in the response-selection step, or in both, thus impacting behavior. Models discussed in the change-point literature, along with new models we introduce, provide a wide range of such deviations from optimality, which we compare to experimental data. This allows us to assess how different sources of suboptimality impact behavior, and to what extent they can capture the salient features in human behavior.

The paper is outlined as follows. We first present the main aspects of our task, in which subjects observe a visual stimulus and infer an underlying, changing, hidden state. The susceptibility of subjects to a new stimulus is shown to differ appreciably between the two conditions (with and without latent temporal structure), and to adapt to the statistics of change points.
We then analyze the variability in the subjects' responses, and show how it is modulated over the course of inference.
After deriving the optimal, Bayesian solution of the inference problem in the context of our task, we examine its behavior in comparison with experimental data. We then turn to investigating a broad family of suboptimal models.
In particular, motivated by the form of the variability present in our human data, we examine stochastic perturbations in both the inference step and in the response-selection step. These models reflect \textit{different forms} of \textit{sampling}: model subjects either perform inference using samples of probability distributions or select responses by sampling; the former option includes models with limited memory as well as sequential Monte Carlo (particle-filter) models. Finally, we discuss model fitting, from which we conclude that humans carry out stochastic approximations of the optimal Bayesian calculations through sampling-based inference (rather than sampling-based response selection).

Our observations confirm and extend the results reported in the change-point literature on human inference in the context of Poisson statistics,
by exploring a more ecological \parencite{Nunes2004, Nakamura2007, Nakamura2008, Anteneodo2009, Hausdorff1995, Griffin2000, Ramus1999, Low2000, Campione2002}, non-Poisson, temporally structured environment. Likewise, our results come to complement those of a number of studies on perception and decision-making that also investigate inference from stimuli with temporal statistics \parencite{Janssen2005, Ghose2002, Li2013, Miyazaki2005, Jazayeri2010, TenOever2014}. Our experimental results demonstrate that humans learn implicitly the temporal statistics of stimuli. Moreover, our work highlights the variability ubiquitous in behavioral data, and shows that it itself exhibits structure: it depends on the temporal statistics of the signal, and it is modulated over the course of inference.
We find that a model in which the Bayesian posterior is approximated with a set of samples captures the behavioral variability during inference. This proposal adds to the growing literature on cognitive `sample-based representations' of probability distributions \parencite{Goodman2008, Moreno-Bote2011, Gershman2012, Vul2014}. Our results suggest that the brain carries out complex inference by manipulating a modest number of samples, selected as a low-dimensional approximation of the optimal, Bayesian posterior.

\section{Results}

\subsection{Behavioral task, and history-independent vs. history-dependent stimuli}

\begin{figure} 
\includegraphics[width=\textwidth]{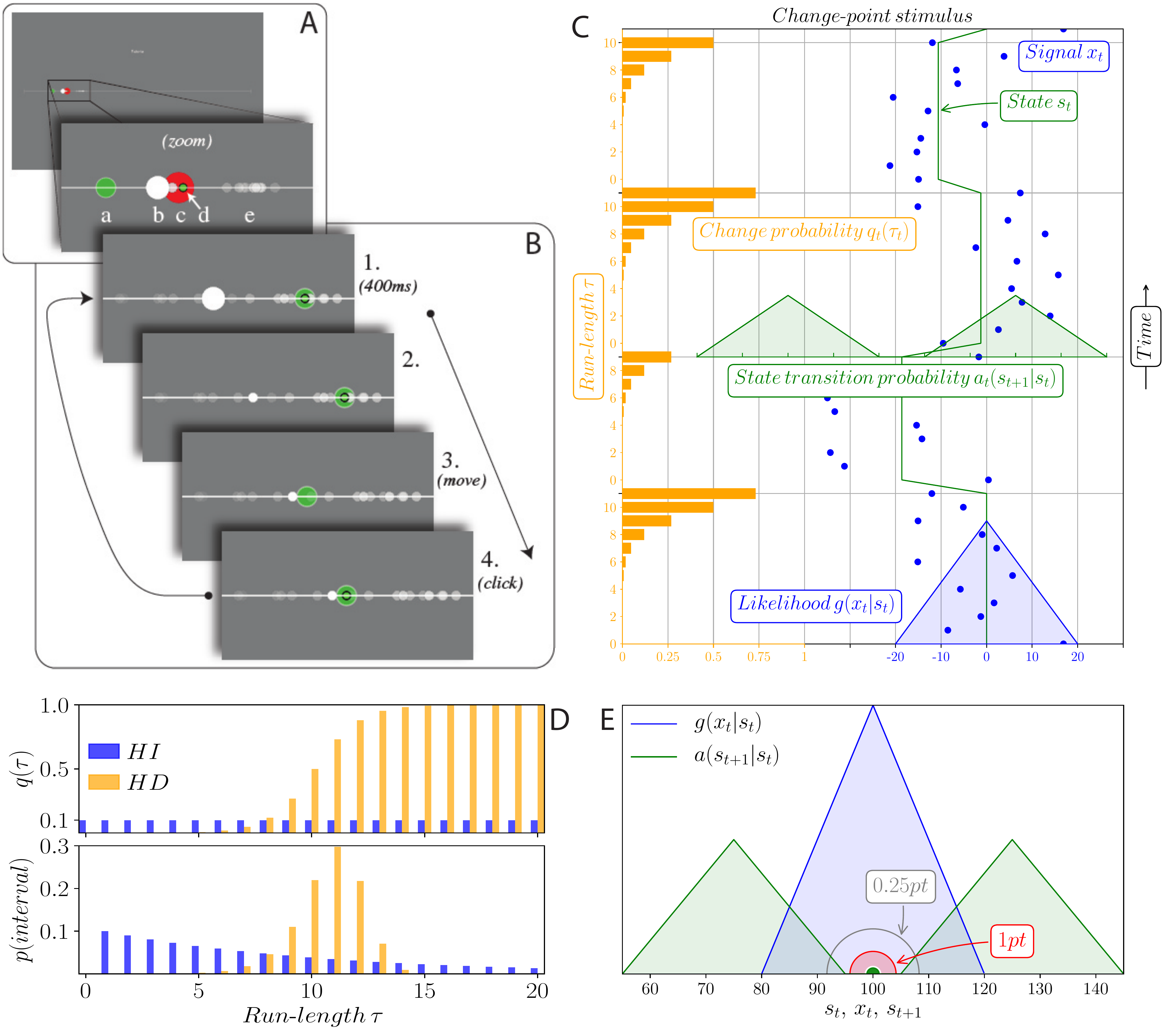}
\caption{\textbf{Inference task and change probability, $q$, in the HI and HD conditions.}
	\textbf{A.} The various elements in the task appear on a horizontal white line in the middle of a gray screen. \textbf{a:} subject's pointer (green disk). \textbf{b:} new stimulus (white disk). \textbf{c:} state (red disk, only shown during tutorial runs). \textbf{d:} position of subject's previous click (green dot). \textbf{e:} for half of subjects, previous stimuli appear as dots decaying with time.
	\textbf{B.} Successive steps of the task: \textbf{1, 2:} a new stimulus is displayed; to attract the subject's attention, it appears as a large, white dot for 400ms, after which it becomes smaller. \textbf{3:} the subject moves the pointer. \textbf{4:} The subject clicks to provide an estimate of the position of the state. After 100ms, a new stimulus appears, initiating the next trial.
	\textbf{C.} The position of the stimulus on the horizontal axis, $x_t$, is generated randomly around the current state, $s_t$, according to the triangle-shaped likelihood function, $g(x_t|s_t)$. The state itself is constant except at change points, at which a new state, $s_{t+1}$, is generated around $s_t$ from the bimodal-triangle-shaped transition probability, $a(s_{t+1}|s_t)$. The run-length, $\tau_t$, is defined as the number of trials since the last change point. Change points occur with the change probability $q(\tau_t)$ (orange bars), which depends on the run-length in the HD condition (depicted here).
	\textbf{D.} Top panel: change probability, $q(\tau)$, as a function of the run-length, $\tau$. It is constant and equal to 0.1 in the HI condition, while it increases with the run-length in the HD condition. Consequently, the distribution of intervals between two consecutive change points (bottom panel) is geometric in the HI condition whereas it is peaked in the HD condition; in both conditions, the average duration of inter-change-point intervals is 10.
	\textbf{E.} Compared extents of the likelihood, $g(x_t|s_t)$ (green), the state transition probability, $a(s_{t+1}|s_t)$ (blue), the `shot' resulting from a click (green dot), and the radii of the 1-point (red disk) and 0.25-point (grey circle) reward areas. A shot overlapping the red (gray) area yields 1 (0.25) point.
	} 
\label{fig:task}
\end{figure}

In our computer-based task, subjects are asked to infer, at successive trials, $t$, the location, on a computer screen, of a hidden point, the state, $s_t$, based on an on-screen visual stimulus, $x_t$, presented as a white dot on a horizontal line (Fig. \ref{fig:task}A,B). Subjects can only observe the white dots, whose positions are generated around the hidden state according to a likelihood probability, $g(x_t | s_t)$ (Fig. \ref{fig:task}C,E, blue distribution).
The state itself, $s_t$, follows a change-point process, i.e., it is constant except when it `jumps' to a new location, which happens with probability $q_t$ (the `hazard rate' or `change probability'). The dynamics of change points are, hence, determined by the change probability, $q_t$.
To examine the behavior of models and human subjects in different `environments', we choose two kinds of signals which differ in their temporal structure.
History-independent (\textbf{HI}) signals are memoryless, Poisson signals: $q_t$ is constant and equal to $0.1$. Consequently, the intervals between two change points last, on average, for 10 trials, and the distribution of these intervals is geometric (Fig. \ref{fig:task}D, blue bars). Conversely, history-dependent (\textbf{HD}) signals are characterized by temporal correlation. Change points also occur every 10 trials, on average, but the distribution of the duration of inter-change-point intervals is peaked around 10. This corresponds to a change probability, $q_t$, that is an increasing function of the number of trials since the last change point — a quantity referred to as the `run-length', $\tau_t$. We thus denote it by $q(\tau_t)$. In HD signals, change points occur in a manner similar to a `jittered periodic' process, though the regularity is not readily detected by subjects.

When a change point occurs, the state randomly jumps to a new state, $s_{t+1}$, according to a state transition probability, $a(s_{t+1} | s_t)$ (Fig. \ref{fig:task}C,E, green distribution). The likelihood, $g$, and the state transition probability, $a$, overlap, thus allowing for ambiguity when a new stimulus is viewed: is it a random excursion about the current state, or has the state changed? At each trial, subjects click with a mouse to give their estimate, $\hat s_t$, of the state. The reward they receive for each response is a decreasing function, $R$, of the distance between the state and the estimate, $|\hat s_t - s_t|$: one reward point if the estimate falls within a given, short distance from the state, 0.25 point if it falls within twice that distance, and 0 point otherwise (Fig. \ref{fig:task}E).
The task is presented as a game to subjects: they are told that someone is throwing snowballs at them. They cannot see this hidden person (whose location is the state, $s_t$), but they observe the snowballs as white dots on the screen (the stimulus, $x_t$). After several tutorial runs (in some of which the state is shown), they are instructed to use the snowballs to guess the location of the person (i.e., produce an estimate, $\hat s_t$). Additional details on the task are provided in \nameref{sec:Methods}.

\subsection{Learning rates adapt to the temporal statistics of the stimulus}

\begin{figure}[!ht]
\centering
       \includegraphics[width=\linewidth]{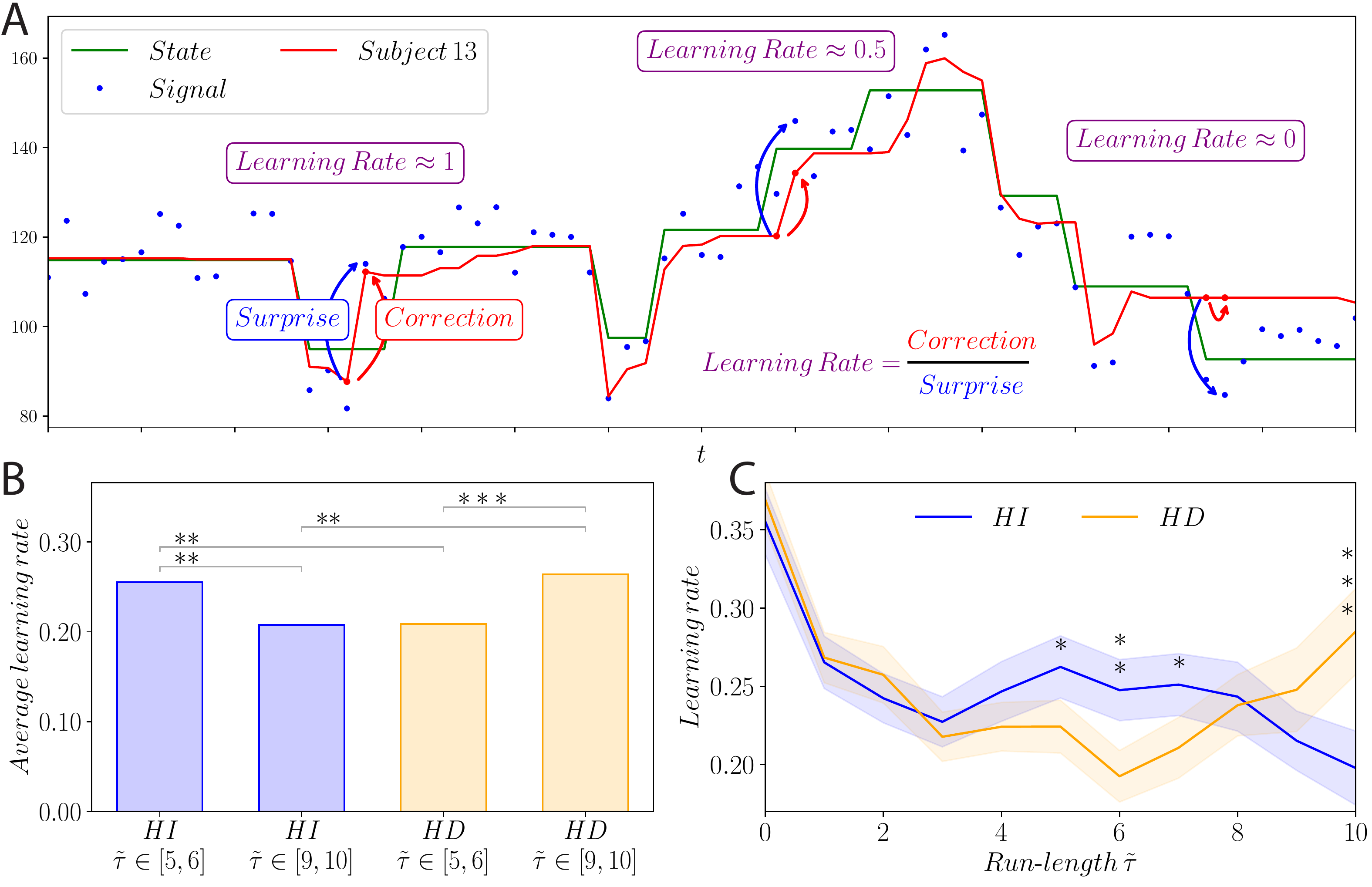}
       \caption{\textbf{Human learning rates depends on the temporal statistics (HI or HD) of the stimulus.}
       \textbf{A.} Illustration of the learning rate using a sample of a subject's responses (red line). The `surprise' (blue arrow) is the difference, $x_{t+1} - \hat s_t$, between the estimate at trial $t$, $\hat s_t$ (red), and the new stimulus at trial $t+1$, $x_{t+1}$ (blue). The `correction' (red arrow) is the difference between the estimate at trial $t$ and the estimate at trial $t+1$, $\hat s_{t+1} - \hat s_t$. The `learning rate' is the ratio of correction to surprise.
       \textbf{B.} Average learning rates in HI (blue) and HD (orange) conditions, at short run-lengths ($\tilde \tau \in [5,6]$) and long run-lengths ($\tilde \tau \in [9,10]$). In the HD condition the change probability increases with the run-length, which advocates for higher learning rates at long run-lengths.
       \textbf{C.} Average learning rates in HI (blue) and HD (orange) conditions, vs. run-length $\tilde \tau$. Shaded bands indicate the standard error of the mean.
       \textbf{B},\textbf{C.} Stars indicate p-values of one-sided Welch's t-tests, which do not assume equal population variance. Three stars: p < 0.01; two stars: p < 0.05; one star: p < 0.1. Bonferroni-Holm correction \parencite{Holm1978} is applied in panel \textbf{B}.
       }
       \label{fig:lrs}
\end{figure}

A typical example of a subject's responses is displayed in Fig. \ref{fig:lrs}A. To describe the data, we focus, throughout this paper, on three quantities: the learning rate, defined as the ratio of the `correction', $\hat s_{t+1} - \hat s_t$, to the `surprise', $x_{t+1} - \hat s_t$; the repetition propensity, defined as the proportion of trials in which the learning rates vanishes ($\hat s_{t+1}=\hat s_t$); and the standard deviation of the responses of the subjects.
The learning rate represents a normalized measure of the susceptibility of a subject to a new stimulus. If the new estimate, $\hat s_{t+1}$, is viewed as a weighted average of the previous estimate, $\hat s_t$, and the new stimulus, $x_{t+1}$, the learning rate is the weight given to $x_{t+1}$. A learning rate of $0$ means that the subject has not changed its estimate upon observing the new stimulus; a learning rate of $0.5$ means that the new estimate is equidistant from the previous estimate and the new stimulus; and a learning rate of $1$ means that the new estimate coincides with the new stimulus, and the past is ignored (Fig. \ref{fig:lrs}A).

Our data show that for human subjects the learning rate is not constant, and can vary from no correction at all (learning rate $\approx$ 0) to full correction (learning rate $\approx$ 1). We investigated how the average learning rate behaved in relation to the run-length, in the HI and HD conditions. As the run-length is not directly accessible to subjects, in our analyses we used the empirical run-length, $\tilde \tau$, a similar quantity derived from the subjects' data (see \nameref{sec:Methods}). Unless otherwise stated, we focus our analyses on cases in which the surprise, $x_{t+1} - \hat s_t$, is in the [8,18] window, in which there is appreciable ambiguity in the signal.

A first observation emerging from our data is that the learning rate changes with the run-length, in a quantitatively different fashion depending on the condition (HI or HD). In the HI condition, learning rates at short run-length ($\tilde \tau \in [5,6]$) are significantly higher than at long run-length ($\tilde \tau \in [9,10]$), i.e., the learning rate decreases with run-length (Fig. \ref{fig:lrs}B, blue bars). In the HD condition, the opposite occurs: learning rates are significantly higher at long run-lengths (Fig. \ref{fig:lrs}B, orange bars), indicating that subjects modify their inference depending on the temporal structure of the signal. In addition, at short run-lengths, learning rates are significantly lower in the HD condition than in the HI condition; this suggests that subjects take into account the fact that a change is less likely at short run-lengths in the HD condition. The opposite holds at long run-lengths: HD learning rates are markedly larger than HI ones (Fig. \ref{fig:lrs}B).

Inspecting the dependence of the learning rate on the run-length (Fig. \ref{fig:lrs}C), we note that the HD learning-rate curve adopts a `smile shape', unlike the monotonic curve in the HI condition. (A statistical analysis confirms that these curves have significantly different shapes; see \nameref{sec:Methods}.)
The HI curve is consistent with a learning rate that simply decreases as additional information is accumulated on the state. In the HD condition, initially the learning rate is suppressed, then boosted at longer run-lengths, reflecting the modulation in the change probability.

These observations demonstrate that subjects adapt their learning rate to the run-length, and that in the HD condition subjects make use of the temporal structure in the signal. These results are readily intuited: shortly after a change point, the learning rate should be high, as little is known about the new state, while at longer run-lengths the learning rate should tend to zero as the state is more and more precisely inferred. This decreasing behavior is observed, but only in the HI condition. The HD condition introduces an opposing effect: as the run-length grows, new stimuli are increasingly likely to divulge the occurrence of a new state, which advocates for adopting a higher learning rate. This tradeoff is reflected in our data in the `smile shape' of the HD learning-rate curve (Fig. \ref{fig:lrs}C; these trends subsist at longer run-lengths, see Supplementary Fig. \ref{fig:lr-vs-rl-longer-rls}.). The increase in learning rate at long run-lengths is reminiscent of the behavior of a driver waiting at a red light: as time passes, the light is increasingly likely to turn green; as a result, the driver is increasingly susceptible to react and start the car.

\begin{figure}[hbt]
\centering
	\includegraphics[width=0.5\linewidth]{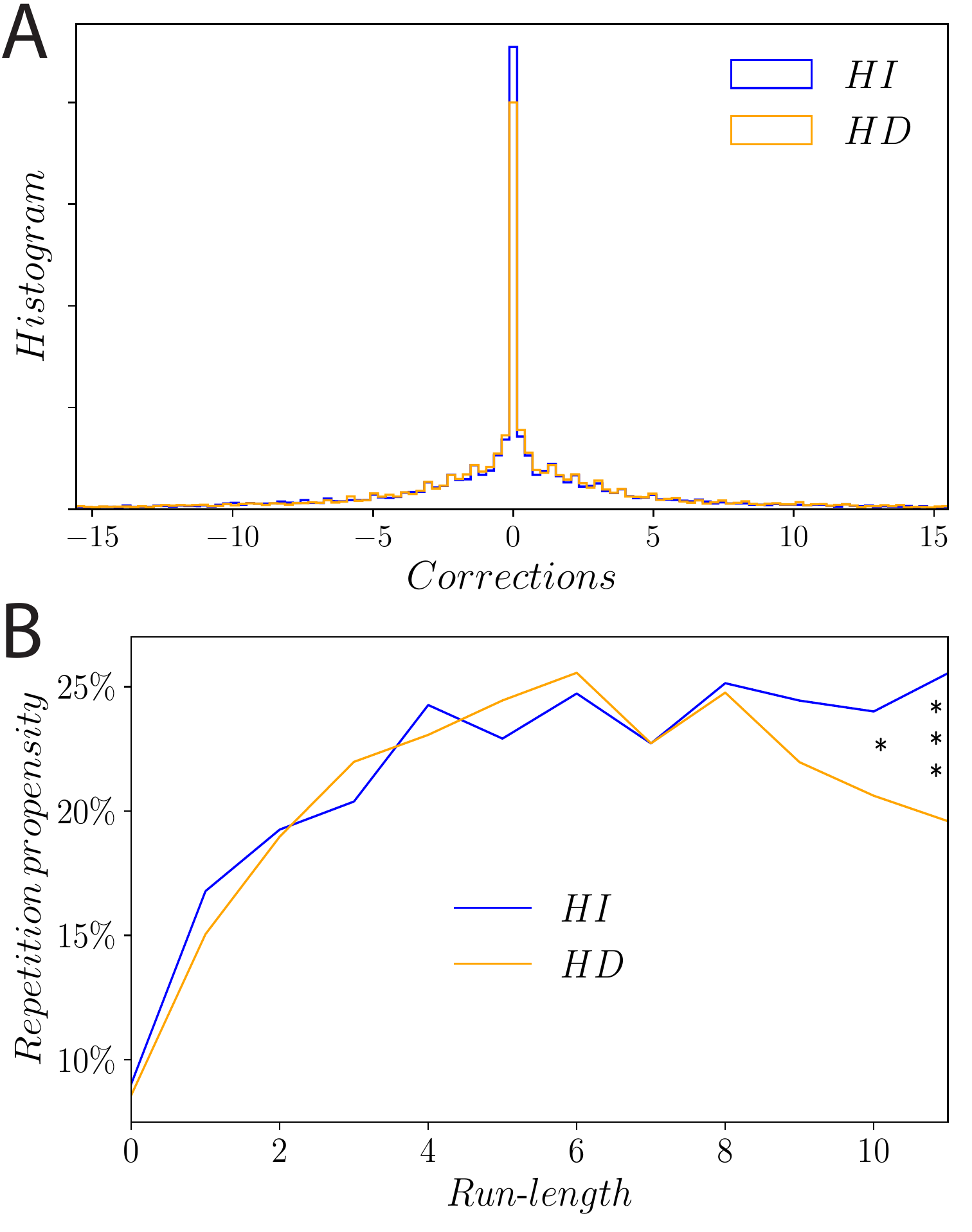}
	\caption{\textbf{Human repetition propensity depends on the temporal statistics, and dynamically on the run-length.}
	\textbf{A.} Histogram of subject corrections (difference between two successive estimates, $\hat s_{t+1} - \hat s_t$), in the HI (blue) and HD (orange) conditions. The width of bins corresponds to one pixel on the screen, thus the peak at zero represents the repetition events ($\hat s_{t+1} = \hat s_t$).
	\textbf{B.} Repetition propensity, i.e., proportion of occurrences of repetitions in the responses of subjects, as a function of run-length, in the HI (blue) and HD (orange) conditions. Stars indicate p-values of Fisher’s exact test of equality of the repetition propensities between the two conditions, at each run-length.
	}
	\label{fig:repeat}
\end{figure}

\subsection{Human repetition propensity}

A closer look at the data presented in the previous section reveals that in a number of trials the learning rate vanishes, i.e., $\hat s_{t+1}=\hat s_t$. The distribution of the subjects' corrections, $\hat s_{t+1} - \hat s_t$, exhibits a distinct peak at zero (Fig. \ref{fig:repeat}A). In other words, in a fraction of trials, subjects click twice consecutively on the same pixel. We call such a response a `repetition', and the fraction of repetition trials the `repetition propensity'. The latter varies with the run-length: it increases with $\tau$ in both HI and HD conditions, before decreasing in the HD condition for long run-lengths (Fig. \ref{fig:repeat}B). 

What may cause the subjects' repetition behavior? The simplest explanation is that, after observing a new stimulus, a subject may consider that the updated best estimate of the state lands on the same pixel as in the previous trial. The width of one pixel in arbitrary units of our state space is 0.28. As a comparison, the triangular likelihood, $g$, has a standard deviation, $\sigma_g$, of 8.165. An optimal observer estimating the center of a Gaussian density of standard deviation $\sigma_g$, using 10 samples from this density, comes up with a posterior density with standard deviation $\sigma_g/\sqrt{10} \approx 2.6$. Therefore, after observing even 10 successive stimuli, the subjects' resolution is not as fine as a pixel (it is, in fact, 10 times coarser). This indicates that the subjects' repetition propensity is higher than the optimal one (the behavior of the optimal model, presented below, indeed exhibits a lower average repetition propensity than that of the subjects). Another possible explanation is that even though the new estimate falls on a nearby location, a motor cost prohibits a move if it is not sufficiently extended to be `worth it' \parencite{Morasso1981, Wolpert1997, Shadmehr2009, Rigoux2012}. A third, heuristic explanation is that humans are subject to a `repetition bias' according to which they repeat their response irrespective of their estimate of the state.

Regardless of its origin, the high repetition propensity in data raises the question of whether it dominates the behavior of the average learning rate. As a control, we excluded all occurrences of repetitions in subjects' data and carried out the same analyses on the truncated dataset. We reached identical conclusions, namely, significant discrepancies between the HI and HD learning rates at short and long run-lengths, albeit with, naturally, higher average rates overall (see Supplementary Fig. \ref{fig:results_wo_repeats}).

\subsection{The variability in subjects' responses evolves over the course of inference}

\begin{figure}[!ht]
\centering
	\includegraphics[width=0.5\textwidth]{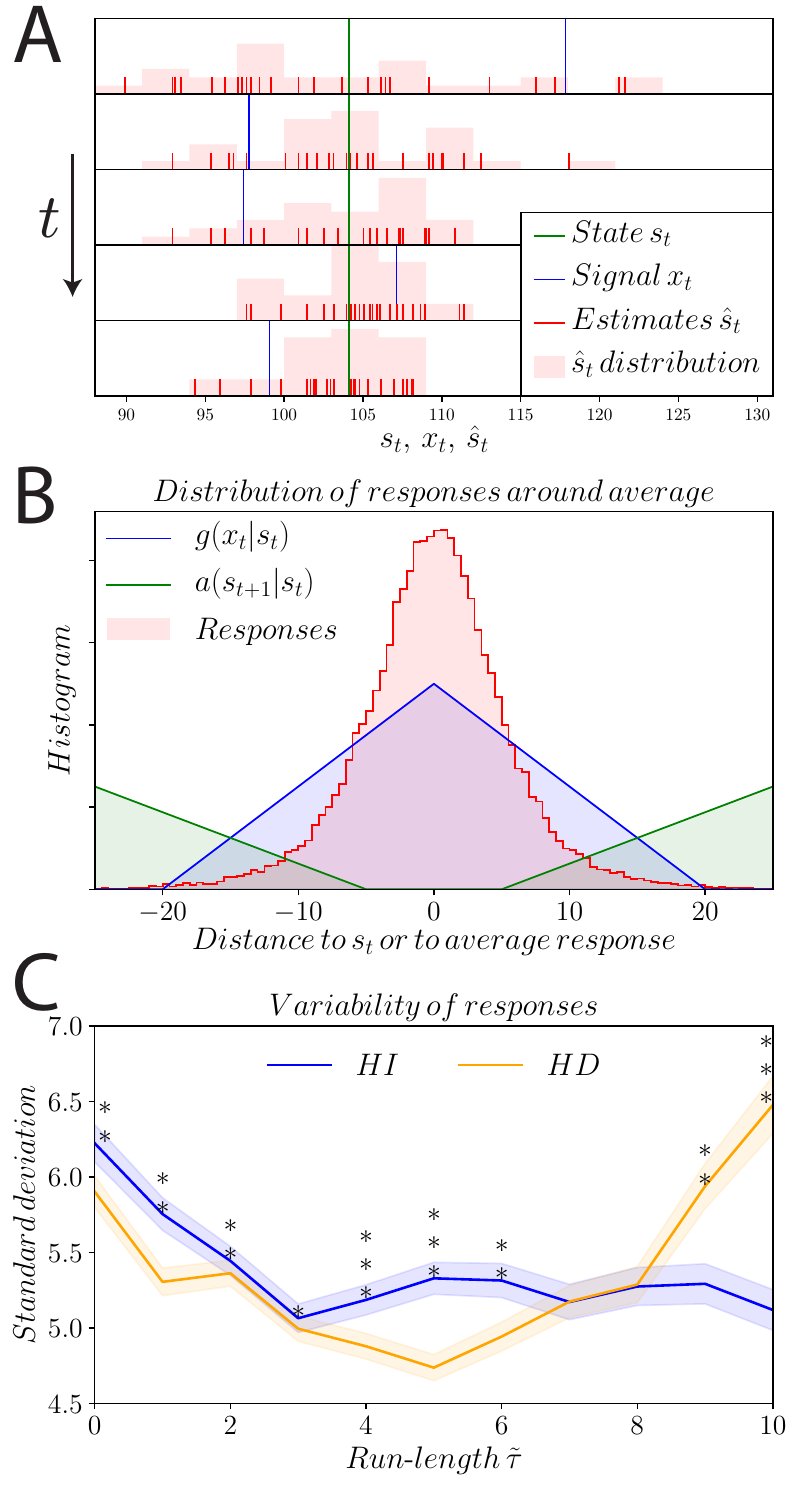}
	\caption{\textbf{The variability in subjects' responses is modulated during inference, and these modulations depend on the temporal statistics of the stimulus.}
	\textbf{A.} Responses of subjects in an example of 5 consecutive stimuli. In this example, there is no change point and the state (green) is constant. At each trial (from top to bottom), subjects observe the stimuli (blue) and provide their responses (red bars). A histogram of the locations of the responses is obtained by counting the number of responses in bins of width 3 (light red).
	\textbf{B.} Distribution of the responses of subjects around their average (red), compared to the likelihood, $g$ (blue), and the state transition probability, $a$ (green).
	\textbf{C.} Standard deviation of the responses of subjects vs. run-length, $\tilde \tau$, in the HI (blue) and HD (orange) conditions. Stars indicate p-value of Levene's test of equality of variance between the two conditions, at each $\tilde \tau$. Shaded bands indicate the standard error of the standard deviation \parencite{Ahn2003}.}
	\label{fig:variability}
\end{figure}

In the previous two sections, we have examined two aspects of the distribution of responses: the average learning rate and the probability of response repetition. We now turn to the variability in subjects' responses. Although all subjects were presented with identical series of stimuli, $x_t$, their responses at each trial were not the same (Fig. \ref{fig:variability}A). This variability appears in both HI and HD conditions. The distribution of responses around their averages at each trial has a width comparable to that of the likelihood distribution, $g(x_t|s_t)$ (Fig. \ref{fig:variability}B). More importantly, the variability in the responses (as measured by the standard deviation) is not constant, but decreases with successive trials following a change point, at short run-lengths (Fig. \ref{fig:variability}C). Comparing the HI and HD conditions, we observe that for run-lengths shorter than 7, the standard deviation in the HD condition is significantly lower than that in the HI condition. At longer run-lengths, the two curves cross and the variability in the HD condition becomes significantly higher than in the HI condition. The HD curve adopts, again, a `smile shape' (Fig. \ref{fig:variability}C). What is the origin of the response variability? Because it changes with the run-length and the HI vs. HD condition, it cannot be explained merely be the presence of noise independent from the inference process, such as pure motor noise. In order to encompass human behavior in a theoretical framework and to investigate potential sources of this inference-dependent variability, we start by comparing the recorded behavior with that of an optimal observer.

\subsection{Optimal estimation: Bayesian update and maximization of expected reward}

We derive the optimal solution for the task of estimating the hidden state, $s_t$, given random stimuli, $x_t$. The first step (the `inference step') is to derive the optimal posterior distribution over the state, $s_t$, using Bayes' rule. Because the state is a random variable coupled with the run-length, $\tau_t$, another random variable, it is convenient to derive the Bayesian update equation for the $(s_t, \tau_t)$ pair (more precisely, the $(s_t, \tau_t)$ pair verifies the Markov property, whereas $s_t$ alone does not, in the HD condition). We denote by $x_{1:t}$ the string of stimuli received between trial 1 and trial $t$, and by $p_t(s, \tau | x_{1:t})$ the probability distribution over $(s, \tau)$, at trial $t$, after having observed the stimuli $x_{1:t}$. At trial $t+1$, Bayes' rule yields
$p_{t+1}(s, \tau | x_{1:t+1}) \propto g(x_{t+1} | s) p_{t+1}(s, \tau | x_{1:t})$.
Furthermore, we have the general transition equation,
\begin{equation}
p_{t+1}(s, \tau | x_{1:t}) = \sum_{\tau_t} \int_{s_t} p_{t+1}(s, \tau | s_t, \tau_t ) p_t( s_t, \tau_t | x_{1:t} ) \D s_t,
\end{equation}
given by the change-point statistics.
As the transition probability, $p_{t+1}(s, \tau | s_t, \tau_t )$, can be expressed using $q(\tau_t)$ and $a(s|s_t)$ (see \nameref{sec:Methods} for details), we can reformulate the update equation as
\begin{equation}\label{eq:HDupdate}
\begin{split}
p_{t+1}(s, \tau | x_{1:t+1}) = \frac{1}{Z_{t+1}} g(x_{t+1} | s ) \Bigg[
    & \mathbbm{1}_{\tau=0} \, \sum_{\tau_t}  q(\tau_t) \int_{s_t} a(s | s_t) p_t( s_t, \tau_t | x_{1:t} ) \D s_t \\
 + & \mathbbm{1}_{\tau>0} \, ( 1 - q(\tau-1)) p_t( s, \tau - 1 | x_{1:t} ) \Bigg],
\end{split}
\end{equation}
where $\mathbbm{1}_{C} = 1$ if condition $C$ is true, 0 otherwise; and $Z_{t+1}$ is a normalization constant. This equation includes two components: a `change-point' one ($\tau=0$) and a `no change-point' one ($\tau>0$). We call the model that performs this Bayesian update of the posterior the \textbf{\textit{OptimalInference}} model.

Finally, following the inference step just presented (i.e., the computation of the posterior), a `response-selection step' determines the behavioral response.
At trial $t$ and for a response $\hat s_t$, the expected reward is $\E_s R = \int R(|\hat s_t - s|) p_t(s | x_{1:t}) \D s$. The optimal strategy selects the response, $\hat s_t$, that maximizes this quantity. Before exploring the impact of relaxing the optimality in the inference step, in the response-selection step, or both, we examine, first, the behavior of the optimal model.

\begin{figure}
        \includegraphics[width=\linewidth]{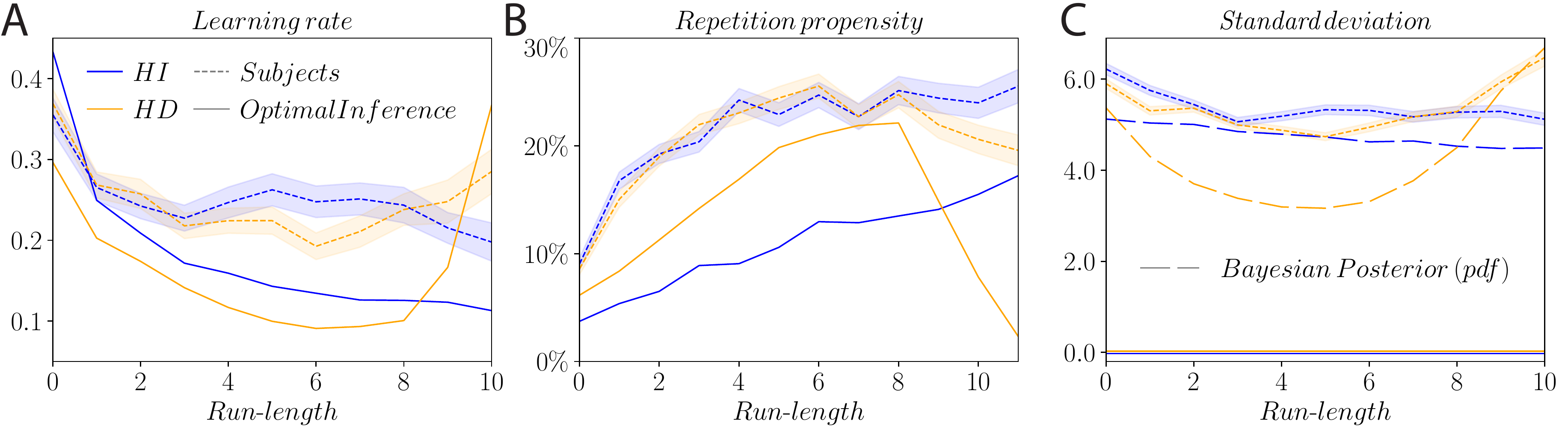}
	\caption{\textbf{The optimal model captures qualitatively the behavior of the learning rate and of the repetition propensity in subjects, but does not account for their variability.
	}
	\textbf{A.} Average learning rate as a function of the run-length. In the HI condition, the learning rate decreases with the run-length, for both the optimal model and the subjects. In the HD condition, learning rates in the optimal model are lower than in the HI condition, for short run-lengths, and higher for long run-lengths. The learning rate of subjects exhibits a similar smile shape, in the HD condition.
	\textbf{B.} Repetition propensity, i.e., proportion of repetition trials, as a function of the run-length.
	\textbf{C.} Standard deviation of the responses of the subjects (dashed lines) and of the optimal model (solid lines), and standard deviation of the optimal, Bayesian posterior distribution (long dashes), as a function of the run-length. The optimal model is deterministic and, thus, exhibits no variability in its responses. The optimal posterior distribution, however, has a positive standard deviation which decreases with the run-length, in the HI condition, and exhibits a smile shape, in the HD condition.
	}\label{fig:metrics}
\end{figure}

\subsection{The optimal model captures qualitative trends in learning rate and repetition propensity}

Equipped with the optimal model for our inference task, we compare its output to experimental data. For short run-lengths ($\tau < 8$), the learning rates in both HI and HD conditions decrease as a function of the run-length, and the HD learning rates are lower than their HI counterparts. They increase, however, at longer run-lengths ($\tau \geq 8$) and ultimately exceed the HI learning rates; these, by contrast, decrease monotonically (Fig. \ref{fig:metrics}A, solid line). These trends are similar to those observed in behavioral data (Fig. \ref{fig:metrics}A, dashed line). Hence, the modulation of the subjects' learning rates with the temporal statistics of the stimuli, and over the course of inference, is consistent, at least qualitatively, with that of a Bayesian observer.

Although a Bayesian observer can, in principle, hold a continuous posterior distribution, we discretize, instead, the posterior, in order to reproduce the experimental condition of a pixelated screen. This discretization allows for repetitions. The repetition propensity of the optimal model varies with the run-length: it increases with $\tau$ in both HI and HD conditions, and decreases in the HD condition for long run-lengths, a pattern also found in experimental data (Fig. \ref{fig:metrics}B).

Hence, the optimal model captures the qualitative trends in learning rate and repetition propensity present in the responses of the subjects. Quantitative differences, however, remain. The learning rates of the subjects, averaged over both HI and HD conditions, are 43\% higher than the average learning rate in the optimal model, and the average repetition propensity of the subjects is 9 percentage points higher than that in the optimal model.

\subsection{Relation between human response variability and the Bayesian posterior}

\begin{figure}
        \includegraphics[width=\linewidth]{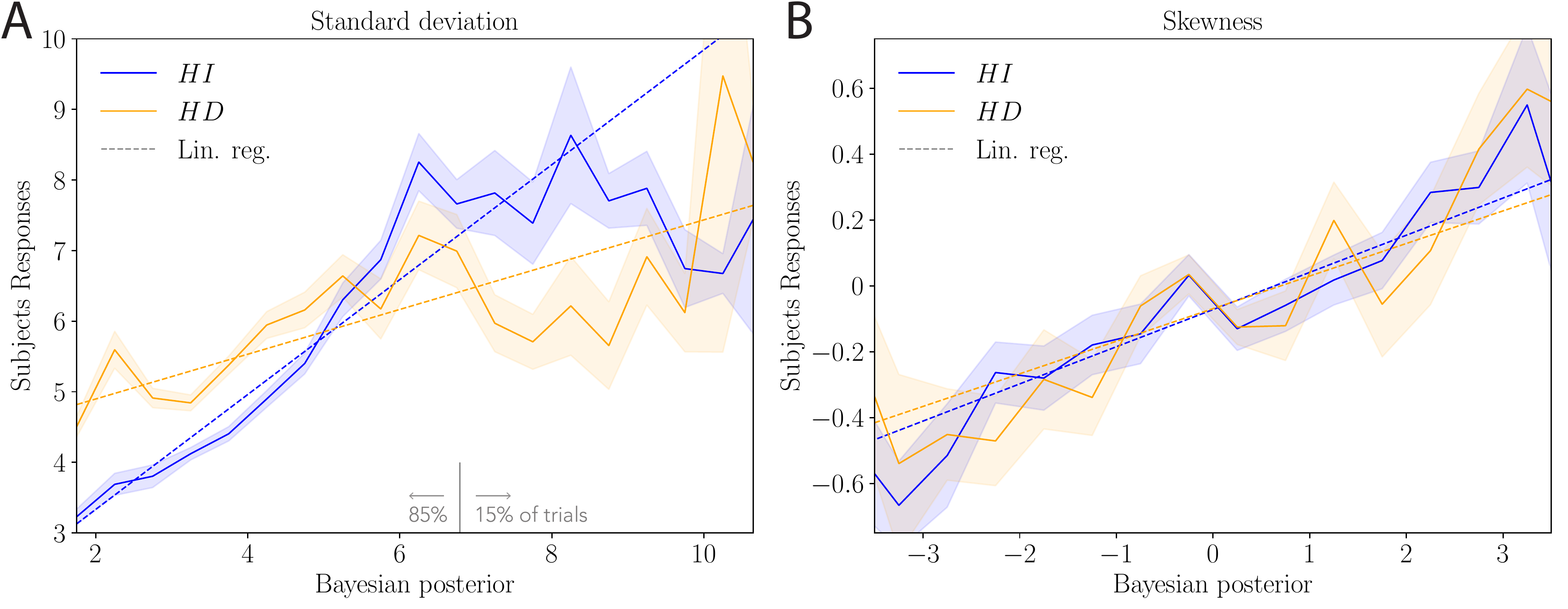}
	\caption{\textbf{Both width and skewness of the distribution of subjects' responses are correlated with those of the Bayesian posterior.} Empirical standard deviation \textbf{(A)} and skewness \textbf{(B)} of subjects' responses as a function of the standard deviation and skewness of the Bayesian posterior, in the HI (blue) and HD (orange) conditions, and linear regressions (ordinary least squares; dashed lines). On $85\%$ of trials, the standard deviation of the Bayesian posterior is lower than $6.8$ (vertical grey line). Shaded bands indicate the standard error of the mean.
	}\label{fig:std_skewness}
\end{figure}

The optimal model captures qualitatively the modulations of learning rate and repetition propensity in subjects, but it is deterministic (at each trial, the optimal estimate is a deterministic function of past stimuli) and, as such, it does not capture the variability inherent to the behavior of subjects. The modulations of the behavioral variability as a function of the run-length and of the temporal structure of the signal (Fig. \ref{fig:variability}C) is a sign that the variability evolves as the inference process unfolds. The standard deviation of the optimal Bayesian posterior decreases with the run-length, in the HI condition: following a change point, the posterior becomes narrower as new stimuli are observed. In the HD condition, the standard deviation of the posterior exhibits a `smile shape' as a function of the run-length: it decreases until the run-length reaches 5, then increases for larger run-lengths (Fig. \ref{fig:metrics}C). This behavior is similar to that of the standard deviation of the responses of the subjects. In fact, the standard deviation of the Bayesian posterior and that of subjects' responses across trials are significantly correlated, both in the HI condition (Pearson's $r=.53$, $p<.0001$) and in the HD condition ($r=.25$, $p<.0001$). In other words, when the Bayesian posterior is wide there is more variability in the responses of subjects, and vice-versa (Fig. \ref{fig:std_skewness}A).

Turning to higher moments of the distribution of subjects' responses, we find that the skewness of this distribution appears, also, to grow in proportion to the skewness of the Bayesian posterior (Fig. \ref{fig:std_skewness}B). The correlation between these two quantities is positive and significant in the two conditions (HI: $r=.21$, $p<.0001$; HD: $r=.14$, $p<.0001$. These results are not driven by the boundedness of the response domain, which could have artificially skewed the distribution of response; see Supplementary Fig. \ref{fig:subj_skew_vs_p_skew_restricted}.). Thus, not only the width, but also the asymmetry in the distribution of subjects' responses is correlated with that of the Bayesian posterior. These observations support the hypothesis that the behavioral variability in the data is at least in part related to the underlying inference and decision processes.

In what follows, we introduce an array of suboptimal models, with the aim of resolving the qualitative and quantitative discrepancies between the behavior of the optimal model and that of the subjects. In particular, we formulate several stochastic models which include possible sources of behavioral variability. Two scenarios are consistent with the modulations of the magnitude and asymmetry of the variability with the width and skewness of the Bayesian posterior: stochasticity in the inference step (i.e., in the computation of the posterior) and stochasticity in the response-selection step (i.e., in the computation of an estimate from the posterior). The models we examine below cover both these scenarios.

\subsection{Suboptimal models reflecting cognitive limitations}

In the previous sections, we have examined the learning rate of the subjects, their repetition propensity, and the variability in their responses; comparison of the behaviors of these quantities to that of the Bayesian, optimal model, revealed similarities (namely, the qualitative behaviors of the learning rate and of the repetition propensity) and discrepancies (namely, quantitative differences in these two quantities, and lack of variability in the optimal model). Although the latter call for a non-Bayesian account of human behavior, the former suggest not to abandon the Bayesian approach altogether (in favor, for instance, of \textit{ad hoc} heuristics). Thus, we choose to examine a family of sub-optimal models obtained from a sequence of deviations away from the Bayesian model, each of which captures potential cognitive limitations hampering the optimal performance.

In the Bayesian model, three ingredients enter the generation of a response upon receiving a stimulus: first, the belief on the structure of the task and on its parameters; second, the inference algorithm which produces a posterior on the basis of stimuli; third, the selection strategy which maps the posterior into a given response.
The results presented above, exhibiting the similarity between the standard deviation of the Bayesian posterior and the standard deviation of the responses of the subjects (Fig. \ref{fig:metrics}C), points to a potential departure from the optimal selection strategy, in which the posterior is sampled rather than maximized. This sampling model, which we implement (see below), captures qualitatively the modulated variability of responses; sizable discrepancies in the three quantities we examine, however, remain (see \nameref{sec:Methods}). Hence, we turn to the other ingredients of the estimation process, and we undertake a systematic analysis of the effects on behavior of an array of deviations away from optimality.

Below, we provide a conceptual presentation of the resulting models; we fit them to experimental data, and comment on what the best-fitting models suggest about human inference and estimation processes. For the detailed mathematical descriptions of the models, and an analysis of the ways in which their predictions depart from the optimal behavior, we refer the reader to the \nameref{sec:Methods} section.

\begin{figure}[!ht]
        \includegraphics[width=\linewidth]{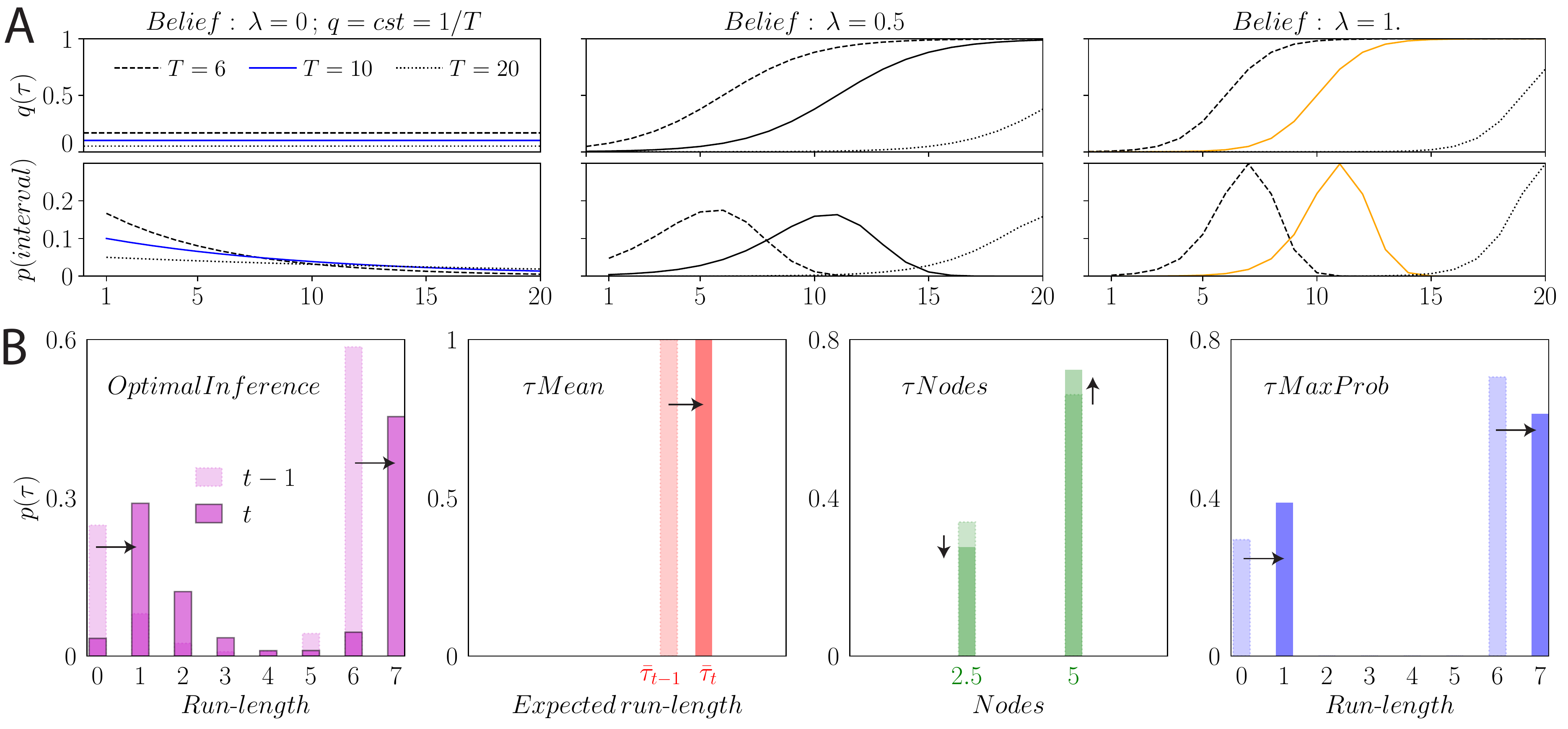}
	\caption{\textbf{Illustration of the erroneous beliefs in the \textit{IncorrectQ} model and of the approximations made in the $\tau$\textit{Mean}, $\tau$\textit{Nodes}, and $\tau$\textit{MaxProb} models.}
	\textbf{A.} Change probability, $q(\tau)$, as a function of the run-length (first row), and distribution of intervals between two consecutive change points (second row), for various beliefs on the parameters of the change probability: the slope, $\lambda$, and the average duration of intervals, $T$. For a vanishing slope ($\lambda=0$), the change probability is constant and equal to $1/T$ (first panel). With $T=10$ this corresponds to the HI condition (blue lines). For a positive slope ($\lambda>0$), the change probability increases with the run-length (i.e., a change-point becomes more probable as the time since the last change-point increases), and the distribution of intervals between two successive change-points is peaked. The HD condition (orange lines) corresponds to $\lambda=1$ and $T=10$.
	\textbf{B.} Schematic illustration of the marginal distribution of the run-length, $p(\tau)$, in each model considered. The \textit{OptimalInference} model assigns a probability to each possible value of the run-length, $\tau$, and optimally updates this distribution upon receiving stimuli (first panel). The $\tau$\textit{Mean} model uses a single run-length which tracks the inferred expected value, $\bar\tau_t$ (second panel). The $\tau$\textit{Nodes} model holds in memory a limited number, $N_\tau$, of fixed hypotheses on $\tau$ (``nodes''), and updates a probability distribution over these nodes; $N_\tau=2$ in this example (third panel). The $\tau$\textit{MaxProb} model reduces the marginal distribution by discarding less likely run-lengths; in this example, 2 run-lengths are stored in memory at any given time (fourth panel).
	}\label{fig:subopt_models}
\end{figure}

\paragraph{Models with erroneous beliefs on the statistics of the signal}
Our first model challenges the assumption, made in the optimal model, of a perfectly faithful representation of the set of parameters governing the statistics of the signal. Although subjects were exposed in training phases to blocs of stimuli in which the state, $s_t$, was made visible, they may have learned the parameters of the generative model incorrectly. We explore this possibility, and, here, we focus on the change probability, $q(\tau)$, which governs the dynamics of the state. (We found that altering the value of this parameter had a stronger impact on behavior than altering the values of any of the other parameters.)
In the HD condition, $q(\tau)$ is a sigmoid function shaped by two parameters: its slope, $\lambda=1$, which characterizes `how suddenly' change points become likely, as a function of $\tau$; and the average duration of inter-change-points intervals, $T=10$. In the HI condition, $q=0.1$ is constant; it can also be interpreted as an extreme case of a sigmoid in which $\lambda=0$ and $T=1/q=10$. We implement a suboptimal model, referred to as \textbf{\textit{IncorrectQ}}, in which these two quantities, $\lambda$ and $T$, are treated as free parameters, thus allowing for a broad array of different beliefs in the temporal structure of the signal (Fig. \ref{fig:subopt_models}A).

\paragraph{Models with limited memory}
Aside from operating with an inexact representation of the generative model, human subjects may use a suboptimal form of inference. In the HD condition, the optimal model maintains `in memory' a probability distribution over the entire $(s, \tau)$-space (see Eq. (\ref{eq:HDupdate})), thus keeping track of a rapidly increasing number of possible histories, each characterized by a sequence of run-lengths. Such a process entails a large computational and memory load. We explore suboptimal models that alleviate this load by truncating the number of possible scenarios stored in memory; this is achieved through various schemes of approximations of the posterior distribution.
More specifically, in the following three suboptimal models, the true (marginal) probability of the run-lengths, $p_{t}(\tau|x_{1:t})$, is replaced by an approximate probability distribution.

A first, simple way of approximating the marginal distribution of the run-lengths is to consider only its mean, i.e., to replace it by a Kronecker delta which takes the value $1$ at an estimate of the expected value of the run-lengths. \textcite{Nassar2010} introduce a suboptimal model based on this idea, some details of which depend on the specifics of the task; we implement a generalization of this model, adapted to the parameters of our task. We call it the \textbf{$\tau$\textit{Mean}} model. While the optimal marginal distribution of the run-lengths, $p_t(\tau|x_{1:t})$, spans the whole range of possible values of the run-length, it is approximated, in the $\tau$\textit{Mean} model, by a delta function parameterized by a single value, which we call the `approximate expected run-length' and which we denote by $\bar \tau_t$. Upon the observation of a new stimulus, $x_{t+1}$, the updated approximate expected run-length, $\bar \tau_{t+1}$, is computed as a weighted average between two values of the run-length, $0$ and $\bar \tau_t + 1$, which correspond to the two possible scenarios: with and without a change point at trial $t+1$. Each scenario is weighted according to the probability of a change point at trial $t+1$, given the stimulus, $x_{t+1}$. This model has no free parameter (Fig. \ref{fig:subopt_models}B, second panel).

In a second limited-memory model, contrary to the $\tau$\textit{Mean} model just presented, the support of the distribution of the run-lengths is not confined to a single value. This model generalizes the one introduced by \textcite{Wilson2013}. In this model, the marginal distribution of the run-lengths, $p_t(\tau|x_{1:t})$, is approximated by another discrete distribution defined over a limited set of constant values, called `nodes' (Fig. \ref{fig:subopt_models}B, third panel). We call this model \textbf{$\tau$\textit{Nodes}}. A difference with the previous model ($\tau$\textit{Mean}) is that the support of the distribution is fixed, i.e., the set of nodes remains constant as time unfolds, whereas in the $\tau$\textit{Mean} model the single point of support, $\bar \tau_t$, depends on the stimuli received.
The details of the implementation of this algorithm, and, in particular, of how the approximate marginal distribution of the run-lengths is updated upon receiving a new stimulus, are provided in \nameref{sec:Methods}. The model is parameterized by the number of nodes, $N_\tau$, and the values of the nodes. We implement it with up to five nodes.

The two models just presented are drawn from the literature. We propose a third suboptimal model that relieves the memory load in the inference process. We also approximate, in this model, the marginal distribution of the run-lengths, $p_t(\tau | x_{1:t})$, by another, discrete distribution. We call $N_\tau$ the size of the support of our approximate distribution, i.e., the number of values of the run-length at which the approximate distribution does not vanish. A simple way to approximate $p_t(\tau | x_{1:t})$ is to identify the $N_\tau$ most likely run-lengths, and set the probabilities of the other run-lengths to zero. More precisely, if, at trial $t$, the run-length takes a given value, $\tau_t$, then, upon the observation of a new stimulus, at trial $t+1$ it can only take one of two values: $0$ (if there is a change point) or $\tau_t+1$ (if there is no change point). Hence, if the approximate marginal distribution of the run-lengths at trial $t$ is non-vanishing for $N_\tau$ values, then the updated distribution is non-vanishing for $N_\tau+1$ values. We approximate this latter distribution by identifying the most unlikely run-length, $\arg\min p_{t+1}(\tau | x_{1:t+1})$, setting its probability to zero, and renormalizing the distribution. In other words, at each step, the $N_\tau$ most likely run-lengths are retained while the least likely run-length is eliminated. We call this algorithm \textbf{$\tau$\textit{MaxProb}} (Fig. \ref{fig:subopt_models}B, fourth panel). It is parameterized by the size of the support of the marginal distribution, $N_\tau$, which can be understood as the number of `memory slots' in the model.

\paragraph{A model with limited run-length memory through sampling-based inference}
The five models considered hitherto (\textit{OptimalInference}, \textit{IncorrectQ}, $\tau$\textit{Mean}, $\tau$\textit{Nodes}, and $\tau$\textit{MaxProb}) are deterministic: a given sequence of stimuli implies a given sequence of responses, in marked contrast with the variability exhibited in the responses of subjects. To account for this experimental observation, we suggest several models in which stochasticity is introduced in the generation of a response. Response stochasticity can stem from the inference step, the response-selection step, or both. We present, first, a model with stochasticity in the inference step.

\begin{figure} 
	\thisfloatpagestyle{empty}
        \includegraphics[width=\linewidth]{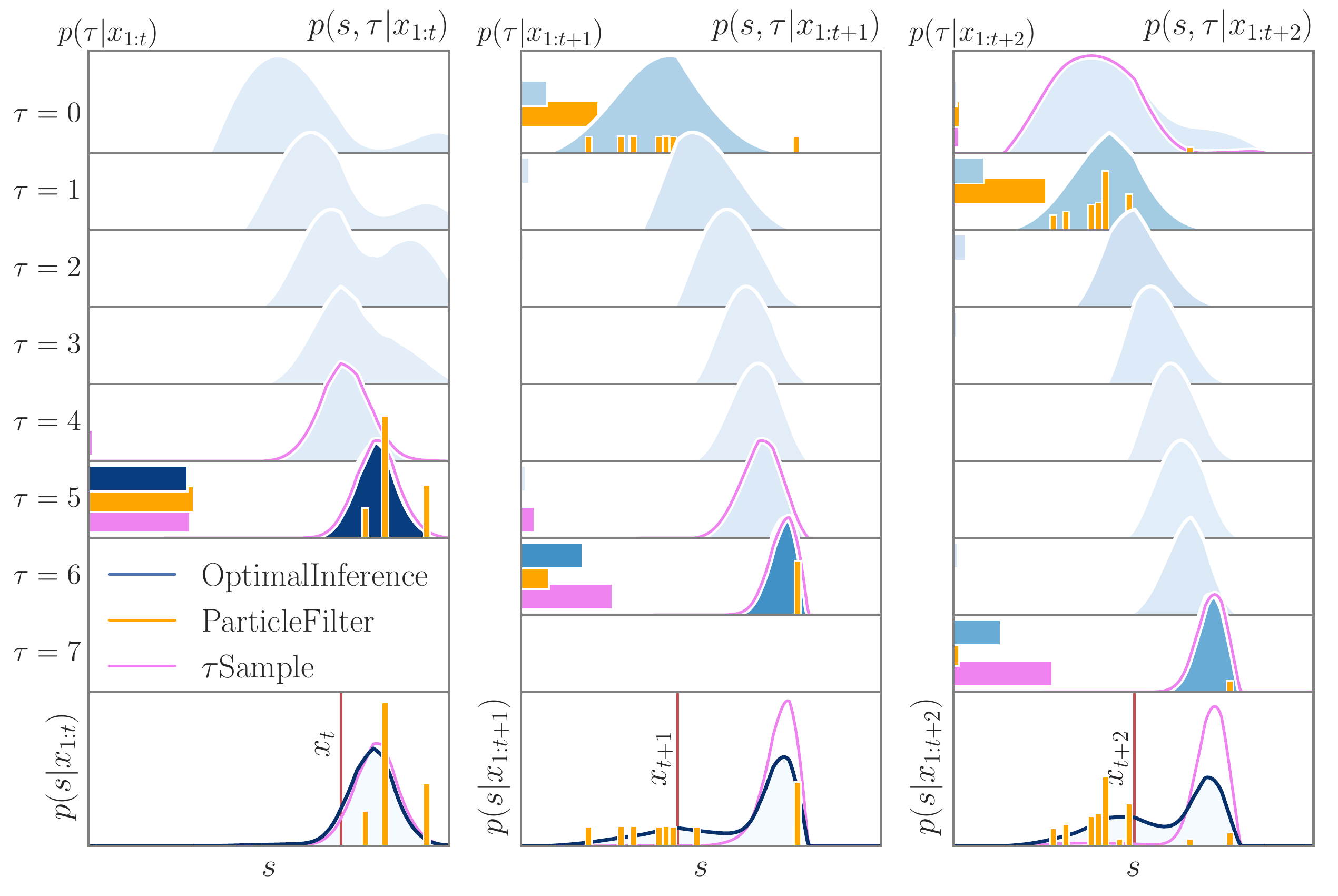}
	\caption{\textbf{Posterior density over three successive trials for the \textit{OptimalInference} model, the $\tau$\textit{Sample} model with $N_\tau=2$, and the \textit{ParticleFilter} model with ten particles.}
	The three panels correspond to the three successive trials.
	Each row except the last one corresponds to a different run-length, $\tau$. In these rows, the \textit{horizontal} bars show the marginal probability of the run-length, $p(\tau|x_{1:t})$.
The posterior (i.e., the joint distribution of the run-length and the state, $p(s, \tau|x_{1:t})$)
is shown as a function of the state, $s$, for the \textit{OptimalInference} model (blue shaded curve), the $\tau$\textit{Sample} model (pink line), and the \textit{ParticleFilter} model (orange vertical bars). The marginal probability of the run-length, $p(\tau|x_{1:t})$, for the \textit{OptimalInference} model, is additionally reflected in the hue of the curve (darker means higher probability). For the \textit{ParticleFilter} model, the heights of the bars are proportional to the weights of the particles. When the state, $s$, of two or more particles coincide, a single bar is shown with a height proportional to the sum of the weights. The last row shows the marginal distributions of the states, $p(s|x_{1:t}) = \sum_\tau p(s, \tau |x_{1:t})$, along with the location of the stimulus at each trial (red vertical line).
At trial $t$ (left panel), the probability of the run-length $\tau=5$ dominates in the three models. In the $\tau$\textit{Sample} model, it vanishes at run-lengths from $0$ to $3$, and it is very small for $\tau=4$. In the \textit{ParticleFilter} model, the run-lengths of the ten particles are all $5$, and thus the probability of all other run-lengths is zero.
At trial $t+1$ (middle panel), upon observation of the new stimulus, $x_{t+1}$, the marginal probability of the vanishing run-length ($\tau=0$), which corresponds to a `change-point' scenario, becomes appreciable in the \textit{OptimalInference} model (top row). The probability of the run-length $\tau=6$ (a `no change-point scenario' ) is however higher. As a result, a `bump' appears in the marginal distribution of the state, around the new stimulus (bottom row).
In the $\tau$\textit{Sample} model, the optimal update of the posterior results in a non-vanishing probability for three run-lengths ($\tau=0$, $5$, and $6$), more than the number of `memory slots' available ($N_\tau=2$). One run-length is thus randomly chosen, and its marginal probability is set to zero; in the particular instantiation of the model presented here, the run-length $\tau=0$ is chosen, and thus the resulting marginal probability of run-length is non-vanishing for $\tau=5$ and $6$ only.
In the \textit{ParticleFilter} model, the stochastic update of the particles results in seven particles adopting a vanishing run-length, and the probability of a `change-point' scenario ($\tau=0$) becomes higher than that of the `no change-point' scenario ($\tau=6$) supported by the remaining three particles.
The various marginal distributions of the states obtained in these three models (bottom row) illustrate how the $\tau$\textit{Sample} model and the \textit{ParticleFilter} model approximate the optimal posterior: the $\tau$\textit{Sample} model assigns a negligible probability to a set of states whose probability is substantial under the \textit{OptimalInference} model, while the \textit{ParticleFilter} yields a coarse approximation reduced to a support of ten states (as opposed to a continuous distribution).
	}
\label{fig:opt_TS_pf}
\end{figure}

This model, which we call \textbf{$\tau$\textit{Sample}}, is a stochastic version of the $\tau$\textit{MaxProb} model: instead of retaining deterministically the $N_{\tau}$ most likely run-lengths at each trial, the $\tau$\textit{Sample} model samples $N_{\tau}$ run-lengths using the marginal distribution of the run-lengths, $p_{t}(\tau |x_{1:t})$. More precisely, if at trial $t+1$ the marginal distribution of the run-lengths, $p_{t+1}(\tau | x_{1:t+1})$, is non-vanishing for $N_\tau+1$ values, then a run-length is sampled from the distribution $\left[ 1-p_{t+1}(\tau |x_{1:t+1})\right] /z_{t+1}$, where $z_{t+1}$ is a normalization factor, and the probability of this run-length is set to zero (Fig. \ref{fig:opt_TS_pf}). In other words, while the $\tau$\textit{MaxProb} model eliminates the least likely run-length deterministically, the $\tau$\textit{Sample} model eliminates one run-length stochastically, in such a fashion that less probable run-lengths are more likely to be eliminated. The $\tau$\textit{Sample} model has one parameter, $N_\tau$, the size of the support of the marginal distribution of the run-lengths.

\paragraph{Stochastic inference model with sampling in time \textit{and} in state space: the particle filter}
Although the $\tau$\textit{Mean}, $\tau$\textit{Nodes}, $\tau$\textit{MaxProb}, and $\tau$\textit{Sample} models introduced above relieve the memory load by prescribing a form of truncation on the set of run-lengths, inference in these models is still executed on a continuous state space labeled by $s$ (or, more precisely, on a discrete space with resolution as fine as a pixel). Much as subjects may retain only a compressed representation of probabilities along the $\tau $ axis, it is conceivable that they may not maintain a full probability function over the 1089-pixel-wide $s$ axis, as they carry out the behavioral task. Instead, they may infer using a coarser spatial representation, in order to reduce their memory and computational loads. Monte Carlo algorithms perform such approximations by way of randomly sampling the spatial distribution; sequential Monte Carlo methods, or `particle filters', were developed in the 1990s to address Hidden Markov Models, a class of hidden-state problems within which falls our inference task \parencite{Gordon1993, Arulampalam2002, Doucet2008}. Particle filters approximate a distribution by a weighted sum of delta functions. In our case, a \textit{particle} $i$ \textit{at trial} $t$ is a triplet, $(s_{t}^{i},\tau _{t}^{i},w_{t}^{i})$, composed of a state, a run-length, and a weight; a particle filter with $N_{P}$ particles approximates the posterior, $p_{t}(s,\tau|x_{1:t})$, by the distribution
\begin{equation}
\label{eq:pfdeltas}
\tilde p_t(s,\tau | x_{1:t}) = \sum_{i=1}^{N_{P}}w_{t}^{i} \delta(s-s_t^i) \delta _{\tau,\tau_{t}^{i}},
\end{equation}
where $\delta(s-s_t^i)$ is a Dirac delta function, and $\delta _{\tau,\tau_{t}^{i}}$ a Kronecker delta.
In other words, a distribution over the $(s,\tau )$ space is replaced by a (possibly small) number of points, or samples, in that space, along with their probability weights.

To obtain the approximate posterior at trial $t+1$ upon the observation of a new stimulus, $x_{t+1}$, we note, first, that the Bayesian update (Eq. (\ref{eq:HDupdate})) of the approximate posterior, $\tilde p_t(s,\tau | x_{1:t})$, is a mixture (a weighted sum) of the $N_P$ Bayesian updates of each single particle (i.e., Eq. (\ref{eq:HDupdate}) with the prior, $p_{t}(s,\tau |x_{1:t})$, replaced, for each particle, by $\delta (s-s_{t}^{i})\delta _{\tau ,\tau _{t}^{i}}$). Then, we sample independently each component of the mixture (i.e., each Bayesian update of a particle), to obtain stochastically the updated particles, $(s_{t+1}^{i},\tau _{t+1}^{i})$, and to each particle is assigned the weight of the corresponding component in the mixture. The details of the procedure just sketched, in particular the derivation of the mixture and of its weights, and how we handle the difficulties arising in practical applications of the particle filter algorithm, can be found in \nameref{sec:Methods}. This model, which we call \textbf{\textit{ParticleFilter}}, has a single free parameter: the number of particles, $N_{P}$ (Fig. \ref{fig:opt_TS_pf}).

\paragraph{Models with variability originating in the response-selection step}
The $\tau$\textit{Sample} and \textit{ParticleFilter} models presented above reduce the dimensionality of the inference problem by pruning stochastically the posterior, in the inference step. But, as we pointed out, the behavior of the standard deviation of the responses of the subjects, as compared to that of the width of the Bayesian posterior (Fig. \ref{fig:metrics}C), hints at a more straightforward mechanism at the origin of response variability. The model we now introduce features stochasticity not in the inference step, but rather in the response-selection step of an otherwise optimal model. In this model, the response is sampled from the marginal posterior on the states, $p_{t}(s|x_{1:t})$, i.e., the response, $\hat s_t$, is a random variable whose density is the posterior. This contrasts with the optimal response-selection strategy, which maximizes the expected score based on the Bayesian posterior, and which was implemented in all the models presented above.
Henceforth, we denote the optimal, deterministic response-selection strategy by \textbf{\textit{Max}}, and the suboptimal, stochastic strategy just introduced by \textbf{\textit{Sampling}}. It has no free parameter.

Another source of variability in the response-selection step might originate in a limited motor precision, in the execution of the task. To model this motor variability, in some implementations of our models we include a fixed, additive, Gaussian noise, parameterized by its standard deviation, $\sigma_{m}$, to obtain the final estimate. Both this motor noise and the \textit{Sampling} strategy entail stochasticity in response selection. The former, however, has a fixed variance, $\sigma_{m}^{2}$, while the variance of the latter depends on the posterior which varies over the course of inference (Fig. \ref{fig:metrics}C). When we include motor noise in the \textit{Max} or in the \textit{Sampling} strategies, we refer to these as \textbf{\textit{NoisyMax}} and \textbf{\textit{NoisySampling}}, respectively.

In sum, we have described four response-selection strategies (\textit{Max}, \textit{Sampling}, \textit{NoisyMax}, and \textit{NoisySampling}), and seven inference strategies, of which five are deterministic (\textit{OptimalInference}, \textit{IncorrectQ}, $\tau$\textit{Mean}, $\tau$\textit{Nodes}, and $\tau$\textit{MaxProb}) and two are stochastic (\textit{$\tau$Sample} and \textit{ParticleFilter}). We can combine any inference strategy with any response-selection strategy: thus, we have at hand 4$\times$7 $=$ 28 different models, 27 of which are suboptimal, obtained from pairings of the inference and selection strategies. We label each of the 28 models by the combination of the two names referring to the two steps in the process, e.g., \textit{ParticleFilter+Sampling}.

\subsection{Fitting models to experimental data favors sample-based inference}

\begin{table}[th]
\includegraphics[width=\linewidth]{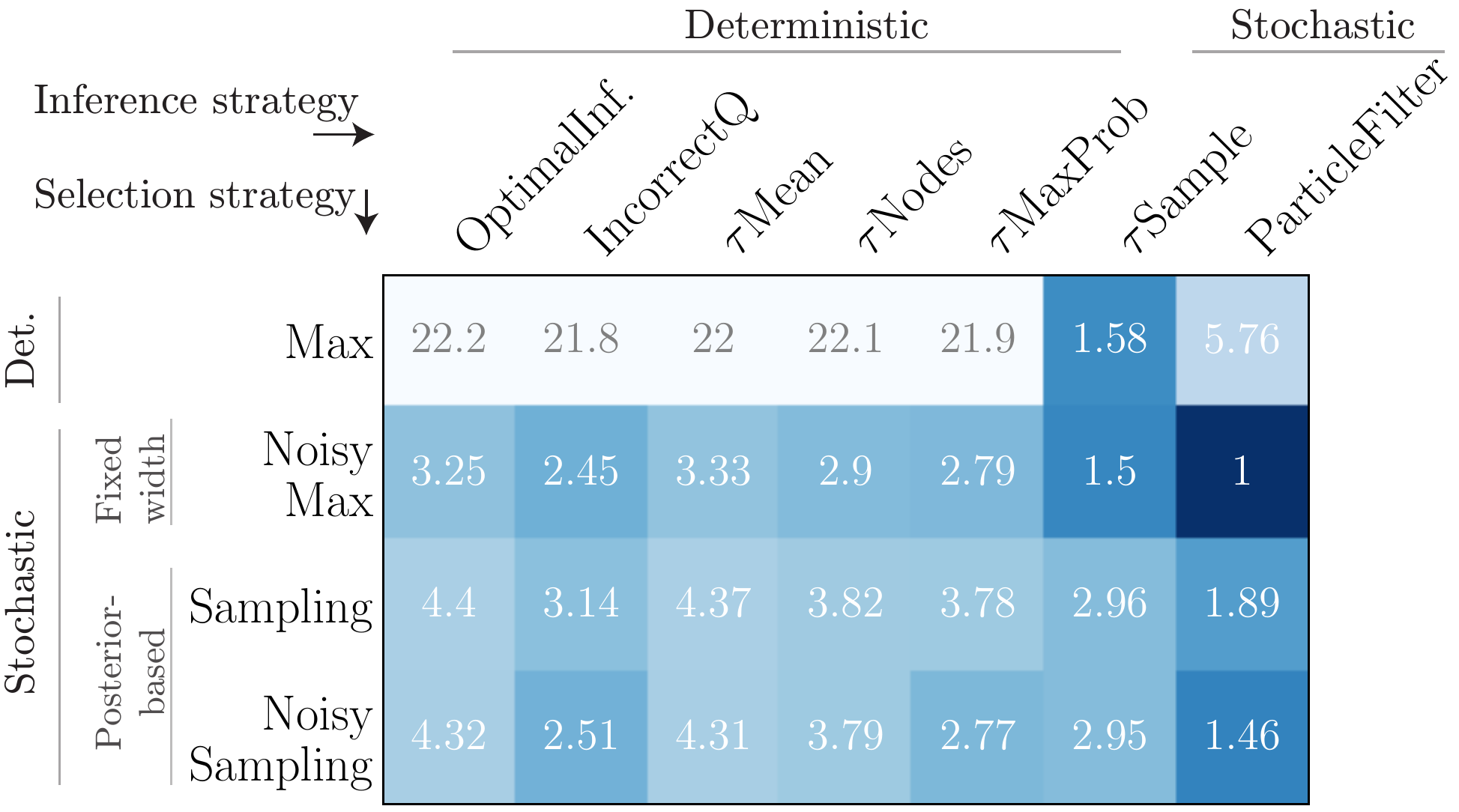}
\caption{\normalfont \textbf{Model fitting favors the \textit{ParticleFilter} inference strategy with {NoisyMax} response selection.}
\textnormal{Ratios of the normalized mean squared error (NMSE) in each model to that of the best-fitting model, \textit{ParticleFilter+NoisyMax}. Each model is a combination of an inference strategy (columns) with a response-selection strategy (rows). The second best model, also a \textit{ParticleFilter} but with a \textit{NoisySampling} response-selection strategy, yields an NMSE 46\% higher than the best-fitting model.}}
\label{tbl:checker}
\end{table}

The 27 suboptimal models introduced in the previous section yield a range of discrepancies from the optimal behavior. The ways in which each deviation from the optimal model impacts behavior is examined in \nameref{sec:Methods}; here, we ask how well these models account for the behavior of human subjects. Whereas the optimal model, \textit{OptimalInference+Max}, computes the Bayesian posterior (\textit{OptimalInference}) and selects the maximizing response (\textit{Max}), the suboptimal models mimic cognitive limitations that may prevent the brain from reaching optimality: incorrect belief in the temporal structure of the signal (\textit{IncorrectQ}), compressed representation of the Bayesian posterior, either deterministically ($\tau$\textit{Mean}, $\tau$\textit{Nodes}, and $\tau$\textit{MaxProb}) or stochastically (\textit{$\tau$Sample} and \textit{ParticleFilter}), and noise introduced in the response-selection step, with a width either scaling with that of the posterior (\textit{Sampling}), or constant (\textit{NoisyMax}), or a combination of the two (\textit{NoisySampling}). 

To evaluate the ability of each of these models to account for human behavior, we compare quantitatively their respective outputs with the responses of human subjects. For the three quantities we examine (the learning rate, the repetition propensity, and the standard deviation of the responses), we compute the normalized mean squared error (NMSE) (sometimes referred to as the `Fraction of Variance Unexplained' in the context of linear regressions). It is defined, for a given model and for each quantity, as the ratio of the mean squared error in the model output as compared to data, and the variance of the quantity under scrutiny in the behavioral data. We fit each of our models to human data, using the average of the three NMSEs as our error measure. (We note that the \textit{OptimalInference} inference strategy is a special case of all the other inference strategies, except $\tau$\textit{Mean}, thus its NMSE cannot be lower than that of these strategies. Likewise, the \textit{Max} and \textit{Sampling} response-selection strategies are special cases of the \textit{NoisyMax} and \textit{NoisySampling} strategies, respectively.)

\begin{figure}[!ht]
        \includegraphics[width=\linewidth]{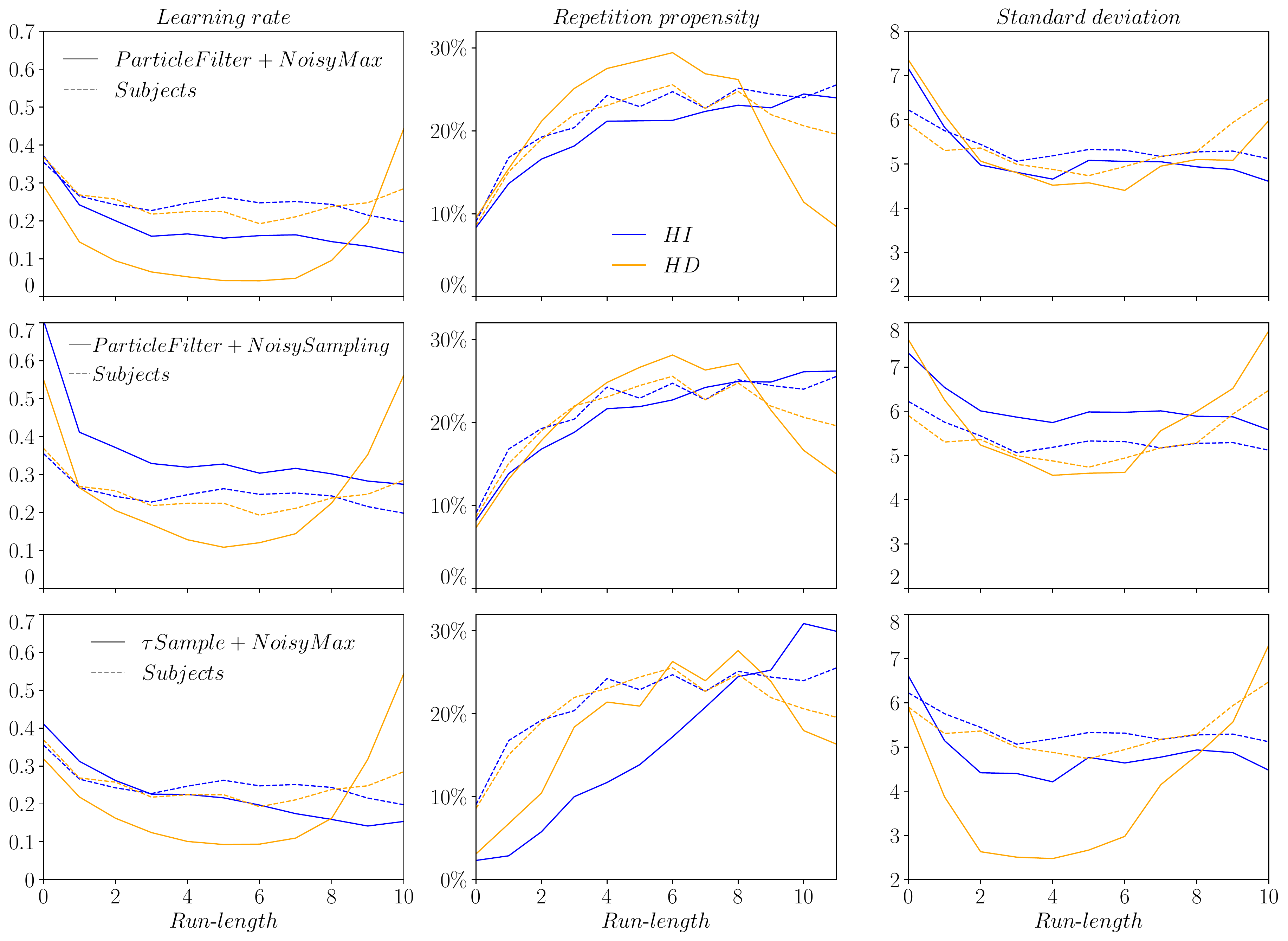}
	\caption{\textbf{Behavior of the three best-fitting models.}
	In HI (blue curves) and HD (yellow curves) conditions, average learning rate (first column), repetition propensity (second column), and standard deviation of responses (third column),  as a function of run-length, for the subjects (solid lines) and the three best-fitting models (dashed lines): \textit{ParticleFilter+NoisyMax} (first row), \textit{ParticleFilter+NoisySampling} (second row), and $\tau$\textit{Sample+NoisyMax} (third row).
	}\label{fig:winners}
\end{figure}

We find that the five best-fitting models make use of stochastic compression in the inference step, in either the $\tau$\textit{Sample} approximation or the \textit{ParticleFilter} approximation
(Table \ref{tbl:checker}). These models all reproduce the qualitative trends in the behavior of subjects with respect to our three measures: for the learning rate and the standard deviation, the `smile shape' of the HD curve, which crosses a decreasing HI curve; for the repetition propensity, conversely, an inverted U shape of the HD curve which crosses an increasing HI curve (Fig. \ref{fig:winners}, results from the $\tau$\textit{Sample+Max} and \textit{ParticleFilter+Sampling} models are not shown, but the corresponding curves
are similar).

The $\tau$\textit{Sample} and \textit{ParticleFilter} strategies have one or two parameters, depending on whether they include motor noise or not. Other models, including all models with a deterministic inference step, have an error at least 30\% higher than the best five models (and 2.45 times higher than the best model), despite the fact that other strategies come with up to five parameters
(Table \ref{tbl:checker}). The best-fitting model is \textit{ParticleFilter+NoisyMax} with $N_P = 9 $ particles. The fitted standard deviation, $\sigma _{m}$, of the Gaussian motor noise is approximately equal to $0.77$ pixels; as a consequence, in about half of the trials, the noise component is within the width of a pixel, and thus has no impact. The second best model also follows a \textit{ParticleFilter} inference strategy, with $N_{P}=14$, combined with a \textit{NoisySampling} response selection (with $\sigma _{m}=0.70$ pixels). 

The third and fourth best-fitting models use the $\tau$\textit{Sample} inference strategy, with $N_\tau=1$, and the \textit{NoisyMax} (with $\sigma_m=0.45$ pixels, for the third one) and the \textit{Max} (for the fourth one) selection strategies. At any given trial, these two models retain only a single assumption, $\tau_t$, on the run-length. Upon receiving a new stimulus, $x_{t+1}$, a model subject computes $p_{change} = p_{t+1}(\tau=0 | x_{1:t+1})$ and $1 - p_{change} = p_{t+1}(\tau=\tau_t + 1 | x_{1:t+1})$, and decides whether there was a change-point by sampling this simple, Bernoulli distribution. This sampling process, over a marginalization of the posterior, is similar to that in the particle filter model, which samples over the full $(s,\tau)$-dependent posterior. As a consequence of sampling, the $\tau$\textit{Sample} strategy also exhibits variability, which behaves in a fashion similar to the variability in the \textit{ParticleFilter} strategy (Fig. \ref{fig:winners}, bottom right). As for response selection, we note that with the \textit{Sampling} and \textit{NoisySampling} selection strategies (instead of the \textit{Max} and \textit{NoisyMax} strategies), these models do not perform as well, and result in errors larger by 86\% (\textit{Sampling} vs. \textit{Max}) and 96\% (\textit{NoisySampling} vs. \textit{NoisyMax}). In fact, for all the seven inference models, the \textit{NoisyMax} response-selection strategy results in errors lower or equal (but more often, lower) than the other three selection strategies (\textit{Max}, \textit{Sampling} and \textit{NoisySampling}) (Table \ref{tbl:checker}).
This suggests that the variability in human responses does not originate from a posterior-sampling strategy in the response-selection step, but, rather, from an intrinsically stochastic inference process. In order to seek further validation of this finding, we explore, below, a generalization of the \textit{Sampling} strategy.

\paragraph{Robustness of the results}
To substantiate the picture that emerges from the results summarized above, we perform two supplementary analyses. First, we investigate whether a generalized \textit{Sampling} strategy yields smaller errors than the \textit{NoisyMax} strategy. Second, we consider our choice of fitting-performance measure (the average of the NMSEs in the three quantities we examine), and we check for the robustness of model fitting to changes in the relative weights of each quantity in the fitting performance measure.

Sampling from the posterior function is only one of many possible sampling strategies for response selection. Furthermore, in practice sampling may be difficult to tease apart from maximizing a perturbed posterior function. \textcite{Acerbi2014} argue that, for some forms of random perturbations of a posterior probability density, maximizing the randomly perturbed function yields similar results to sampling from a modified posterior density function obtained as a power of the correct posterior: $p_{\kappa }(s)\propto p(s|x_{1:t})^{\kappa }$. To establish the equivalence, the exponent, $\kappa $, is chosen as inversely related to the magnitude of the perturbing noise. Sampling from the modified posterior yields a behavior that interpolates between posterior sampling (for $\kappa =1$) and maximizing (for $\kappa \rightarrow \infty $); it yields a family of softmax operations over the posterior \parencite{Vul2011, Yu2014}. Another interpretation of this sampling strategy is proposed by \textcite{Battaglia2011}: in the case of an integer $\kappa $ and a Gaussian posterior, the mean of $\kappa$ samples drawn from the posterior is a Gaussian random variable, with a standard deviation equal to that of the posterior scaled by $1/\sqrt{\kappa}$; i.e., a distribution equal to the posterior raised to the power $\kappa $, and normalized. Hence, in the Gaussian case, sampling from the exponentiated posterior can be interpreted as drawing $\kappa$ samples from the unexponentiated posterior, and taking the mean.

\begin{figure}[!ht]
        \includegraphics[width=\linewidth]{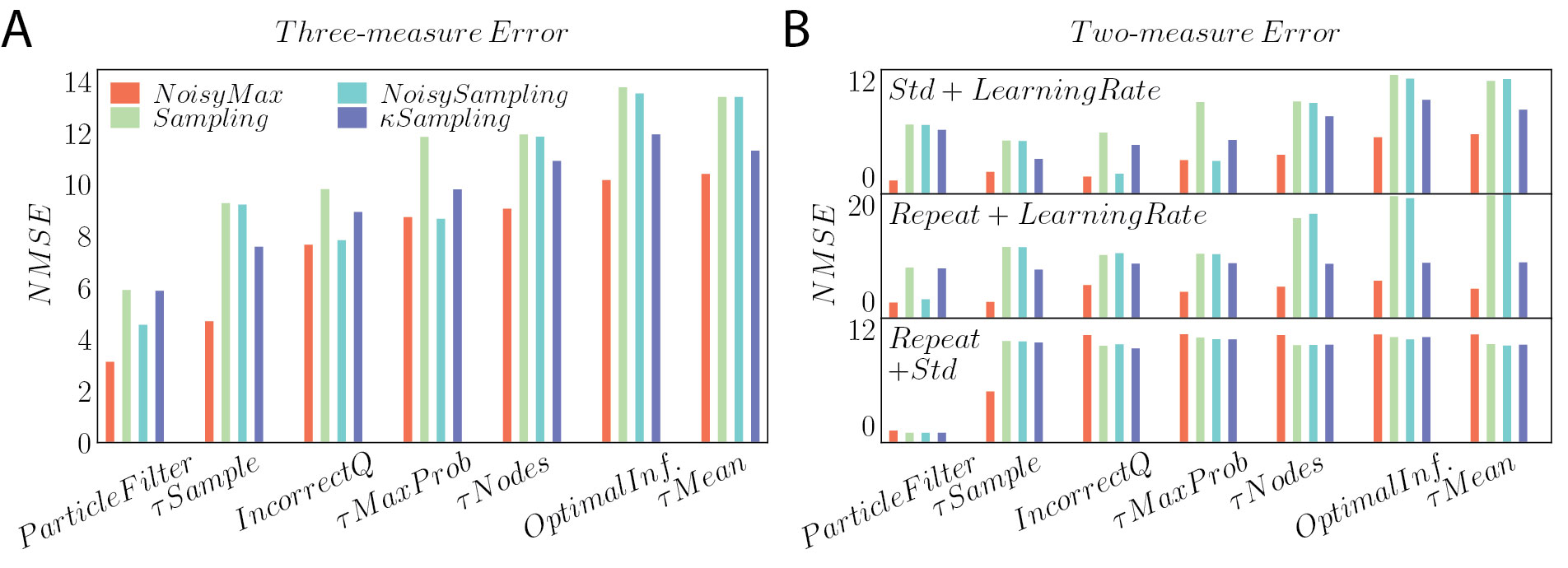}
	\caption{\textbf{Model fitting is robust to the measure used for model comparison.}
	Normalized Mean Squared Error of fitted models. 
	\textbf{A.} NMSE of models fitted to subjects data, averaged over the three measures (learning rate, repetition propensity, and standard deviation of responses), grouped by inference models.
	\textbf{B.} NMSE between fitted models and subjects data, averaged over two out of the three measures.
	}\label{fig:fit-errors}
\end{figure}

We implement this strategy of response selection by sampling a modified posterior, which we denote $\kappa$\textit{Sampling}. We find that it performs better than the \textit{Sampling} strategy, as expected since the \textit{Sampling} strategy is a special case of the $\kappa$\textit{Sampling} (with the parameter, $\kappa$, set to unity). However, in the case of all seven inference models, the $\kappa$\textit{Sampling} strategy, which has one free parameter, performs worse than the \textit{NoisyMax} strategy, which has, also, a single parameter (Fig. \ref{fig:fit-errors}B). Hence, a random, additive perturbation of the maximization strategy remains a better account of human behavior than a posterior-sampling strategy.

Our results, which suggest that the variability in the responses of subjects originate in the inference step rather than in the response-selection step, rely upon the fitting performance measure used for model comparison. We chose a measure that weighted equally the three NMSEs (on the learning rate, repetition propensity, and standard deviation), so as to obtain a model performing well on all fronts, but that choice was arbitrary.
Hence, one may be concerned, for instance, that the goodness-of-fit of the \textit{ParticleFilter} model be due to our choice of weighing errors. As a control, we computed the three `two-measure errors', each excluding one of the three measures and averaging the errors in the two remaining ones. We found that, regardless of the choice of the combination, the relative order of the models in terms of performance stays identical, with only a few exceptions. Most importantly, the \textit{ParticleFilter} remains, in all three cases of error combinations, the best-fitting model (Fig. \ref{fig:fit-errors}B).

\begin{figure}[!ht]
        \centering
        \includegraphics[width=0.7\linewidth]{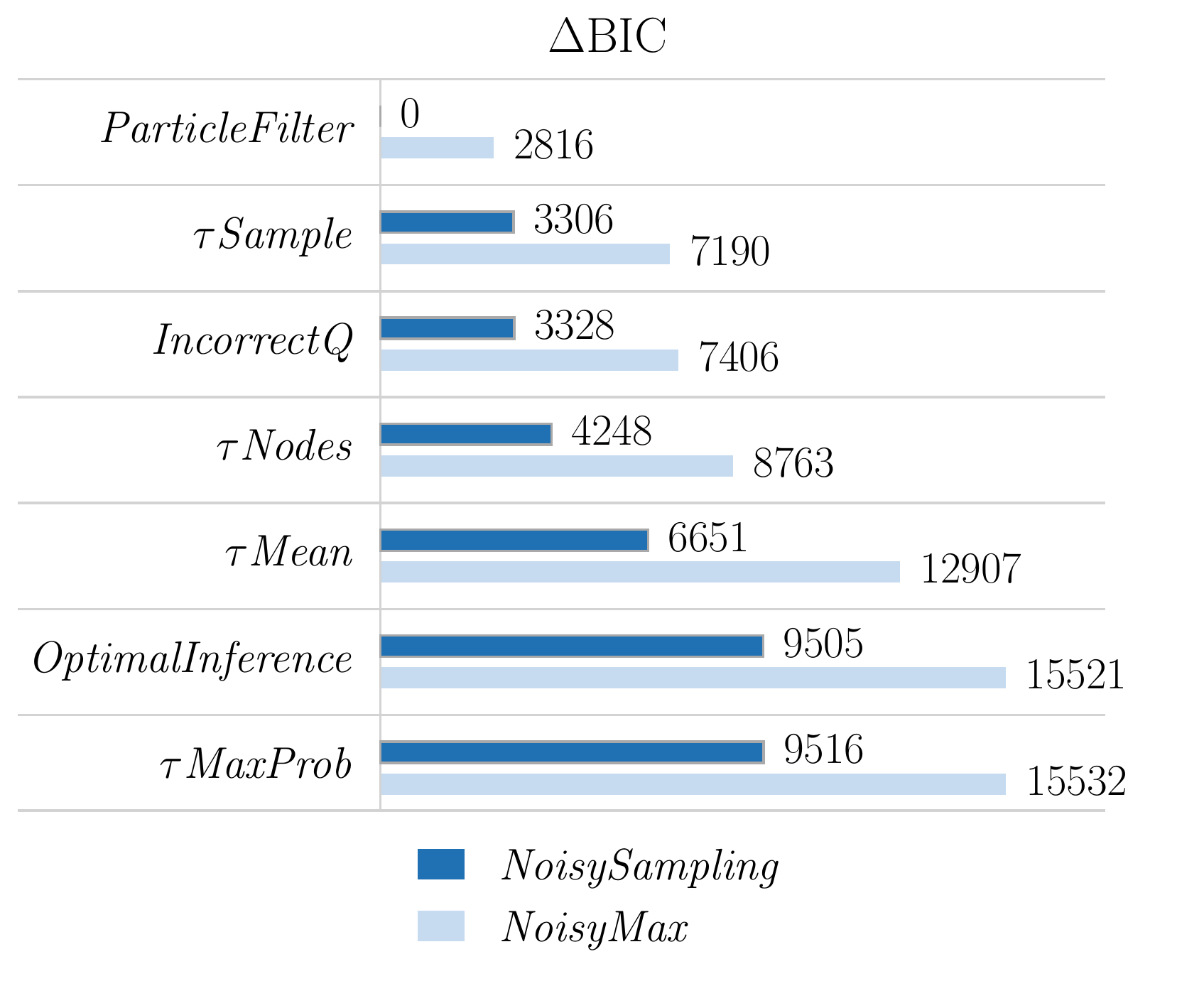}
	\caption{
	\textbf{Bayesian model selection favors the \textit{ParticleFilter} inference strategy.}
	Difference between the BIC of each model and that of the best-fitting model, for the models combining one of the seven inference strategies with the $NoisyMax$ or the $NoisySampling$ response-selection strategies. The two best-fitting models make use of the \textit{ParticleFilter} inference strategy.
	}\label{fig:bics}
\end{figure}

\subsection{Bayesian model selection also favors sample-based inference}

Another concern regarding the choice of the NMSE as our performance measure for model comparison is that it does not take into account the number of parameters in the models. A standard method to fit and compare models is to maximize the log-likelihood of each model and compute its Bayesian Information Criterion (BIC), which includes a penalty as a function of the number of parameters in the model \parencite{Schwarz1978}. In several of our models (and in many models in the literature), the responses in successive trials, conditioned on the stimuli presented to the subject, are independent; as a result, the log-likelihood over all trials is the sum of the log-likelihoods for each trial, taken separately. This obtains for all the models in which the inference strategy is deterministic (\textit{OptimalInference}, \textit{IncorrectQ}, $\tau$\textit{Mean}, $\tau$\textit{Nodes}, and $\tau$\textit{MaxProb}). It does not apply, however, for the models with stochastic inference strategies ($\tau$\textit{Sample} and \textit{ParticleFilter}): in these models, successive responses, conditional on observed stimuli, are \textit{not} independent as they depend on the realization of the stochastic process that governs inference. To compute the BIC, it is therefore necessary to compute, first, the distribution of the possible realizations of the stochastic inference process. The difficulty, here, lies in the fact that the space of these realizations grows exponentially with the number of trials in an experimental run.

In the context of our task, in which the subjects undergo 1000 trials in a run, an exact computation of the BIC is prohibitive.
In order to circumvent this problem, we propose to approximate the log-likelihood of a model by way of a Monte-Carlo estimation of the log-likelihoods of short sequences of responses. This approach limits the computational load of the estimation while taking into account the sequential dependence of responses. We report, here, the results of this estimation scheme using short sequences of 10 successive trials, but in our investigations we repeated the calculations for different choices, which yielded comparable results. We detail the procedure in \nameref{sec:Methods}. Here, we mention that, even though models with temporal correlation such as the particle filter have been used to capture cognitive processes, to the best of our knowledge Bayesian model selection using the BIC has not been applied to them, except in the case of a binary categorization task
\parencite{Lloyd2019}. The approximate approach we propose may thus be of use beyond the confines of the specifics of our problem.

In the models that do not feature a (Gaussian) motor noise, some responses of the subjects have a vanishing probability, and thus these models have an infinite BIC. Hence, we look at the BICs of the models equipped with the \textit{Noisy} or the \textit{NoisySampling} response-selection strategy. We find that the three best-fitting models involve a stochastic approximation of the Bayesian inference: the two best-fitting models make use of the \textit{ParticleFilter} inference strategy, and the third best-fitting model has a $\tau$\textit{Sample} inference strategy (Fig. \ref{fig:bics}). We note that with the NMSE metric the three best-fitting models were also the two \textit{ParticleFilter} models followed by a $\tau$\textit{Sample} model. Thus, both model-fitting approaches suggest that human inference evolves according to a stochastic compression of the posterior. The best-fitting model is \textit{ParticleFilter+NoisySampling}, with $N_P=8$ particles, and its BIC is smaller than that of the second best-fitting model, the \textit{ParticleFilter+NoisyMax} model with $N_P=4$ particles, by $2,816$. This result is consistent with the best-fitting numbers of particles obtained when minimizing the NMSE, which were also relatively modest, although slightly larger ($N_P=14$ with \textit{NoisySampling} and $N_P=9$ with \textit{NoisyMax}).

Taken together, our results suggest that variability in human behavior, at least in the context of our task, is dictated primarily by stochasticity in the inference step --- i.e., in the manipulation and update of probabilities --- rather than by `output noise' such as stochasticity in the response-selection step or motor noise. This view agrees with the conclusion of a recent study of a cue combination task \parencite{Drugowitsch2016}; its authors argue that a ``dominant fraction'' of human choice suboptimality arises from random fluctuations in the inference step.

\section{Discussion}

\subsection{Summary}

This study investigates the behavior of human subjects in an online inference task, and examines mechanisms that can account for behavioral trends found in experimental data. An important aspect of this task is that it makes use of both a history-independent (HI) condition with no temporal structure, and a history-dependent (HD) condition in which a hidden state is almost periodical and, hence, highly structured in time (Fig. \ref{fig:task}D). We find that subjects display different behaviors in the two conditions, adapting their learning rate to the temporal structure of the hidden state. We also note a propensity in subjects to repeat their response in consecutive trials; this repetition propensity increases with the run-length, and in the HD condition drops again for larger run-lengths. Moreover, we observe that subjects exhibit a greater variability in their responses shortly after a change point, in both conditions, and at long run-lengths in the HD condition, i.e., the \textit{variability} in behavior also depends on the temporal statistics of the stimuli.

The distinctive behaviors of the learning rate and the repetition propensity in the HI and HD conditions are reproduced qualitatively by a Bayesian model of inference which yields optimal updates of the probability density of the hidden state. As for the variability in subjects' responses, we find that its behavior is similar to that of the standard deviation of the Bayesian posterior. We therefore use the Bayesian model as a starting point to elaborate variant models which can account for the trends exhibited in human responses.
We find that the variability in human behavior, and its dynamics, can be reproduced by supoptimal models in which inference is executed in a stochastic manner. Specifically, the $\tau$\textit{Sample} and the \textit{ParticleFilter} models alter the optimal inference step by maintaining an approximate version of the posterior, by means of random sampling. This alteration of the optimal model at once introduces variability in the behavior and relieves the memory capacity, through sampling either in the `time dimension' (in the $\tau$\textit{Sample} model) or in the `time and space dimensions' (in the \textit{ParticleFilter} model).

The behavioral patterns that arise in our task in the HI condition are also found in other experiments. \cite{Gallistel2014} and \cite{Khaw2017} both conducted an online inference task, with change points that occurred with constant probability (similarly to our HI condition). We examined the responses of their subjects in the context of the respective tasks, and we observed very similar behavioral trends: the learning rate decreases as a function of the run-length, while the repetition propensity increases.
As for the empirical standard deviation of subjects' responses, we note that in these studies some subjects were presented several times with the same sequence of stimuli, in different sessions, thus allowing for the examination of the variability of responses within subjects. We find, here also, that the within-subject standard deviation of responses shows the same modulations than the standard deviation of the Bayesian posterior (Supplementary Fig. \ref{fig:metrics-h-GallistelKSW}).

\subsection{Temporal structures in nature and their behavioral and neural counterparts}

In order to make appropriate decisions in relation to their environment, humans and animals must infer the state of the surrounding world on the basis of the sensory signals they receive. If these signals are noisy and if the environment is changing, their inference task is complicated by the fact that a new stimulus may reflect either noise or a change in the underlying state. However, if events in the world present some kind of temporal structure, such as in our HD signal, it is possible to use this structure to refine one's inference. Conversely, if events follow a Poisson process, as in the HI signal, their occurrences present no particular temporal structure, and what just happened conveys no information on what is likely to happen next. Hence, there is a fundamental difference between the HI and HD conditions, which impacts the inference of an optimal observer.

Many natural events are not Poisson-distributed in time, and exhibit strong regularities.
\textcite{Nunes2004}, \textcite{Nakamura2007, Nakamura2008}, and \textcite{Anteneodo2009} have recorded the motor activity of both rodents and human subjects over the course of several days. In both species, they found that the time intervals between motion events were distributed as a power law, a distribution characterized by a long tail, leading to bursts, or clusters, of events followed by long waiting epochs. The durations of motion episodes also exhibited heavy tails. These kinds of distribution are incompatible with Poisson processes, which yield exponentially distributed inter-event epochs. Moreover, both rodent and human activity exhibited long-range autocorrelations, another feature that cannot be explained by a Poisson process.
A particular form of autocorrelation is periodicity, which occurs in a wide range of phenomena. In the context of human motor behavior, walking is a highly rhythmical natural activity \parencite{Hausdorff1995, Griffin2000}. More complex patterns exist (neither clustered nor periodic), such as in human speech which presents a variety of temporal structures, whether at the level of syllables, stresses, or pauses \parencite{Ramus1999, Low2000, Campione2002}. In all these examples, natural mechanisms produce series of temporally structured events. The ubiquity of history-dependent statistics of events in nature begs for explorations of inference mechanisms in their presence.
For the purposes of our experiment, we chose an idealized temporal signal that combined several advantages: it featured a prominent form of history dependence, approximate periodicity; it was not easily distinguishable from the other, history-independent signal used in the task; and it was amenable to modeling.

In the case of studies of perception and decision-making, in both humans and animals, history-dependent signals have been used widely.
In a number of experiments \parencite{Janssen2005, Ghose2002, Li2013, Miyazaki2005, Jazayeri2010}, a first event (a sensory cue, or a motor action such as a lever press) is followed by a second event, such as the delivery of a reward, or a `go' signal triggering the next behavior. The time elapsed between these two events – the `reward delay' or the `waiting time' – is randomized and sampled from distributions that, depending on the studies, vary in mean, variance, or shape. For instance, both \textcite{Janssen2005} and \textcite{Ghose2002} use unimodal and bimodal temporal distributions.
Because of the stochasticity of the waiting time, the probability of occurence of the second event varies with time, similarly to the probability of a change point in our HD condition; these studies explore whether variations of this probability are used by human and animal subjects.
\textcite{Janssen2005}'s recordings from the V4 cortical area in rhesus monkey indicate that, for both unimodal and bimodal waiting times distributions, the attentional modulation of sensory neurons varies consistently with the event probability.
\textcite{Ghose2002} note that the reaction times of macaques are inversely related to the event probability, for both unimodal and bimodal distributions,
and that the activity of neurons in the lateral intraparietal (LIP) area is correlated with the evolution of this probability over time. \textcite{Li2013} manipulate another aspect of the distribution of reward delays: between blocks of trials, the standard deviation of this distribution is varied, while the mean is left unchanged. Mice, in this situation, are shown to adapt their waiting times to this variability of reward delays, consistently with a probabilistic inference model of reward timing.

Akin to the tasks just outlined are `ready-set-go time-reproduction tasks', in which subjects are asked to estimate the random time interval between `ready' and `set' cues, and to reproduce it immediately afterwards. \textcite{Miyazaki2005} and \textcite{Jazayeri2010} show that human subjects combine optimally the cue (consisting in the perceived ready-set interval) with their prior on the interval length. Different priors are learned in training runs: they differ by the variances of the interval distributions \parencite{Miyazaki2005} or by their means \parencite{Jazayeri2010}. In both cases, subjects integrate the prior in a fashion consistent with Bayesian inference. Adopting a different approach, \textcite{TenOever2014} show that attentional resources can be dynamically allocated to points in time at which input is expected: when asked to detect auditory stimuli (beeps) of low intensity embedded in a continuous white noise, human subjects perform better when detecting periodic beeps rather than random beeps, suggesting that they are able to identify the temporal regularity and use it in their detection process.

In all these studies, the event of interest has a probability of occurrence that varies with time. The resulting temporal structure in the signal appears to be captured by human and animal subjects, and reflected in behavior and in its neural correlate. Various probability distributions used in the reported tasks can be compared directly to our HD sigmoid-shaped change probability, with adjusted parameters. In line with these studies, our results confirm that human subjects adapt their behavior depending on the temporal structure of stimuli. Additionally, we provide a comparison between two different conditions, a HD condition akin to a `jittered periodic' process, and the Poisson, HI condition; the latter produces a memoryless process. Importantly, it plays the role of a benchmark from the point of view of probability theory: in discrete time it yields a geometric distribution, and in continuous time it yields an exponential distribution; both distributions maximize the entropy, subject to the constraint of a fixed event rate. In this study, we compared a specific, temporally structured HD condition to this benchmark, HI condition.

\subsection{Online Bayesian inference}

Our first observation is that the average learning rate of subjects and their repetition propensity are captured by a Bayesian model. The Bayesian paradigm has been viewed as an extension of logic that enables reasoning with propositions whose truth or falsity is uncertain \parencite{Cox1946,Jaynes2003}. In cognitive science, it has successfully accounted for a wide range of observations, including cue combination in humans \parencite{Jacobs1999, vanBeers1999, Ernst2002, Knill2003, Battaglia2003, Hillis2004, Knill2007, Battaglia2011}, sensorimotor control \parencite{Kording2004, Kording2006, Berniker2011}, integration of temporal statistics \parencite{Ghose2002, Janssen2005, Miyazaki2005, Jazayeri2010, Li2013}, perceptual multistability \parencite{Sundareswara2008, Moreno-Bote2011, Gershman2012}, and various aspects of cognition \parencite{Griffiths2006, Blaisdell2006, Stocker2008, Goodman2008, Griffiths2011}.

The literature on Bayesian \textit{online} inference in cognition, where belief is updated iteratively as a function of incoming information, is growing; examples can be found in word segmentation \parencite{Pearl2011}, sentence processing \parencite{Levy2008}, conditioning \parencite{Daw2008}, as well as in the change-point literature \parencite{Nassar2010, Wilson2013, Wilson2010, Glaze2015, Glaze2018, Piet2018, Gallistel2014, Khaw2017}. In change-point tasks, subjects are presented with a long sequence of consecutive inference problems (1000 of them, in our case). Each trial is a slightly different task, in which one has to handle the uncertainty resulting from the belief distribution, from the signal likelihood, and from the possibility of a change point. The latter, in the HD condition, bears the added complexity of a change-point probability, $q(\tau)$, that depends on the time of the last change point. How these uncertainties are handled determines the behavior, and in particular to what extent an observer reacts to a new stimulus: either shift the estimate towards it, or not move at all. We quantify this response through the learning rate and the repetition propensity, and we find that the ideal Bayesian observer and the subjects obey similar trends (Fig. \ref{fig:metrics}A-B).

The success of the Bayesian paradigm, however, is limited, and comes with three shortcomings. First, subjects do not behave quantitatively like the ideal Bayesian observer, and hence there remains unexplained suboptimality. Second, we find variability in the responses of subjects (Fig. \ref{fig:metrics}C), an observation incompatible with optimal Bayesian inference. Third, inference problems in the real world are complex and high-dimensional, rendering Bayesian reasoning computationally heavy and memory-intensive. This suggests that humans use approximations when carrying out inference and estimation tasks \parencite{Gershman2015, Gershman2016, Sanborn2016}. These three observations call for the investigation of alternatives to the optimal Bayesian paradigm. Our study explores several scenarios.

\subsection{Sampling versus noisy maximization}

Although the behavior of the subjects and of the Bayesian model differ in that the former exhibits variability while the latter is deterministic, the temporal modulation of the human variability follows a similar course to that of the standard deviation of the Bayesian posterior (Fig. \ref{fig:metrics}C), and both the standard deviation and the skewness of the distribution of subjects' responses are correlated with those of the Bayesian posterior (Fig. \ref{fig:std_skewness}). Therefore, it is natural to propose that response selection operates by sampling the Bayesian posterior instead of maximizing the expected score. Decision by posterior sampling, or `probability matching', has been suggested by other decision-making experiments \parencite{Herrnstein1961, Koehler2009, Denison2013} and, more recently, perceptual experiments \parencite{Wozny2010, Battaglia2011, Moreno-Bote2011}. Although close to optimal in some specific paradigms \parencite{Kaufmann2012}, sampling is suboptimal in the context of our behavioral task. When fitting models to human data, we observe that the \textit{ParticleFilter+Sampling} and the \textit{ParticleFilter+NoisySampling} models yield larger NMSE than the \textit{ParticleFilter+NoisyMax} model (the best-fitting model), by 90\% and 48\%, respectively. More generally, we observe that for each of our seven inference strategies, the \textit{NoisyMax} response-selection strategy results in better fits (lower NMSE) than the \textit{Sampling} and the \textit{NoisySampling} strategies (Fig. \ref{fig:fit-errors}A). Relaxing the \textit{Sampling} model by allowing the posterior to be exponentiated before sampling, as in the $\kappa$\textit{Sampling} strategy, does not yield better fits than the \textit{NoisyMax} strategy either. We conclude that posterior sampling accounts less successfully for our data than a simple perturbation of the optimal maximization strategy by an additive, fixed-width, Gaussian noise.

\subsection{Alternative models of response selection: `rational inattention' and \textit{ad~hoc} repetition probability}

Aside from posterior sampling and motor noise, the so-called `rational inattention' approach which has been gaining grounds in economics \parencite{Sims2003,Woodford2009,Sims2011}, suggests a different account of the variability of responses in decision-making tasks. In complement to the study of Bayesian and approximate Bayesian models, we have examined models inspired by that approach. We summarize, here, our results, and provide a more detailed discussion in \nameref{sec:Methods}. Rational-inattention models posit the existence of a cognitive cost which prevents subjects from making optimal decisions. In a standard formulation of the approach, this cost is assumed to be proportional to the mutual information between a subject's mental representation (of quantities relevant to produce responses) and the external variables relevant to the decision (here, the sequence of presented stimuli). The subject optimizes the `information structure', i.e., the distribution of the mental representation conditional on the observed stimuli, under the cognitive cost. The optimal distribution of responses depends on both the (prior) distribution of stimuli and the form of the reward function. We implement this model and compute its BIC (see \nameref{sec:Methods} for details). We find that it is much larger (by 29,656) than the BIC of the \textit{OptimalInference+NoisyMax} model, which itself has a larger BIC than most of our other models (Fig. \ref{fig:bics}). Hence, a direct application of a rational inattention approach does not provide a better account of behavioral data than the addition of a Gaussian noise in response selection following optimal inference.

Faced with a similar issue, \cite{Khaw2017} introduced a variant of the rational-inattention model that is particularly relevant to our study, as it applies to a sequential inference task with (history-independent) change points. In this variant, the response selection is split into a two-stage decision process: first, the subject decides whether to repeat the previous response; second, only if the decision is made not to repeat, then the subject chooses the location of a new response; and both decision stages are subject to cognitive costs. This presents a difference with the models that we have analyzed so far, in that an \textit{ad~hoc} probability of repetition is included explicitly in the model, whereas our approach, instead, was to study deviations from optimality that resulted from deterministic or stochastic approximations of a Bayesian scheme.

We have analyzed our data using a model similar to the one proposed by \cite{Khaw2017}. In order to evaluate the relevance of a rational-inattention information structure, we have also studied a model that include a two-stage decision process, but does not involve cognitive costs. Specifically, this model combines the \textit{OptimalInference} strategy with a strategy of response selection in which at each trial the model subject chooses, with fixed probability, whether to repeat the previous response. The probability of a repetition is constant, in this last model, whereas in the rational-inattention model it depends on the stimulus history and on the location of the slider at the beginning of the trial.

These two models yield a BIC lower than that of our previously best-fitting model, suggesting that a two-stage response-selection process is worth considering as a candidate mechanism for sequential decision-making. The previously best-fitting model, however, makes use of the \textit{ParticleFilter} inference strategy, whereas the models just presented rely on the \textit{OptimalInference} strategy.
Hence, we implement an array of models that combine the same two-stage response-selection processes with, instead, the \textit{ParticleFilter} inference strategy.
With this inference strategy, the model in which the subject chooses whether to repeat with a fixed repetition probability results in a lower BIC (by 1,398) than the model in which the repetition probability is governed by a rational-inattention cognitive cost. Moreover, as in the other analyses conducted above, the models with a \textit{ParticleFilter} inference strategy all yield substantially lower BICs than their counterparts that make use of the \textit{OptimalInference} strategy. It appears, thus, that while the introduction of an explicit repetition probability in models improves their explanatory power, deriving this repetition probability from a simple form of cognitive cost does not provide a better account of behavioral data than positing a fixed repetition probability.

An explicit repetition probability alone is insufficient to capture human inference in our task. Instead, stochastic compression of beliefs, as illustrated by stochastic pruning or particle filtering, results in a closer match with experimental observations. (We provide details on the rational-inattention models, the fixed-repetition-probability models, and their BICs in \nameref{sec:Methods}.)

\subsection{Memory load and stochastic pruning during inference}

After rejecting the rational inattention and the sampling hypotheses for response selection, we are left with an unexplained modulation of the variability -- specifically, the relation between the magnitude of behavioral variability and the width of the Bayesian posterior at successive times (Fig. \ref {fig:variability}). Noisy maximization, which makes use of an additive random perturbation with fixed variance, leads to behavioral variability with constant variance; it is thus insufficient to explain the experimental observations. If modulated variability does not originate in the response-selection step, it must derive from the inference step. Out of the $28$ models we consider, the five best-fitting models implement either the \textit{ParticleFilter} or the $\tau$\textit{Sample} inference strategy, both of which rely on sampling \textit{during inference}. These strategies capture the trends in the variability of human responses, in both HI and HD conditions (Fig. \ref{fig:winners}).

Both these inference strategies reduce the memory load in the inference problem by stochastically trimming the posterior, in a fashion akin to the `pruning' model proposed by \textcite{Huys2012} in the context of a decision-tree task. In their decision model, the evaluation of a possible sequence of decisions is more likely to be curtailed (thus alleviating the dimensionality of the problem) if it appears to have a low value. Similarly, the \textit{ParticleFilter} and the $\tau$\textit{Sample} inference strategies ignore with a higher probability possible run-lengths and states that are less likely to be correct. Furthermore, we note that the $\tau$\textit{MaxProb} inference strategy also relies on pruning unlikely run-lengths, but it deterministically eliminates the most unlikely, in contrast to its stochastic counterpart, $\tau$\textit{Sample} --- which yields a better fit of the data. In explore-exploit problems, `Thompson sampling' \parencite{Thompson1933} refers to a strategy in which one `explores' by randomly choosing an action with the probability that this action maximizes the reward (instead of deterministically choosing the action most likely to maximize the reward). Several studies have reported that the responses of human subjects in explore-exploit tasks appeared consistent with Thompson sampling \parencite{Schulz2015,Speekenbrink2015,Gershman2018}. In this perspective, the stochastic pruning of the posterior in our best-fitting models appears as an exploration strategy, deployed during inference. Beyond the specifics of the pruning or exploration mechanisms, the main conceptual point, here, is that behavioral biases may result from an approximation to a Bayesian scheme that relieves memory load. A similar picture has been advanced in the context of prediction tasks, where `over-reaction' -- effectively a biased, enhanced learning rate -- results from the compression of information stored in memory \parencite{AzeredodaSilveira2019,Neligh2019,Afrouzi2020}.

Aside from stochasticity in the inference step and in the response-selection step, noise in the sensory observation is a possible alternate account of behavioral variability, discussed, among others, by \textcite{Stocker2006}  and \textcite{Drugowitsch2016}. The design of our experiment, however, minimized perceptual ambiguity: the stimulus we presented to the subject at each trial, in our task, was a white dot that clearly contrasted with the background, and which remained on the screen until the subject responded (the subject was thus free to look at it for as long as he or she wished). By contrast, in the two studies just mentioned, the stimuli consisted of low-contrast gratings presented for one second or less. We presume that our experimental design limited perceptual noise.

It would nevertheless be interesting to know whether and how perceptual noise may contribute to behavioral variability in a sequential experiment, and how it may couple with stochasticity in inference. A possibility is that the magnitude of perceptual noise is constant throughout the task, in which case it would be expected to contribute an equal amount of variability at all run-lengths; if so, it would not account for the modulations of variability that we record in our task. Another possibility is that perceptual noise itself adapts dynamically during the task. Under this hypothesis, we speculate that the magnitude of perceptual noise would decrease if uncertainty increases; if so, it would result in an effect on behavioral variability \textit{opposite} to the observed effect. Although we cannot exclude that an effect of this nature is at play, it does not appear to offset completely the modulations of the variability which can be understood in terms of an approximate Bayesian inference. In sum, in the present setting of the experiments, a natural explanation of the behavioral variability in terms of perceptual ambiguity seems unlikely.

\subsection{Incorrect belief about the temporal statistics of the signal}

Comparing the behaviors of the best-fitting model and that of the subjects, we note that there remain discrepancies between the two, particularly at long run-lengths in the HD condition. The increase in the subjects’ learning rate, and the reduction in their repetition propensity, at these run-lengths and in this condition, are not as sharp as those of the best-fitting model (Fig. \ref{fig:winners}). A candidate explanation of these deviations is that the subjects hold an inexact belief on the shape of the change probability, $q(\tau)$, as a function of the run-length. The analysis of the \textit{IncorrectQ} inference strategy, in \nameref{sec:Methods}, examines the case of a model subject who believes that the change probability increases more slowly ($\lambda < 1$) than it actually does and that the average interval length is greater ($T>10$) than it actually is in our task. The learning rate of this model subject does not increase as abruptly, and the repetition propensity does not decrease as quickly, as those of the best-fitting model, similarly to the behavior of actual subjects (Fig. \ref{fig:qprior}, middle panels).
Moreover, among the five deterministic inference strategies considered, \textit{IncorrectQ} is the best-fitting strategy, regardless of the response-selection it is combined with, and with both model-comparison measures --- NMSE (Table \ref{tbl:checker}) and BIC (Fig. \ref{fig:bics}). This suggests that, aside from the stochastic compression of the posterior, the subjects’ deviations from optimality may also result, to some extent, from an incorrect belief in the temporal structure of the signal.

\subsection{Inference through sample-based representations of probability}

The \textit{ParticleFilter} strategy is noteworthy in a number of respects. First, it is our best-fitting model. Second, it constitutes a generic approach to inference; it was reported to account successfully for other inference and learning behaviors, such as category learning \parencite{Sanborn2006, Sanborn2010}, conditioning in pigeons \parencite{Daw2008}, sentence processing \parencite{Levy2008}, hidden state inference \parencite{Brown2009}, and visual tracking of multiple objects \parencite{Vul2009}. Third, out of all the models we consider, it is by far the less demanding on memory: with nine particles, one needs to store 27 numbers (for $s$, $\tau$, and the weight of each particle) in memory. As a comparison, the optimal model stores a discretized probability distribution over the $(s,\tau)$ space, which amounts to about 16000 numbers (the optimal posterior could be well approximated with less memory-intensive methods, but this would require further hypotheses.) Previous uses of particle-filter methods in the context of a variety of cognitive tasks yielded best-fitting numbers of particles which ranged from one to several hundreds: from one to 400 particles, with a mean of 56, in \textcite{Brown2009}; 130 particles (but 70 when subjects simultaneously perform a distractor task) in \textcite{Thaker2017}; around 20 particles in \textcite{Levy2008}; 20 particles in \textcite{Glaze2018}; and as few as one particle in \textcite{Daw2008} and \textcite{Sanborn2010}. In addition, \textcite{Bramley2017} consider only the case of a single particle. We note that our analysis of the \textit{ParticleFilter} inference strategy, detailed in \nameref{sec:Methods}, reveals that a model with just one particle fails to reproduce the decreasing learning rates, in the HI condition, and the smile shape of the learning rates, in the HD condition, while models with two or more particles do capture these behavioral trends (Fig. \ref{fig:pf}B). Hence, in contrast to the last three studies cited, we find that a particle-filter model with a single particle is qualitatively inconsistent with the behavior of human subjects, at least in the context of our task.

A fourth aspect of particle filters is that they provide a natural interpretation of the high repetition propensity observed in subjects (Fig. \ref{fig:metrics}B). As the support of the probability distribution is reduced to $N_{P}=9$ points on the $(s,\tau)$ plane, there is a fair chance that the posterior-maximizing particle at trial $t$, $(s_{t},\tau _{t})$, remains the posterior-maximizing particle at trial $t+1$. Hence, the response $s_{t}$ is likely to be repeated. In a similar spirit, particle filters have been shown to account for order effects in category learning \parencite{Sanborn2010} and observations about online sentence comprehension (such as the processing of `garden-path sentences' \parencite{Levy2008}).

The success of particle filters, also known as Sequential Monte Carlo method, in accounting for human behavior in an online inference task adds to a growing literature on sample-based representations in cognitive processes \parencite{Goodman2008, Moreno-Bote2011, Gershman2012, Vul2014}. Monte Carlo methods, which approximate probability distributions with sets of samples, constitute a major element of a family of techniques used in machine learning to address a wide range of problems (inference, optimization, numerical integration, etc); they have also been put forth as candidate cognitive algorithms \parencite{Sanborn2015, Gershman2016}. Moreover, they account for a range of cognitive biases in the laboratory, such as base-rate neglect, conjunction fallacy, and the unpacking effect, as well as for human performance in complex, real-world tasks, and specific observations such as response variability and autocorrelation in perception and reasoning tasks \parencite{Gershman2012, Sanborn2016}. At the implementation level, sample-based representations are well suited to learning in neural networks \parencite{Fiser2010}. Here, the variability in neural activity can be interpreted in terms of sampling-based representations of probability \parencite{Hoyer2003, Fiser2010, Buesing2011, Gershman2012}, and a number of neural network models performing probability sampling have been proposed \parencite{Shi2009, Moreno-Bote2011, Hennequin2014, Savin2014, Aitchison2016}.

\section{Methods}
\label{sec:Methods}

\subsection{Details of the behavioral task}
The computer-based task was programmed and run with Psychopy \parencite{Psychopy}. In this task, white dots appeared on a horizontal line in the middle of a grey screen. Subjects were told that these white dots were snowballs thrown by a hidden person, the `enemy' (also located on the horizontal line). The horizontal location of a snowball was the stimulus, $x_t$, and the position of the hidden person was the state, $s_t$. The state space was arbitrarily chosen to be [0, 300]; this scale did not appear on the screen. By clicking with a mouse (whose pointer moved on the horizontal axis only), subjects could indicate where they thought the hidden person was (i.e., give their estimate, $\hat s_t$, of the state). The time of response was not constrained. A green dot provided a visual feedback of the location of the click. After 100ms, a new white dot appeared, starting the next trial (Fig. \ref{fig:task}A-B). If a subject's `shot' was within a fixed distance around the state (the radius of the enemy), the subject was rewarded with 1 point. If the shot was `outside the enemy' but within a distance equal to twice the enemy radius, the reward was 0.25 point (Fig. \ref{fig:task}E). Otherwise, the reward was zero. Subjects were not informed of the reward immediately after each shot, as this would have provided additional information on the location of the state. The total score was given every 100 trials, to allow for an assessment of average self-performance and to foster motivation.

\subsection{Subjects}

We ran the computer-based task on 30 paid subjects; all gave informed consent. The study was approved by Princeton University's Institutional Review Board for Human Subjects. The sample size was determined so as to be comparable to that used in similar experiments \parencite{Nassar2010, Nassar2012}. Four subjects performed significantly worse than the other ones: their average error, defined as the absolute difference between their estimate and the state, $|\hat s_t - s_t|$, was 10.4 (standard deviation (s.d.): 0.93), while the average error of the other 26 subjects was 6.5 (s.d.: 0.62). Because of this difference of more than 5 standard deviations, these four subjects were excluded from the analyses. Hence, a total of 26 subjects were included in the analyses. Our conclusion remain unchanged if all 30 subjects are included in the analyses (see Supplementary Fig. \ref{fig:metrics-h-4bg}).

\subsection{Details of the signal}

The stimulus, $x_t$, was generated around the state, $s_t$, according to the likelihood probability, $g(x_t | s_t)$, which was chosen to be triangular, centered at $s_t$, and of half-width 20. The state, $s_t$, was piecewise-constant with respect to time, i.e., constant in the absence of a change point. In the HI condition, the probability of a change point, $q_t$, was constant and equal to 10\%. In the HD condition, $q_t$ depended on the run-length, $\tau_t$, defined as the number of trials since the last change point, and had a sigmoid shape: $q_t(\tau_t) = 1/(1+e^{-(\tau_t-10)})$. At $\tau_t=0$ (i.e., immediately after a change point), the probability of another change point was very small. Six trials after a change point it was still small, less than 2\%, before growing appreciably (50\% at$\tau_t=10$, 95\% at $\tau_t=13$). This led to more regular intervals between change points than in the HI condition, with a change point roughly every 10 trials (Fig. \ref{fig:task}C-D). The average number of trials between two change points in both conditions was 10. At a change point, the state randomly jumped to a new state, $s_{t+1}$, according to the state transition probability, $a(s_{t+1} | s_t)$. This distribution was chosen to be bimodal, symmetric, and centered at $s_t$ (two triangles of half-width 20 each, centered at $s_t \pm d$, where $d=25$, Fig. \ref{fig:task}E). This prevented new states to be too close or too far from the previous state, which would have made change-point detection too difficult or too obvious.

\subsection{Training runs}

All subjects did the task in both HI and HD conditions. 14 started with the HD task and 16 started with the HI task (no significant difference were found in results between these two groups).
Subjects were not told the specificity of each situation. An explanatory text indicated that there were ``differences'' between each condition but no further indications were given. Each condition started with a series of explanations and tutorial runs.
In a first tutorial run, the enemy (i.e., the state) was visible and moved according to the current (HI or HD) condition, and successive snowballs appeared without any action from the user (as in passive video viewing).
In a second run, the enemy was still visible and subjects had to click at each trial, after which the next snowball would appear. This run was a very simple version of the actual task, because subjects were seeing the state.
In a third run, the half-width of the triangular likelihood, $g(x_t|s_t)$, was 10, i.e., half the value it took in the actual task. In this run, the state was not visible, except after a change point: in the occurrence of a change point, the position of the state before the change point was shown, along with the shots of the subject since the previous change point. This run had two goals: first, to emphasize the timing of change points, and, second, to allow for self-performance assessment and to illustrate that a strategy consisting in `following the white dots', i.e., clicking on the stimuli, was inefficient.
A fourth tutorial run was an `easy' version of the actual task: the state was always hidden, but the likelihood, $g$, had a half-width of 10.
A fifth and last tutorial run reproduced the third run, but with the likelihood, $g$, with half-width 20.
During the task, 15 subjects (7 amongst the HD-first group and 8 amongst the HI-first group) were also shown the positions of past stimuli, as white dots with decreasing contrast, gradually merging with the grey background (Fig. \ref{fig:task}). The other 15 subjects were not shown past stimuli. No significant differences were found in data between the two groups. The number of stimuli presented in the tutorial runs totaled 297 for each condition. During the actual task there were 1000 trials in each condition, leading to a total of 2000 data points per subjects.

\subsection{Empirical run-length}

As subjects did not know the true run-length, $\tau$, we computed an empirical run-length, $\tilde \tau$, based on the responses of subjects. Whereas the true run-length is defined as the number of trials since the last change point, the empirical run-length is defined as the number of trials since the last `large correction'; a large correction is defined as a correction with absolute value larger than the 90th percentile of corrections. This percentile level is chosen in relation to the average frequency of change points, 1 for every 10 trials, in both HI and HD conditions. In some occasions, a subject ``misses'' a change point: the run-length and the empirical run-length, consequently, differ. For instance, $\tilde \tau = 10$, while $\tau = 0$ or 1. In such a case, because the change point did occur, the subject experiences a large surprise and is thus likely to subsequently opt for a large correction, i.e., to increase the learning rate. In the HD condition, because of the temporal statistics of change points, this situation is more likely to occur at empirical run-lengths around 10. Hence, this effect could bias the learning rates to higher values at these empirical run-lengths, in this condition. This effect, however, does not originate in the inference process, but rather in the temporal statistics of the HD signal. In other words, even an observer whose inference algorithm is not adapted to the HD condition would have higher learning rates at empirical run-lengths around 10. In the results presented, we removed all trials with a (true) run-length of 0 or 1, in order to avoid this artifact.

\subsection{Regression of the learning rate on run-lengths}

In order to provide statistical evidence of the smile shape of the learning rates in the HD condition (and the absence of a smile shape in the HI condition), we regress the learning rate on the run-lengths, with a quadratic term. For the HD condition, we find that the coefficient for the quadratic term is positive and significantly different from zero (0.0046; p-value = 1e-11). For the HI condition, this coefficient is smaller and we cannot reject at a significance level of 5\% the null hypothesis that it is zero (0.0013; p-value = 0.068). Moreover, the difference between the two quadratic coefficients is statistically significant (F-test p-value = 5.7e-4). The coefficient for the linear term is significantly negative (p-value < 1e-2).

\subsection{Bayesian update equation}

We derive the Bayesian update equation for a general case with $q = q(s_t, \tau_t)$, $p(x_t|s_t, \tau_t) = g(x_t|s_t, \tau_t)$, and $p(s_{t+1}|\tau_{t+1}=0, s_t, \tau_t) = a(s_{t+1}|s_t, \tau_t)$, which includes the case used in our task with $q=q(\tau_t)$, $g=g(x_t|s_t)$, and $a=a(s_{t+1}|s_t)$.

Our goal is to obtain an update rule for the posterior, $p_t(s, \tau | x_{1:t})$ upon the observation of a new stimulus, $x_{t+1}$. Bayes' rule yields
\begin{equation}\label{eq:bayes}
p_{t+1}(s, \tau | x_{1:t+1}) = \frac{1}{Z_{t+1}} g(x_{t+1} | s, \tau) p_{t+1}(s, \tau | x_{1:t}),
\end{equation}
where $Z_{t+1} = p_{t+1}(x_{t+1} | x_{1:t})$ is a normalization constant. The third term in this product can be written as
\begin{equation}\label{eq:margin}
p_{t+1}(s, \tau | x_{1:t}) = \sum_{\tau_t} \int_{s_t} p_{t+1}(s, \tau | s_t, \tau_t ) p_t( s_t, \tau_t | x_{1:t} ) \D s_t .
\end{equation}
The transition probability, $p_{t+1}(s, \tau | s_t, \tau_t )$, is determined by $q$ and $a$.
An absence of change point occurs with probability $1-q(s_t, \tau_t)$, and in such a case a state $(s_t, \tau_t)$ evolves into the state ${(s_{t+1} = s_t, \tau_{t+1} = \tau_t + 1)}$. In the case of a change point, an event which occurs with probability $q(s_t, \tau_t)$, possible states at $t+1$ have the form $(s_{t+1}, \tau_{t+1} = 0)$. Hence the transition probability from $(s_t, \tau_t)$ to $(s, \tau)$ at $t+1$ is given by
\begin{equation}\label{eq:transition}
p_{t+1}(s, \tau | s_t, \tau_t ) = \mathbbm{1}_{\tau=0} \, q(s_t, \tau_t) a(s_t, \tau_t, s ) + \mathbbm{1}_{\tau = \tau_t + 1, s = s_t } ( 1 - q(s_t, \tau_t)).
\end{equation}
Combining Equations (\ref{eq:bayes}), (\ref{eq:margin}), and (\ref{eq:transition}), we obtain the Bayesian update equation, as
\begin{equation}\label{eq:update}
\begin{split}
p_{t+1}(s, \tau | & x_{1:t+1}) = \frac{1}{Z_{t+1}} g(x_{t+1} | s, \tau) \Bigg[  \\ 
    & \mathbbm{1}_{\tau=0} \, \sum_{\tau_t} \int_{s_t} q(s_t, \tau_t) a(s_t, \tau_t, s ) p_t( s_t, \tau_t | x_{1:t} ) \D s_t \\
 + & \mathbbm{1}_{\tau>0} \, ( 1 - q(s, \tau-1)) p_t( s, \tau - 1 | x_{1:t} ) \Bigg].
\end{split}
\end{equation}

In the special case with $q = q(\tau_t)$, $a = a(s_{t+1}|s_t)$, and $g = g(x_t | s_t)$, we obtain the slightly simpler Eq. (\ref{eq:HDupdate}). In addition, we note that, in the HI condition, the change probability is constant, $q(\tau)=q$; in this condition, we can marginalize over the variable $\tau$ to obtain a closed recursion over the state posterior, as
\begin{equation} \label{eq:HIupdate}
p_{t+1}(s | x_{1:t+1}) = \frac{1}{Z_{t+1}} g(x_{t+1} | s ) \Bigg[
    q  \int_{s_t} a(s | s_t) p_t( s_t | x_{1:t} ) \D s_t 
 + ( 1 - q) p_t( s  | x_{1:t} ) \Bigg].
\end{equation}

\subsection{Derivation of the suboptimal models and analysis of their behaviors}

\subsubsection{\textit{IncorrectQ} model}

In the \textit{IncorrectQ} model, the two quantities governing the shape of $q(\tau)$, $\lambda$ and $T$, are treated as free parameters, and we explore how varying these parameters impacts behavior, as compared to the \textit{OptimalInference} model.

\begin{figure}[thb]
\centering
        \includegraphics[width=\textwidth]{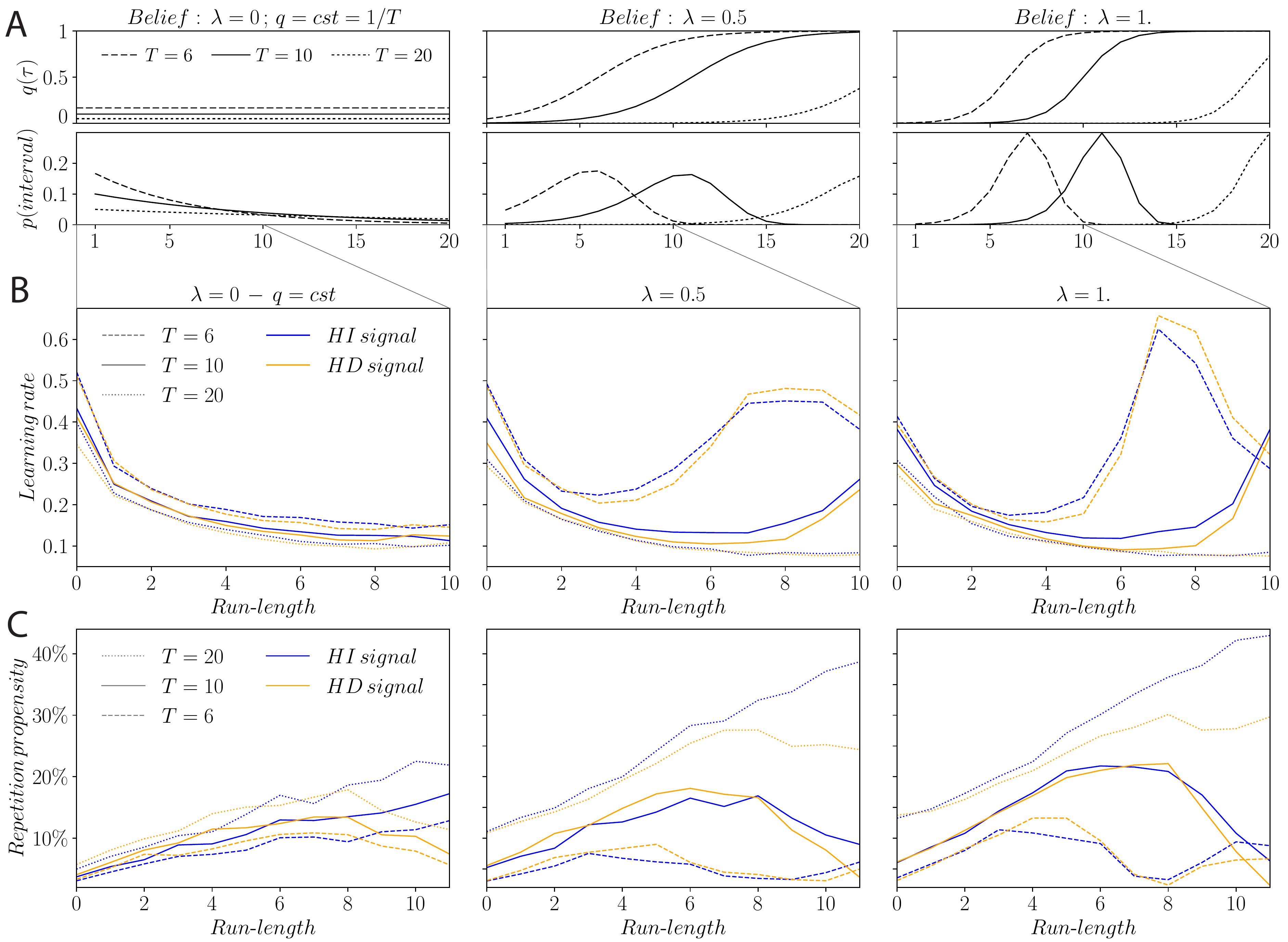}
	\caption{\textbf{Illustration of the \textit{IncorrectQ} model with various beliefs on the shape of the change probability.}
	\textbf{A.} Examples of beliefs in \textit{IncorrectQ} models. Change probability, $q(\tau)$ (top line), and resulting inter-change-point interval distribution (bottom line), for constant $q$ (left column), sigmoid-shaped $q(\tau)$ with slope parameter $\lambda=0.5$ (middle) and $\lambda=1$ (right); and for average interval length, $T$, of 6 (dashed line), 10 (solid), and 20 (dotted). The `true' HI signals used in the task correspond to  $\lambda=0$, $T=10$, while the HD signals correspond to $\lambda=1$, $T=10$.
	\textbf{B, C.} Average learning rate (B) and repetition propensity (C) as a function of the run-length, in \textit{IncorrectQ} models performing optimal inference with various beliefs on the change probability, $q(\tau)$, and presented with HI signals (blue) and HD signals (orange).
	}\label{fig:qprior}
\end{figure}

Keeping $T$ constant at $10$, and varying $\lambda$ from $0$ (HI condition) to $1$ (HD condition), we find that the learning rate as a function of the run-length gradually morphs from the HI, monotonically decreasing curve, to the HD, non-monotonic and `smile-shaped' curve (Fig. \ref{fig:qprior}B). A similar behavior obtains at any fixed value of $T$, with the difference that the minimum of the HD curve is shifted to smaller $\tau$ for smaller $T$, and to larger $\tau$ for larger $T$. In other words, for fixed $T$, a higher value of $\lambda$, i.e., a sharper slope of the change probability, leads to a higher learning rate at run-lengths comparable to $T$. (Note that, for $T=20$, the minimum of the learning rate occurs at run-lengths larger than $10$, hence the non-monotonicity is not apparent in Fig. \ref{fig:qprior}B.) Conversely, for a fixed $\lambda>0$, the minimum of the learning rate occurs at a run-length comparable to $T$ which, precisely, determines when change points become likely. Finally, for $\lambda=0$, the change probability is constant and there is no increase in the learning rates; these are, however, slightly higher for smaller $T$, because in that case the change probability, $q=1/T$, is larger, so a new stimulus is more likely to be interpreted as stemming from a change point.

A subtlety that any analysis has to grapple with is that the statistics of responses depend not only on the inference process, but also, of course, on the statistics of the stimuli. To tease the two effects apart, for each \textit{IncorrectQ} model subjects (with differing values of $\lambda$ and of $T$) we computed the response behavior in presence of either signal: the HI signal, characterized by a constant change probability, $q=0.1$, and the HD signal, characterized by a change probability, $q(\tau)$, varying with the run-length as a sigmoid with parameters $\lambda=1$ and $T=10$. We note that the impact on behavior of changing the signal is modest, as compared to the impact of changing the model subject's beliefs (Fig. \ref{fig:qprior}B). This indicates that the discrepancy in human behavior in the HI and HD conditions does not originate primarily from the statistics of the signals, but rather from the different beliefs on the temporal statistics of the signals, held by the subjects.

Paralleling the behavior of learning rates, the repetition propensity in the HD condition peaks earlier or later depending on the value of $T$, and its shallowness depend on the value of $\lambda$ (Fig. \ref{fig:qprior}C). A belief in a shorter average inter-change-point interval, $T$, leads to a smaller repetition propensity: assuming frequent change points enhances the frequency of changes in one's estimate.

Human subjects correctly believe that $q$ is not constant in the HD condition, and they use this belief in their inference process, but they may hold an inexact representation of the shape of $q(\tau)$ (Fig. \ref{fig:qprior}). This, however, is not sufficient to capture data quantitatively: subjects exhibit both higher learning rates and more frequent repetitions than in the optimal model (Fig. \ref{fig:metrics}), an observation that cannot be explained by manipulating $\lambda$ and $T$ in the \textit{IncorrectQ} model; in the latter, high learning rates are accompanied by lower repetition propensity, and vice versa. Thus, and letting alone the issue of variability, an erroneous belief on the change probability, $q(\tau)$, is insufficient to model experimental data.

\subsubsection{$\tau$\textit{Mean} model}

\paragraph{Derivation} This model is a generalization of the model introduced by \textcite{Nassar2010}. The approximate joint probability of the state and the run-length, which we denote by $\tilde p_t(s, \tau | x_{1:t})$, is assumed, in this model, to vanish at all values of the run-length, except for one, which we call the `approximate expected run-length' and which we denote by $\bar \tau_t$. Hence,
\begin{equation}
\tilde p_t(s, \tau | x_{1:t}) = \delta_{\tau, \bar \tau_t} \, \tilde p_t(s, \bar \tau_t | x_{1:t}),
\end{equation}
where $\delta_{\tau, \bar \tau_t}$ is the Kronecker delta. As in the optimal model (see Eq. (\ref{eq:HDupdate})), we use Baye's rule and the parameters of the task to derive the update equation, as
\begin{equation}\label{eq:tauMeanUpdate}
\begin{split}
p_{t+1}(s, \tau | x_{1:t+1}) = \frac{1}{Z_{t+1}} g(x_{t+1} | s ) \Bigg[
    & \mathbbm{1}_{\tau=0} \, q(\bar \tau_t) \int_{s_t} a(s | s_t) p_t( s_t, \bar \tau_t | x_{1:t} ) \D s_t \\
 + & \mathbbm{1}_{\tau=\bar \tau_t +1} \, ( 1 - q(\bar \tau_t)) p_t( s, \bar \tau_t | x_{1:t} ) \Bigg].
\end{split}
\end{equation}
This distribution is non-vanishing for two values of the run-length, $0$ and $\bar \tau_t + 1$, which correspond to the two possible scenarios: with and without a change point at trial $t$. We use this distribution to compute the approximate expected run-length at trial $t+1$, $\bar \tau_{t+1}$, and the approximate posterior at trial $t+1$, $\tilde p_{t+1}(s, \tau | x_{1:t+1})$. First, we obtain the probability of a change point at trial $t+1$,
\begin{equation}
\Omega_{t+1} \equiv p_{t+1}(\tau=0 | x_{1:t+1}) = \frac{1}{Z_{t+1}} q(\bar \tau_t) \int_s g(x_{t+1}|s) \int_{s_t}  a(s|s_t) p_t(s_t, \bar \tau_t| x_{1:t}) \D s_t \D s,
\end{equation}
and we use it to compute the approximate expected run-length at trial $t+1$:
\begin{equation}
\bar \tau_{t+1} = \Omega_{t+1} \cdot 0 + (1-\Omega_{t+1}) (\bar \tau_t + 1 ).
\end{equation}
Second, we approximate the posterior (Eq. (\ref{eq:tauMeanUpdate})) by marginalizing it over the run-lengths, and multiplying the result by a Kronecker delta which takes the value $1$ at $\bar \tau_{t+1}$:
\begin{equation}
\label{eq:tauMeanPosteriorApprox}
\begin{split}
p_{t+1}(s | x_{1:t+1}) = \sum_\tau p_{t+1}(s, \tau | x_{1:t+1}) & = p_{t+1}(s, \tau = 0 | x_{1:t+1}) + p_{t+1}(s, \tau = \bar \tau_t + 1 | x_{1:t+1}), \\
\tilde p_{t+1}(s, \tau | x_{1:t+1}) & = \delta_{\tau, \bar \tau_{t+1}} p_{t+1}(s | x_{1:t+1}).
\end{split}
\end{equation}
This model has no parameter.

\paragraph{Behavior} While the optimal marginal distribution of the run-lengths, $p_t(\tau|x_{1:t})$, spans the whole range of possible values of the run-length, it is approximated, in the $\tau$\textit{Mean} model, by a delta function on a single value, $\bar \tau_t$.
In the HI condition, the change probability, $q$, does not depend on the run-length, $\tau$; the approximate joint distribution evaluated at $\tau=\bar \tau_t$, $\tilde p_t(s, \bar \tau_t | x_{1:t})$, is equal to the optimal posterior distribution on the state, $p_t(s|x_{1:t})$. (Compare Eq. (\ref{eq:HIupdate}) to the combination of Eqs. (\ref{eq:tauMeanUpdate}) and (\ref{eq:tauMeanPosteriorApprox}).) As a result, the $\tau$\textit{Mean} model computes the optimal posterior on the state, and, thus, the responses in this model are the same as those in the optimal model, i.e., the $\tau$\textit{Mean} model is optimal in the HI condition. In the HD condition, by contrast, the change probability, $q(\tau)$, depends on the run-length. The $\tau$\textit{Mean} model evaluates this function at only one run-length, $\bar \tau_t$, an approximation of the mean run-length; as compared to the optimal model, it fails to capture fully the consequences of the dependence of the change probability on the run-length (Fig. \ref{fig:task}D). The learning rates in this model are higher than the optimal ones for short run-lengths, and lower than the optimal ones for long run-lengths (Fig. \ref{fig:limitedtaus}B); and the repetition propensities are lower than the optimal ones for short run-lengths, and higher than the optimal ones for long run-lengths (Fig. \ref{fig:limitedtaus}C).

\subsubsection{$\tau$\textit{Nodes} model}
\paragraph{Derivation} This model generalizes the one introduced by \textcite{Wilson2013}. In this paper, the authors interpret a change-point setting similar to ours as a `message-passing' graph where run-lengths are nodes, weighted by their marginal probability $p_t(\tau | x_{1:t})$, edges are characterized by the change probability, $q$, and `messages' are passed along edges from one node to another. More precisely, we compute the marginal probability of the run-length, using Eqs. (\ref{eq:margin}) and (\ref{eq:transition}):
\begin{equation}
p_{t+1}(\tau | x_{1:t}) =
	\begin{cases}
   		\sum_{\tau_t} q(\tau_t)p_t(\tau_t | x_{1:t}) & \text{if } \tau=0\\
    		( 1 - q(\tau-1)) p_t(\tau - 1 | x_{1:t} )          & \text{otherwise}
	\end{cases}.
\end{equation}

Hence, at trial $t+1$, the weight of a node $\tau$ (i.e., the marginal probability of the corresponding run-length) is equal, if $\tau=0$, to a sum of the marginal probabilities of all nodes at trial $t$, $\tau_t$, weighted by their corresponding change probabilities, $q(\tau_t)$; and, if $\tau>0$, it is the probability of the node $\tau-1$, at trial $t$, weighted by the probability that there was no change, $1-q(\tau-1)$. Taking a different view, one can reformulate these weighted sums of probabilities as transfers of probability masses, as follows. Each node $\tau$ sends two `messages': a `no-change-point' message is sent to node $\tau+1$ so as to set its weight to $(1-q(\tau))p_t(\tau | x_{1:t})$, and a `change-point' message is sent to node $\tau=0$ to increase its probability by $q(\tau) p_t(\tau | x_{1:t})$. This is the message-passing algorithm. \textcite{Wilson2013} assume a change probability, $q$, that is constant; a likelihood, $g$, which belongs to the exponential family; and a state transition probability, $a(s_{t+1})$, that does not depend on the previous state, $s_t$, and which is a conjugate prior of the likelihood. They show that in this case each node can be seen as implementing a delta-rule, and the optimal Bayesian model amounts to the weighted sum of these delta-rules.

The authors then `reduce' this model by removing nodes and accordingly revise the message-passing algorithm and each node's update rule. We only focus on the aspects of the model that will be used in our $\tau$\textit{Nodes} implementation. The set of new nodes comprises `virtual' run-lengths $l \in \{l^0, l^1, ..., l^N\}$. A node $l^i$, with $i \neq 0$, now sends three messages: one to $l^0$, one to the next node $l^{i+1}$, and one to itself. The `change-point' message remains the same as in the previous algorithm, i.e., the quantity $q(l^i)p_t(l^i|x_{1:t})$ is sent to $l^0$ (i.e., this quantity is added to the probability of this node). The `no-change-point' message is now split in two, one message being sent to the next node, $l^{i+1}$, and the other one being a self-passing message (i.e., sent to itself, $l^i$). The authors seek the relative weight $w(l^i)$ assigned to the self-passing message which gives an average run-length increase of 1 (i.e. $\E ( l_{t+1} | l^i_t, \mathrm{no\,change}) = l^i + 1$). They find $w(l^i) = \frac{l^{i+1}-l^i-1}{l^{i+1}-l^i}$ for $i \neq N$ and $w(l^N)=1$. The next node, $l^{i+1}$, hence receives the message $(1-w(l^i))(1-q(l^i))p_t(l^i | x_{1:t})$. With the assumptions mentioned above on $q$, $g$, and $a$, the model can again be understood as `a mixture of delta-rules'.

We implement these ideas in our $\tau$\textit{Nodes} model. Instead of $p_t(s, \tau | x_{1:t})$, we consider the probability distribution $p_t(s, l | x_{1:t})$ and apply the same Bayesian and marginalization equations used in Eqs. (\ref{eq:bayes}) and (\ref{eq:margin}). The main difference of the new model resides in the transition probability, $p_{t+1}(s,l | s_t, l_t)$, which becomes
\begin{equation}\label{eq:transition_rlnodes}
\begin{split}
p_{t+1}(s, l | s_t, l_t ) = \mathbbm{1}_{l=l^0} \, & q(s_t, l_t) a(s_t, \tau_t, s ) \\
	+ \mathbbm{1}_{l = l_t, s = s_t } & ( 1 - q(s_t, l_t)) w(l_t) \\
	+ \mathbbm{1}_{l = l_t + 1, s = s_t } & ( 1 - q(s_t, l_t)) (1 - w(l_t)).
\end{split}
\end{equation}

Combining Eq. (\ref{eq:transition_rlnodes}) to Eqs. (\ref{eq:bayes}) and (\ref{eq:margin}) adapted with $l$, we obtain an update equation similar to the full Bayesian update equation (Eq. (\ref{eq:update})), with an additional term corresponding to the ability of nodes for self-passing messages. The model is parameterized by the number of nodes, $N_\tau$, but also by the values of the nodes. We chose the possible values of the nodes to be in the set \{0, 2.5, 5, 7.5, 10, 12.5\}. When fitting the model for a given $N_\tau$, all the models corresponding to every possible choice of $N_\tau$ nodes within these values, were computed, and the best-fitting one was chosen.

\begin{figure}[!b]
	\centering
        \includegraphics[width=\textwidth]{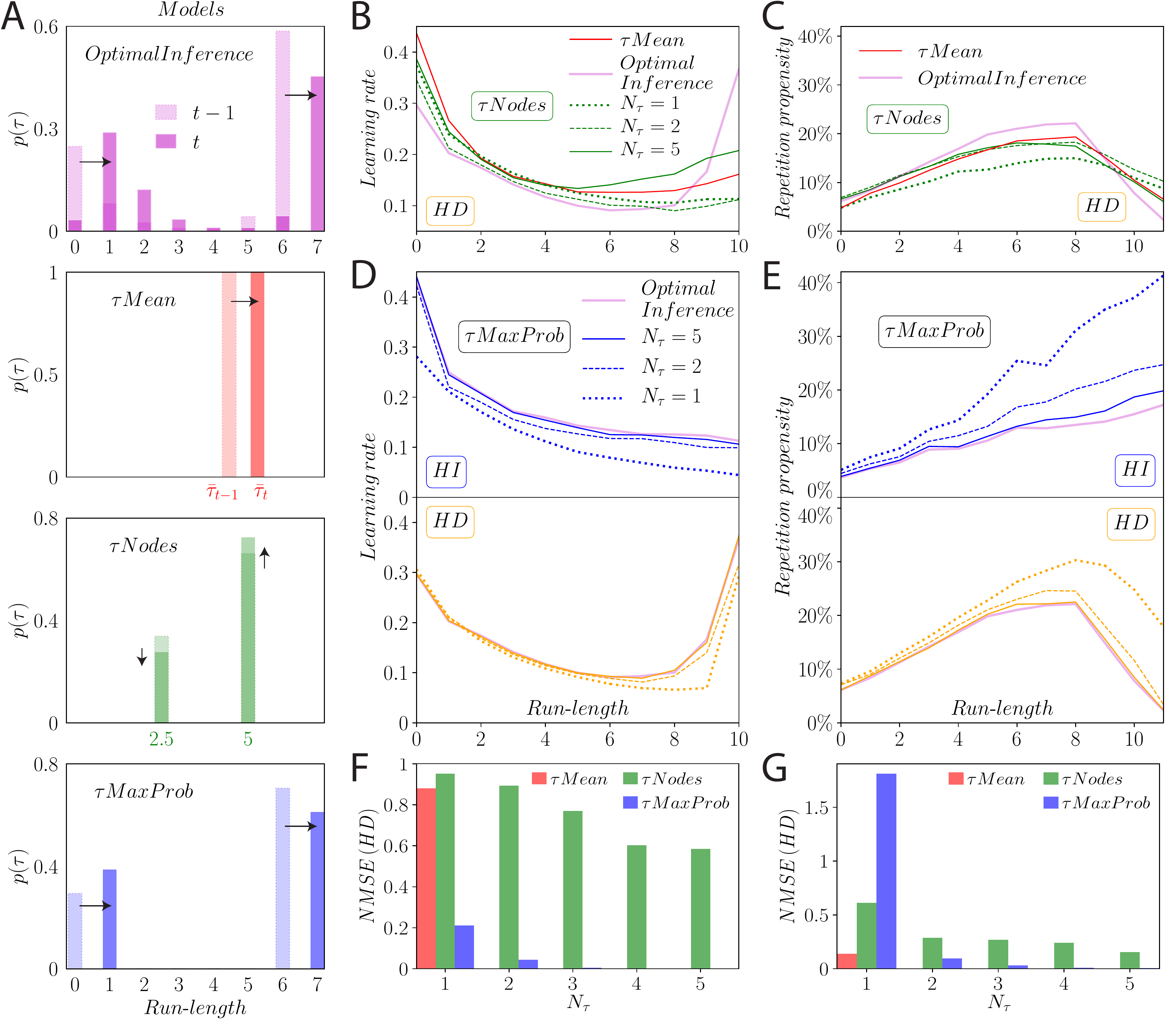} 
	\caption{\textbf{Behavior of the limited-memory models, as compared to the \textit{OptimalInference} model.}
	\textbf{A.} Schematic illustration of the marginal distribution of the run-length, $p(\tau)$, in each model considered. The \textit{OptimalInference} model assigns a probability to each possible value of the run-length, $\tau$, and optimally updates that distribution upon receiving stimuli (first panel). The $\tau$\textit{Mean} model uses a single run-length which tracks the inferred expected value, $\langle\tau\rangle$ (second panel). The $\tau$\textit{Nodes} model holds in memory a limited number, $N_\tau$, of fixed hypotheses on $\tau$ (``nodes''), and updates a probability distribution over these nodes; $N_\tau=2$ in this example (third panel). The $\tau$\textit{MaxProb} model reduces the marginal distribution by discarding less likely run-lengths; in this example, 2 run-lengths are stored in memory at any given time (fourth panel).
	\textbf{B, C.} Average learning rate (B), and repetition propensity (C), as functions of the run-length, in the \textit{OptimalInference} model, the $\tau$\textit{Mean} model, and the $\tau$\textit{Nodes} model with $N_\tau=$ 1, 2, and 5, in the HD condition. The HI condition is not displayed as the $\tau$\textit{Mean} and $\tau$\textit{Nodes} models do not differ from the \textit{OptimalInference} model in this condition.
	\textbf{D, E.} Average learning rate (D), and repetition propensity (E), as functions of the run-length, in the \textit{OptimalInference} model and the $\tau$\textit{MaxProb} model with $N_\tau=$ 1, 2, and 5, in the HI condition (top panels) and in the HD condition (bottom panels).
	\textbf{F,G.} Normalized Mean Squared Error relative to the learning rate (F) and the repetition propensity (G), as compared to the \textit{OptimalInference} model, for the $\tau$\textit{Mean} model, which has no free parameter, and for the $\tau$\textit{Nodes} and $\tau$\textit{MaxProb} models with $N_\tau=$ 1 to 5.
} 
\label{fig:limitedtaus}
\end{figure}

\paragraph{Behavior} In the HI condition, the model computes the optimal posterior on the state, $p_t(s|x_{1:t})$. Thus, as for the $\tau$\textit{Mean} model above, the responses in the $\tau$\textit{Nodes} model are the same as those in the optimal model, in the HI condition.
In the HD condition, the greater the number of nodes, the more faithfully the model approximates optimal behavior (Fig. \ref{fig:limitedtaus}B,C). The learning rates are higher than the optimal ones for short run-lengths, and lower than the optimal ones for long run-lengths (Fig. \ref{fig:limitedtaus}B). The repetition propensities are higher than the optimal ones for long run-lengths; for short run-lengths, they are appreciably closer to the optimal ones, in the model with five nodes, than in the model with one node (Fig. \ref{fig:limitedtaus}C).

\subsubsection{$\tau$\textit{MaxProb} model}

\paragraph{Derivation} In the $\tau$\textit{MaxProb} model, we assume that we have, at trial $t$, an approximation of the joint distribution of the state and the run-length, which we denote by $\tilde p_t(s, \tau | x_{1:t})$, and we assume that the approximate marginal distribution of the run-lengths, $\tilde p_t(\tau|x_{1:t})$, is non-vanishing for no more than $N_\tau$ values of the run-length. Upon receiving a new stimulus, we perform a Bayesian update of the approximate joint distribution, $\tilde p_t(s, \tau | x_{1:t})$, as in Eq. (\ref{eq:HDupdate}), and obtain the posterior, $p_{t+1}(s, \tau | x_{1:t+1})$, from which we derive the marginal distribution of the run-lengths, $p_{t+1}(\tau | x_{1:t+1})$. If, at trial $t$, the run-length takes a given value, $\tau_t$, then, at trial $t+1$, it can only take one of two values: $0$ (if there is a change point) or $\tau_t+1$ (if there is no change point). Hence, if the marginal distribution at trial $t$, $\tilde p_t(\tau|x_{1:t})$, is non-vanishing for at most $N_\tau$ values, as we assume, then the updated distribution, $p_{t+1}(\tau | x_{1:t+1})$, is non-vanishing for at most $N_\tau+1$ values. In the case that this distribution is non-vanishing for \textit{less} than $N_\tau+1$ values, we do not perform further approximations, at this stage. In the other, more generic case, i.e., if $N_\tau+1$ values of the run-length have a non-zero probability, then we identify the most unlikely run-length, $\tau^{*} = \arg\min p_{t+1}(\tau | x_{1:t+1})$, and we approximate the posterior as
\begin{equation}
\label{eq:maxprob}
\tilde p_{t+1}(s,\tau|x_{1:t+1}) =
  \begin{cases}
    0 & \text{if } \tau = \tau^{*} \\
    \frac{1}{Z} p_{t+1}(s,\tau|x_{1:t+1}) & \text{if } \tau \neq \tau^{*}
  \end{cases},
\end{equation}
where $Z$ is a normalization constant equal to $1-p_{t+1}(\tau^{*}|x_{1:t+1})$.

\paragraph{Behavior} Even with $N_\tau=1$ (one memory slot), the $\tau$\textit{MaxProb} model captures qualitatively the optimal, high learning rates for large run-lengths, in the HD condition (Fig. \ref{fig:limitedtaus}D, second panel, dotted line). However, in other situations (HD condition for shorter run-lengths, and HI condition for all run-lengths), change points are not likely ($q<0.5$). Hence, in most cases, a vanishing run-length, i.e., the hypothesis of a change point, minimizes the marginal distribution, $p_{t+1}(\tau|x_{1:t+1})$, and its probability vanishes in our approximation: $\tilde p_{t+1}(\tau=0|x_{1:t+1}) = 0$. In other words, change points tend to go by undetected. Consequently, suppressed learning rates and enhanced repetition propensity obtain in a model with a single memory slot (Fig. \ref{fig:limitedtaus}D, E).

To compare our suboptimal models to the \textit{OptimalInference} model, we compute their normalized mean squared errors (NMSE) with regard to the responses of the optimal model (as opposed to the responses of the human subjects, as we do in the main text). With a NMSE for learning rates at 0.21, the $\tau$\textit{MaxProb} model with $N_\tau=1$ is closer to optimality than the $\tau$\textit{Mean} model (NMSE of 0.88) and the $\tau$\textit{Nodes} model with one node (NMSE of 0.95), in the HD condition (Fig. \ref{fig:limitedtaus}F). The high repetition propensity of the $\tau$\textit{MaxProb} model, however, leads to a larger error for this measure (NMSE of 1.81), as compared to the $\tau$\textit{Mean}  (0.14) and $\tau$\textit{Nodes} (0.61) models (Fig. \ref{fig:limitedtaus}G). Adding a second memory slot allows for a better approximation of the marginal distribution, $p_t(\tau | x_{1:t})$, in the $\tau$\textit{MaxProb} model, as demonstrated by its close-to-optimal behavior with $N_\tau=2$, both in terms of learning rates (NMSE of 0.043; compare to the $\tau$\textit{Nodes} model: 0.89) and repetition propensity (0.097; compare to the $\tau$\textit{Nodes} model: 0.29) (Fig. \ref{fig:limitedtaus}F, G).

\subsubsection{$\tau$\textit{Sample} model}

\begin{figure}[htb]
        \includegraphics[width=\linewidth]{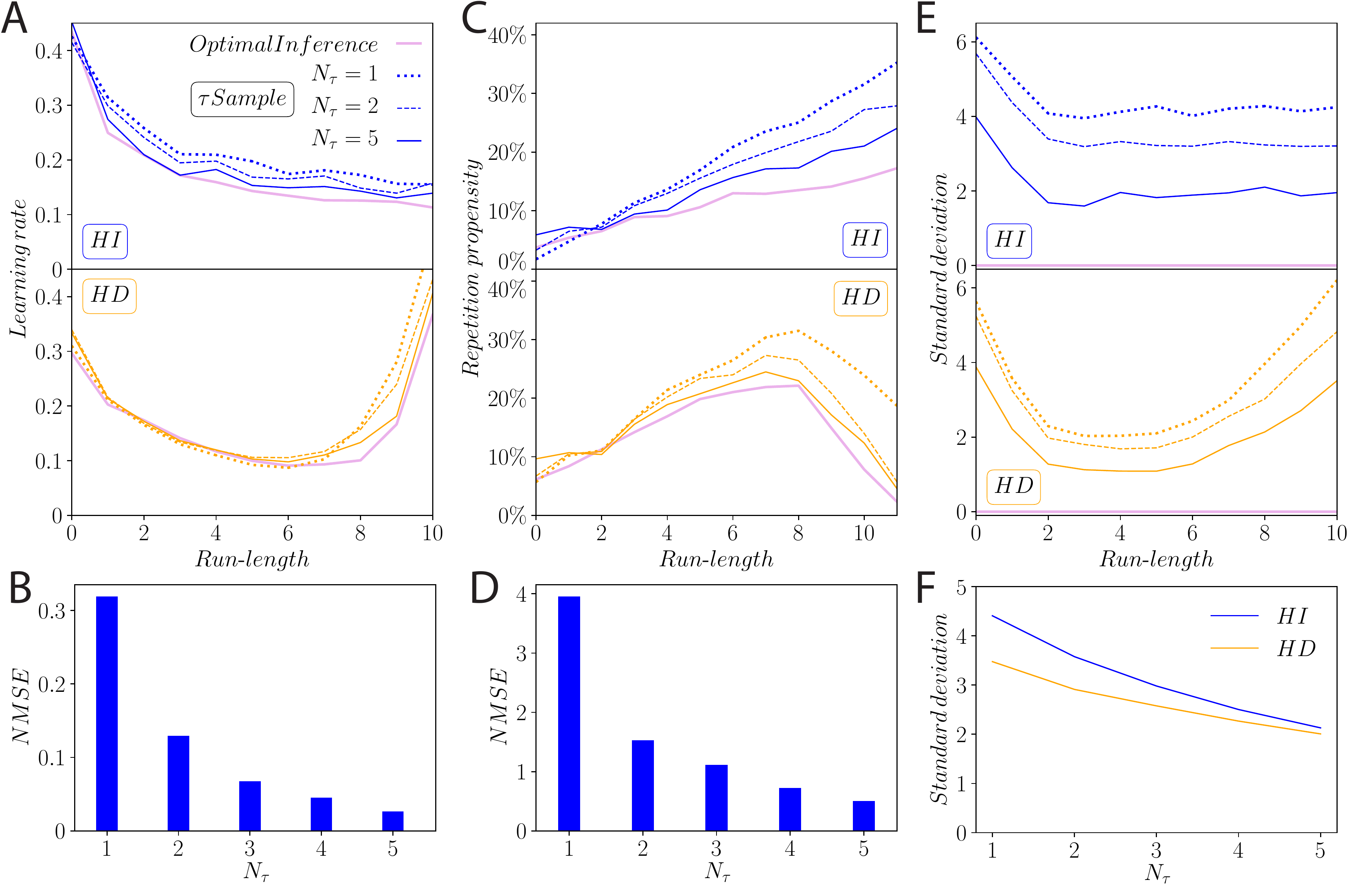}
	\caption{\textbf{Behavior of the $\tau$\textit{Sample} model, as compared to the \textit{OptimalInference} model.}
	\textbf{A, C, E.} Average learning rate (A), repetition propensity (C), and standard deviations of responses (E), as a function of run-length, in the \textit{OptimalInference} model and the $\tau$\textit{Sample} model with $N_\tau=$ 1, 2, and 5, in the HI condition (top panels) and in the HD condition (bottom panels).
	\textbf{B, D.} Normalized Mean Squared Error on learning rates (B), and on repetition propensity (D), as compared to the \textit{OptimalInference} model, for the $\tau$\textit{Sample} model, with $N_\tau=$ 1 to 5.
	\textbf{F.} Standard deviation of the responses of the $\tau$\textit{Sample} model, as a function of $N_\tau$.
	}\label{fig:tausample}
\end{figure}

The $\tau$\textit{Sample} model is identical to the $\tau$\textit{MaxProb} model, except that the run-length $\tau ^{\ast }$ is chosen randomly, i.e., sampled from the distribution $\left[ 1-p_{t+1}(\tau |x_{1:t+1})\right] /z_{t+1}$, where $z_{t+1}$ is a normalization factor. The stochastic nature of the update rule on the probable run-lengths influences the learning rate and the repetition propensity. In the case $N_{\tau }=1$, there is, at trial $t$, a single run-length, $\tau _{t}$, with non-vanishing probability. At trial $t+1$, the model subject chooses randomly between the no-change-point scenario, with $\tau _{t+1}=\tau _{t}+1$, and the change-point scenario, with $\tau _{t+1}=0$. Hence, the model can incur `false positives' (a change-point scenario is opted for in the absence of a true change point) and `false negatives' (a true change point goes undetected by the model subject), and these occur stochastically. In most trials, the change-point scenario is less likely than the no-change-point scenario; in the $\tau$\textit{MaxProb} model, the former would be eliminated, but it occurs with some probability in the $\tau$\textit{Sample} model, leading to false positives, which induce higher learning rates. Similarly, average learning rates of the $\tau$\textit{Sample} model are higher than the optimal ones (Fig. \ref{fig:tausample}A). The false negatives, in which change points go undetected, result, as in the $\tau$\textit{MaxProb} model, in higher repetition propensities (Fig. \ref{fig:tausample}C). With increasing memory capacity, $N_{\tau }$, the behavior of the model approaches optimality, as reflected in the decrease of the NMSEs for the learning rates (Fig. \ref{fig:tausample}B) and the repetition propensities (Fig. \ref{fig:tausample}D).

A qualitatively new aspect brought in by the $\tau$\textit{Sample} model is the stochasticity in the inference step, which is reflected in behavioral variability and measured by the standard deviation of the responses of a model subject. Quantitatively, false negatives have a large impact on the behavioral variability. A model subject can however correct for a false negative during the few trials that follow a true change point, i.e., at short run-lengths. This occurs randomly, in the $\tau$\textit{Sample} model, resulting in variability in responses. At longer run-lengths, the posterior probability of a change point, $p_{t}(\tau=0 |x_{1:t})$, is dominated by the shape of the change probability, $q(\tau)$, rather than by the observed evidence. In the HI condition, $q$ is constant; hence, the variability reaches a plateau for run-lengths larger than about $2$ (Fig. \ref{fig:tausample}E, top panel). In the HD condition, as $q(\tau)$ is an increasing function of the run-length, the variability increases for run-lengths larger than $5$, resulting in the `smile shape' of the curve (Fig. \ref{fig:tausample}E, bottom panel). As the parameter $N_{\tau }$ is increased, the behavior of the model approaches optimality, and, correspondingly, the standard deviation of the responses of the model subject decreases (Fig. \ref{fig:tausample}F).

\subsubsection{\textit{ParticleFilter}}

\paragraph{Derivation} The \textit{ParticleFilter} approximates the posterior by a weighted sum of delta functions (Eq. (\ref{eq:pfdeltas})). To obtain the approximate posterior at trial $t+1$, upon receiving a new observation, $x_{t+1}$, we start by writing the Bayesian update (Eq. (\ref{eq:HDupdate})) of the approximate posterior at trial $t$, $\tilde{p}_{t}(s,\tau |x_{1:t})$, as 
\begin{equation}
p_{t+1}(s,\tau |x_{1:t+1}) = \frac{1}{Z_{t+1}} g(x_{t+1}|s) \sum_{\tau_t} \int_{s_t} p_{t+1}(s,\tau |s_{t},\tau _{t}) \tilde{p}_{t}(s,\tau |x_{1:t}) \D s_{t},
\label{eq:pfintermediateposterior}
\end{equation}
with the transition probability, $p_{t+1}(s,\tau |s_{t},\tau _{t})$, defined in Eq. (\ref{eq:transition}). Injecting the expression of the approximate posterior at trial $t$ (Eq. (\ref{eq:pfdeltas})), we can rewrite the Bayesian update as a sum of $N_{P}$ functions:
\begin{equation}
p_{t+1}(s,\tau |x_{1:t+1}) = \frac{1}{Z_{t+1}} \sum_{i=1}^{N_{P}} w_{t}^{i} g(x_{t+1}|s) p_{t+1}(s,\tau |s_{t}^{i},\tau_{t}^{i}),
\end{equation}
where 
\begin{equation}
p_{t+1}(s,\tau |s_{t}^{i},\tau _{t}^{i}) = \sum_{\tau _{t}} \int_{s_t} p_{t+1}(s,\tau|s_{t},\tau _{t}) \delta (s-s_{t}^{i})\delta _{\tau ,\tau _{t}^{i}} \D s_{t}.
\end{equation}
The interpretation of this form becomes apparent if we introduce, for each particle, a probability distribution over $(s,\tau )$, defined as
\begin{equation}
\pi _{t+1}(s,\tau |s_{t}^{i},\tau _{t}^{i},x_{t+1})\equiv \frac{ g(x_{t+1}|s) p_{t+1}(s,\tau |s_{t}^{i},\tau _{t}^{i})}{p(x_{t+1}|s_{t}^{i},\tau _{t}^{i})},
\end{equation}
where the denominator is obtained by normalization,
\begin{equation}
p(x_{t+1}|s_{t}^{i},\tau _{t}^{i}) = \sum_\tau \int_s g(x_{t+1}|s) p_{t+1}(s,\tau |s_{t}^{i},\tau _{t}^{i}) \D s.
\end{equation}
The distribution $\pi _{t+1}(s,\tau |s_{t}^{i},\tau_{t}^{i},x_{t+1})$ is none other than the Bayesian update of a single particle (i.e., Eq. (\ref{eq:pfintermediateposterior}) with the approximate prior, $\tilde{p}_{t}(s,\tau |x_{1:t})$, replaced by$\ \delta (s-s_{t}^{i})\delta _{\tau ,\tau _{t}^{i}}$), and the full Bayesian update is a weighted sum of these $N_{P}$ functions:
\begin{equation}
p_{t+1}(s,\tau |x_{1:t+1}) = \sum_{i=1}^{N_{P}} \frac{w_{t}^{i}p(x_{t+1}|s_{t}^{i},\tau _{t}^{i})}{Z_{t+1}} \pi _{t+1}(s,\tau|s_{t}^{i},\tau _{t}^{i},x_{t+1}).
\end{equation}
To complete the definition of the particle filter, we have to formulate a prescription for selecting the $N_{P}$ particles at trial $t+1$. Following the literature, instead of sampling the full Bayesian update, $p_{t+1}(s,\tau |x_{1:t+1})$, we sample independently each component of the mixture, $\pi _{t+1}(s,\tau|s_{t}^{i},\tau _{t}^{i},x_{t+1})$, to obtain the updated particles, $(s_{t+1}^{i},\tau _{t+1}^{i})$. To each sample, i.e., to each particle, is assigned the weight of the corresponding component in the mixture, $w_{t+1}^{i}=w_{t}^{i}p(x_{t+1}|s_{t}^{i},\tau _{t}^{i})/Z_{t+1}$. In the rare cases in which $p(x_{t+1} | s^i_t, \tau^i_t)=0$, i.e., if new data invalidate particle $i$, and thus, $\pi _{t+1}(s,\tau |s_{t}^{i},\tau_{t}^{i},x_{t+1})=0$, we resample a new particle $i$ from the other particles.

In practical applications of particle filters, there exists a `weight degeneracy' risk, whereby the weight of one particle may overwhelm the combined weight of the others. A common method to mitigate this shortcoming is called `resampling'. It is a stochastic method in which the particles with high weights are likely to be duplicated, while the particles with low weights are likely to be eliminated. To achieve this, we use the $N_P$-dimensional categorical distribution parameterized by the $N_P$ weights of the particles, i.e., $p(j) = w_t^j$. We sample this distribution $N_P$ times, and obtain, thus, a set of $N_P$ indexes, $\{j_i\}_{i=1}^{N_P}$. We use those to define the new $N_P$ particles: for each particle $i$, we replace $(s_t^i, \tau_t^i)$ by $(s_t^{j_i}, \tau_t^{j_i})$, and we set all the weights to $1/N_P$. In other words, the set of particles is randomly sampled with replacement, $N_P$ times. Particles with low weights are unlikely to survive this scheme, as compared to particles with high weights. For the sake of simplicity, we resample at each trial.

A possible consequence of resampling is the `sample impoverishment' problem, i.e., the loss of particle diversity (all particles end up bearing the same state). A common procedure in the particle-filter literature that addresses this problem is `particle rejuvenation', which increases the variability of particles by `jittering' their parameters. Here, however, this issue is mitigated naturally by the structure of our problem, as a new state is sampled from the distribution $a(s|s_t)$ every time a new particle carries a change-point run-length ($\tau=0$), thus renewing the set of particles. In addition, introducing a rejuvenation step implies choosing arbitrarily a transition kernel and an acceptance rule for candidate particles (usually, the Metropolis-Hastings rule is adopted). Many kernels used in the literature come with additional parameters. While the rejuvenation method would be an interesting addition to our model, the performance of our implementation of the particle-filter model does not warrant the introduction of this new layer of complexity.

\begin{figure}
        \includegraphics[width=\linewidth]{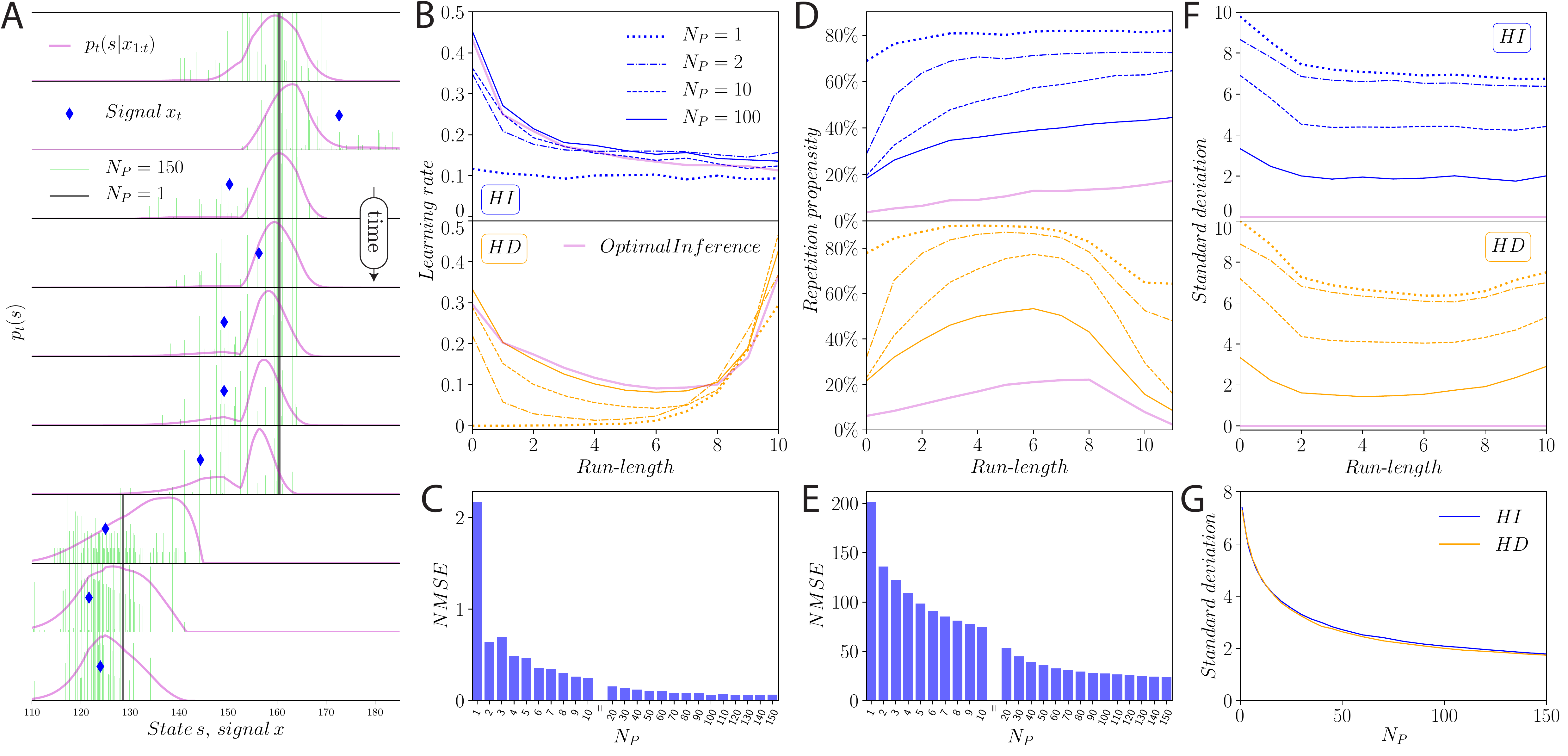}
	\caption{\textbf{Illustration of the \textit{ParticleFilter} model and its behavior.}
	\textbf{A.} Distribution of particles during inference, compared with the optimal posterior distribution, for particle filters with $N_P=$1 and 150. Each particle is a point on the $(s, \tau)$ plane, equipped with a weight. Only the spatial components $s$ are represented here, as vertical bars (grey for $N_P=1$, green for $N_P=150$). Bars heights are proportional to the corresponding weights, but some are truncated due to the choice of scale, which emphasizes weight diversity. Upon receiving a new stimulus, $x_{t+1}$ (blue), a particle $i$ is updated by sampling $p_{t+1}(s,\tau|s^i_t, \tau^i_t, x_{t+1})$. This may or may not involve a change point, in which case $s^i_{t+1} \neq s^i_t$.
	\textbf{B, D, F.} Average learning rate (B), repetition propensity (D), and standard deviations of responses (F), as a function of run-length, in the \textit{OptimalInference} model and the \textit{ParticleFilter} model with $N_P=$ 1, 2, 10, and 100, in the HI condition (top panels) and in the HD condition (bottom panels).
	\textbf{C, E.} Normalized Mean Squared Error on learning rates (C), and on repetition propensity (E), as compared to the \textit{OptimalInference} model, for the \textit{ParticleFilter} model, with $N_P=$ 1 to 150.
	\textbf{G.} Standard deviation of the responses of the \textit{ParticleFilter} model, as a function of number of particles, $N_P$.
	}\label{fig:pf}
\end{figure}

\paragraph{Behavior} With a single particle ($N_{P}=1$), the posterior is reduced to a unique sample, $(s_{t}^{1},\tau_{t}^{1})$, and, thus, the model subject has access to a single `hypothesis' on the change probability, $q(\tau _{t}^{1})$. The particle can then evolve in one of two ways: either it opts, with this probability, for a change-point scenario, in which the new stimulus, $x_{t+1}$, is the only information available on the new state, along with the prior transition probability, and thus the learning rate is close to 1, or, with probability $1-q(\tau _{t}^{1})$, a no-change-point scenario is opted for, and the particle stays put ($s_{t+1}^{1}=s_{t}^{1}$), i.e., the learning rate vanishes. As a result, when averaged over several instantiations of the particle filter, the behavior of the learning rate as a function of run-length resembles that of the change probability, $q(\tau)$, i.e., constant in the HI condition, and increasing in the HD condition (Fig. \ref{fig:pf}B), a behavior qualitatively different from either that of the \textit{OptimalInference} model or the human responses. But it is sufficient to add no more than a second particle for the model to capture the main trends in the learning rate (a decreasing learning rate in the HI condition and smile shape in the HD condition). The NMSE drops sharply from 2.2 for $N_{P}=1$ to less than 0.7 for $N_{P}=2$. As additional particles are included in the model, the latter approaches optimality (the NMSE becomes less than 0.1 for $N_{P}\geq 70$) (Fig. \ref{fig:pf}C).

As mentioned, sampling in the particle filter induces variability in behavior:\ two particle filters receiving the same sequence of observations do not respond with the same sequence of estimates. Since the stochasticity stems from the sampling of an (approximate) posterior, the resulting variability scales with the width of the posterior. As measured by the standard deviation of responses, it decreases with the run-length, in the HI condition. In the HD condition, it decreases at short run-lengths before increasing at longer run-lengths (Fig. \ref{fig:pf}F). This behavior reproduces, at least qualitatively, that of the subjects (compare to Fig. \ref{fig:metrics}C). The greater the number of particles in a particle filter, the closer the latter approximates the \textit{OptimalInference} model; the standard deviation of the responses is a decreasing function of the number of particles (Fig. \ref{fig:pf}G).

Since it operates on a low-dimensional spatial representation, the particle filter naturally predicts a higher repetition propensity than the \textit{OptimalInference} model does. More specifically, the posterior is non-vanishing for only a finite (possibly small) set of values at each trial, and it is more likely than in the optimal case that the subject model's estimate remains unchanged following stimulus presentation. This effect is quantitatively appreciable, and leads to repetition propensities which are multiples of those in the \textit{OptimalInference} model. Again, the repetition propensities decrease toward their optimal values as the number of particles, $N_{P}$, increases. The corresponding NMSE drops from $202$ for $N_{P}=1$ to $24$ for $N_{P}=150$ (Fig. \ref{fig:pf}D, E).

\subsubsection{\textit{Sampling} model}

\begin{figure}
        \includegraphics[width=\linewidth]{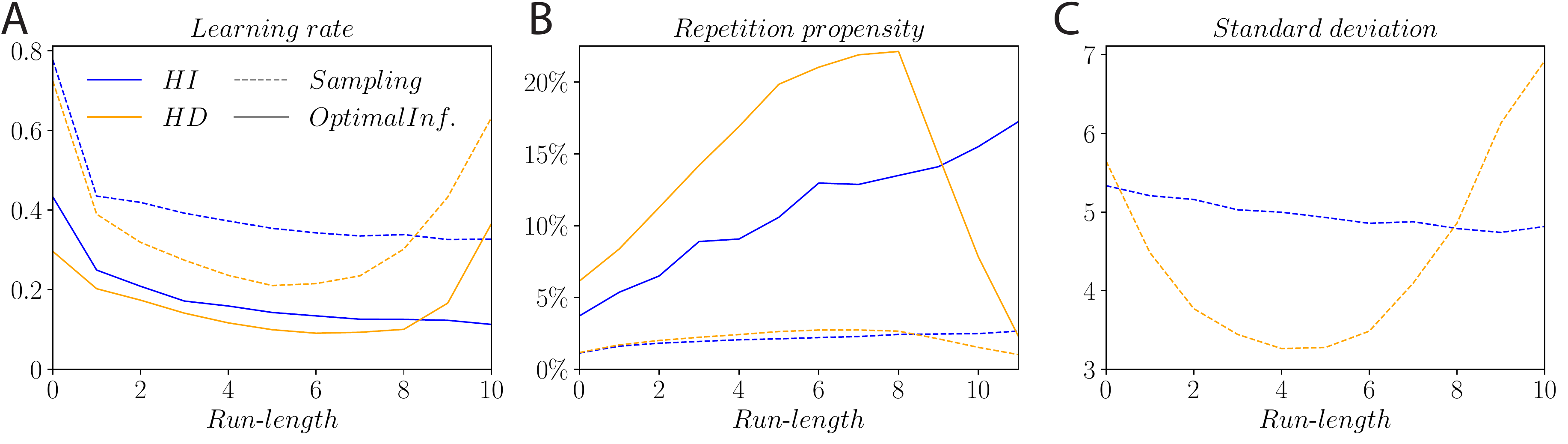}
	\caption{\textbf{Behavior of the \textit{Sampling} model, as compared to the \textit{OptimalInference} model.}
	\textbf{A,B.} Average learning rate (A), and repetition propensity (B), in the \textit{Sampling} model (dashed lines) and the \textit{OptimalInference} model (solid lines), as a function of run-length, in the HI and HD conditions.
	\textbf{C.} Standard deviations of responses, as a function of run-length, of the \textit{Sampling} model in the HI and HD conditions. The \textit{OptimalInference} model exhibits no variability.
	}\label{fig:sampling}
\end{figure}

In the \textit{Sampling} model, instead of using the Bayesian posterior to maximize its expected score, a model subject samples its response from the marginal posterior on the states, $p_{t}(s|x_{1:t})$. In spite of this suboptimal selection rule, the average learning rate as a function of the run-length has a behavior similar to the optimal one (decreasing in the HI condition, smile-shaped in the HD condition), albeit with higher average values (Fig. \ref{fig:sampling}A). The repetition propensity also behaves similarly to the optimal one, but is suppressed in magnitude due to sampling (Fig. \ref{fig:sampling}B). Finally, as expected by construction, the \textit{Sampling} model leads to behavioral variability, and the amplitude of the latter scales with the width of the posterior distribution (Fig. \ref{fig:sampling}C).

\subsection{Normalized Mean Squared Error}

This section provides some details on the Normalized Mean Squared Error we use to compare the results of the various models to the \textit{OptimalInference} model and to human data. Let $y_i(\tau)$ be the value at run-length $\tau$ of the quantity of interest $i$ (learning rate, repetition propensity, or standard deviation of responses), as observed in data or as resulting from the optimal model, and $\hat y_i (\tau)$ the value resulting from a suboptimal model. The mean squared error is $MSE (\hat y_i)=\frac{1}{n} \sum_\tau ( \hat y_i (\tau) - y_i(\tau) )^2$, where $n$ is the number of run-lengths. We want to be able to compare the errors for different quantities of interest. By dividing the $MSE$ by the variance of $y_i$, we obtain the Normalized Mean Squared Error, which is translation-invariant and scale-invariant:
\begin{equation}
	NMSE_i = \frac{MSE(\hat y_i)}{\mathrm{Var} [ y_i ]} = \frac{ \sum_\tau ( \hat y_i (\tau) - y_i(\tau) )^2 }{ \sum_\tau ( \bar y_i - y_i(\tau) )^2 },
\end{equation}
where $\bar y_i$ is the average $y_i(\tau)$. For model fitting, we then use the average of this quantity over the three quantities of interest (``three-error measure'') or over two of them (``two-error measure'').

\subsection{Approximate formulation of the Bayesian Information Criterion}

For a given subject in a given condition (HI or HD), we denote the probability of a sequence of $T$ responses, $\hat s_{1:T}$, by $p(\hat s_{1:T})$. In the models with deterministic inference, the joint probability of responses is the product of the probabilities of each of the responses in the successive trials:
\begin{equation}
p(\hat s_{1:T}) = \prod_{t=1}^T p(\hat s_t).
\end{equation}
This independence condition does not hold for the models with stochastic inference. We describe, here, how we overcome this issue in the case of the \textit{ParticleFilter} model, which is the most involved model we consider and the most costly computationally. We denote by $N$ the number of particles, and we define the `internal state' of a particle filter at time $t$ as the joint states of its $N$ particles, each defined by a location, $s_t$, a run-length, $\tau_t,$ and a weight, $w_t$. We denote the internal state by $\sigma_t$, and the sequence of the internal states throughout a run with $T$ trials by $\sigma_{1:T}$. The probability of a subject's sequence of responses is expressed in terms of the probability of the sequences of internal states, as
\begin{equation}
p(\hat s_{1:T}) = \sum_{\sigma_{1:T}} p(\hat s_{1:T} | \sigma_{1:T}) p(\sigma_{1:T}). 
\end{equation}
We note that conditional on the internal state, the responses are independent:
\begin{equation}
p(\hat s_{1:T} | \sigma_{1:T}) = \prod_{t=1}^T p(\hat s_t | \sigma_t).
\end{equation}
In other words, a \textit{single} realization of the particle filter can be thought of as a model with deterministic inference. To compute the probability of responses, however, we must determine the distribution of the realizations of the internal states of the particle filters, $p(\sigma_{1:T})$.
The support of this distribution is the Cartesian product of the $NT$ internal states of the particle filter, and, hence, its size grows exponentially with $NT$. Estimating a probability distribution over this space seems computationally intractable; we can, however, carry out an approximate calculation of the probability distribution. 

Our method of approximation relies on a Monte-Carlo estimation: we run $M=500$ simulations of the inference model (\textit{ParticleFilter} or $\tau$\textit{Sample}), and consequently we obtain $M$ points $\sigma_{1:T}$ in the space of possible sequences of internal states. For each realization of the internal state, we know the probability distribution of the response, conditional on the state, $p(\hat s_t | \sigma_t)$. A Monte-Carlo approximation of the probability of a sequence of $T$ responses, $\tilde p(\hat s_{1:T})$, is then obtained as
\begin{equation}
\tilde p(\hat s_{1:T}) = \frac{1}{M} \sum_{\sigma_{1:T}} p(\hat s_{1:T} | \sigma_{1:T}),
\end{equation}
i.e.,
\begin{equation}
\tilde p(\hat s_{1:T}) = \frac{1}{M} \sum_{\sigma_{1:T}} \prod_{t=1}^T p(\hat s_t | \sigma_t),
\end{equation}
where the sum is taken over the $M$ sampled sequences of internal states. This empirical approximation is satisfactory provided M is sufficiently large, so as to overcome the exponential growth of the number of possible sequences. 

For any sampled sequence of internal states, $\sigma_{1:T}$, it is extremely likely that at least one response, $\hat s_t$, has vanishingly small probability given the corresponding internal state, $\sigma_t$, i.e., $p(\hat s_t | \sigma_t) \approx 0$. In other words, given a sequence of responses, $\hat s_{1:T}$, it is extremely unlikely that any one of $M$ sequences of internal states produces $\hat s_{1:T}$, i.e., it is prohibitively improbable that any one of $M$ sequences of internal states account \textit{simultaneously} for $1000$ responses. Thus, if we carry out the MonteCarlo approximation naively, we underestimate severely the likelihood of the data. (We emphasize that, in that case, the low value of the likelihood does not reflect an inherent inability of the model to account for the data, but rather the poor sampling of the internal states in the model.)

We can circumvent this practical problem by dividing our experimental runs into shorter sequences: while our sample is too small for obtaining a useful approximation of the density of possible sequences of $1000$ successive internal states, we can treat this density over \textit{shorter sequences}. In the extreme case, we can consider the (Monte-Carlo-approximated) likelihood of just \textit{one} response, $\hat s_t$, at a trial $t$: 
\begin{equation}
\tilde p(\hat s_t ) = \frac{1}{M} \sum_{\sigma_{1:T}} p(\hat s_t | \sigma_t ).
\end{equation}
Here, the $M$ samples, $\sigma_{1:T}$, are used to estimate a one-dimensional density, instead of a $1000$-dimensional density. The joint likelihood of all responses can then be approximated as the product of the likelihoods of each response, i.e.,
\begin{equation}
\tilde p(\hat s_{1:T}) \approx \prod_{t=1}^T \tilde p(\hat s_t).
\end{equation}
This approximation, however, in effect makes the crude assumption that successive responses are independent, and thus neglects the sequential dependence in responses that a model may predict. For instance, this approximation vastly underestimates the likelihood of a model that correctly predicts an appreciable probability of repetitions.

In order to obtain an approximation of the likelihood that takes into account the sequential dependence of responses, and which can be computed on the basis of $M$ samples, we choose to compute the likelihood of responses over sequences of $10$ successive trials,
\begin{equation}
\tilde p(\hat s_{t:t+9})=\frac{1}{M} \sum_{\sigma_{1:T}} p(\hat s_{t:t+9} | \sigma_{t:t+9}).
\end{equation}
We fit the models by evaluating how well they reproduce the responses in all the 10-trial-long sequences. More precisely, we associate to each model a BIC calculated as
\begin{equation}
\operatorname{BIC} = -2 \ln \Bigg[ \prod_{t=1,11, \dots} \tilde p(\hat s_{t:t+9}) \Bigg] + k \ln n,
\end{equation}
where $k$ is the number of parameters in the model under consideration, and $n$ is the number of data points. The specific choice of sequences with 10 trials is arbitrary; in our analyses, we repeated the calculations for different choices, which yielded comparable results. We chose to illustrate this choice as it corresponds to sequences no shorter than the mean inter-change-point duration, that optimizes the precision of the approximation.

We note that our approximation of the \textit{ParticleFilter} model’s BIC could be interpreted as a (possibly still approximate) calculation of the BIC of a different model. Specifically, the latter would include a particular form of a particle-rejuvenation procedure in which all the particles are replaced, every 10 trials, by as many new particles, each randomly sampled from the distribution of possible particles, at these trials. The rejuvenation kernel would thus be independent of the rejuvenated particle (a possibility considered in the particle-filter literature, see \cite{Chopin2002}), and the acceptance rate would be equal to 1, each 10 trials, and to 0 at other trials; thus, it would also be independent of the particle (which is at odds with the usual form of the rejuvenation procedures introduced in the literature).

\subsection{Rational-inattention models and models with fixed repetition probability}

The rational-inattention models we present are inspired by the model introduced by \cite{Khaw2017}. We adopt their notation for the new quantities introduced here to describe the response-selection process.
The major new ingredient in this model is that the subject, after having observed a sequence of stimuli, $x_{1:t}$, is assumed to choose, first, whether to adjust or to repeat the current estimate (the `repetition variable'); the repetition variable is a Bernoulli random variable parameterized by the probability of adjusting, denoted by $\Lambda_t(x_{1:t}, \hat s_{t-1})$ (and the probability of a repetition is thus $1-\Lambda_t(x_{1:t}, \hat s_{t-1})$). Second, conditional on adjusting, the subject randomly chooses a new estimate (the `location variable'), sampled from a distribution which we denote by $\mu_t(\hat s_t | x_{1:t})$. Thus, the model subject's distribution of responses conditional on the observed stimuli and on the preceding response is
\begin{equation}\label{eq:ri_phat}
p(\hat s_t | x_{1:t}, \hat s_{t-1}) = (1 - \Lambda_t(x_{1:t}, \hat s_{t-1})) \delta(\hat s_t - \hat s_{t-1})
	+ \Lambda_t(x_{1:t}, \hat s_{t-1}) \mu_t(\hat s_t | x_{1:t}),
\end{equation}
where $\delta$ is the Dirac delta function.

Following the rational-inattention approach, we assume that the distribution of responses conditional on the observed stimuli and on the preceding response, $p(\hat s_t | x_{1:t}, \hat s_{t-1})$, maximizes, under a constraint detailed below, the expected reward. Although the response at trial $t$, $\hat s_t$, affects the rewards in subsequent trials, through the ensuing distributions of responses (Eq. (\ref{eq:ri_phat})), for the sake of calculations simplicity we will carry out a 'greedy' optimization in the model. Specifically, we will assume that the distribution of responses conditional on the observed stimuli and on the preceding response, $p(\hat s_t | x_{1:t}, \hat s_{t-1})$, is obtained by considering only the immediate reward in expectation over the state, the response, and the sequence of past stimuli, i.e., the quantity
\begin{align}
\begin{split}
\bar R \equiv \int \dots \int p(x_{1:t}) \int p(\hat s_t | x_{1:t}, \hat s_{t-1}) \int p_t(s|x_{1:t}) R(\hat s, s) \D s \D \hat s_t \D x_1 \dots \D x_t,
\end{split}
\end{align}
where $R(\hat s,s)$ is the reward obtained if the estimate is $\hat s$ and the correct state is $s$.

In the absence of a constraint, the optimal response is obtained by maximizing the expected reward implied by the sequence of past stimuli, $\int p_t(s|x_{1:t}) R(\hat s, s) \D s$, which we denote by $r(\hat s_t | x_{1:t})$. We assume, however, that it is costly for the subject to choose with precision the repetition variable and the location variable, and this hampers the ability to obtain this optimal estimate. Following \cite{Khaw2017}, we assume that the repetition variable (distributed according to $1-\Lambda_t(x_{1:t}, \hat s_{t-1})$), and the location variable in trials in which the estimate is not repeated (distributed according to $\mu_t(\hat s_t|x_{1:t})$), each bear a cognitive cost proportional to measures of the amount of information on the sequence of stimuli involved in choosing the repetition variable and the location variable, respectively, defined as
\begin{align}
& I_1 = \int \dots \int p(x_{1:t}) D_{KL} ( \Lambda_t(x_{1:t}, \hat s_{t-1}) || \tilde \Lambda ) \D x_1 \dots \D x_t \\
\text{and }
& I_2 = \int \dots \int p(x_{1:t}) \Lambda_t(x_{1:t}, \hat s_{t-1}) D_{KL} ( \mu_t(.|x_{1:t}) || \tilde \mu ) \D x_1 \dots \D x_t,
\end{align}
where $\tilde \Lambda$ and $\tilde \mu$ are the unconditional (not conditional on $x_{1:t}$) probability of adjusting and the distribution of estimates, respectively.
The distributions $\Lambda_t(x_{1:t}, \hat s_{t-1})$ and $\mu_t(\hat s_t|x_{1:t})$ are obtained as those that maximize the quantity
\begin{align}
\begin{split}
\bar R - \psi_1 I_1 - \psi_2 I_2
\end{split}
\end{align}
which expresses a trade-off between expected reward and cognitive costs,
where $\psi_1$ and $\psi_2$ are numerical coefficients specifying the strength of the information-theoretic costs.

The solution to the optimization problem just posed is given by
\begin{equation}
\mu_t(\hat s_t|x_{1:t}) = \frac{1}{Z(x_{1:t})} \tilde \mu(\hat s_t) \exp( \psi_2^{-1} r(\hat s_t|x_{1:t}) ),
\end{equation}
and
\begin{equation}
\ln \frac{\Lambda_t(x_{1:t}, \hat s_{t-1})}{1-\Lambda_t(x_{1:t}, \hat s_{t-1})} = \ln \frac{\tilde \Lambda}{1-\tilde \Lambda}
					+ \psi_1^{-1} ( \psi_2 \ln Z(x_{1:t}) - r(\hat s_{t-1} | x_{1:t}) ),
\end{equation}
where
\begin{equation}
Z(x_{1:t}) = \int \tilde \mu(\hat s) \exp( \psi_2^{-1} r(\hat s|x_{1:t}) ) \D \hat s.
\end{equation}
The unconditional distribution of estimates, $\tilde \mu(\hat s)$, can be approximated by a uniform distribution on the space of responses; for the sake of simplicity, we use this approximation in our calculations.

\begin{table}[!tb]
\begin{center}
\begin{tabular}{c|c|c|c}
\textbf{Repetition}	& \textbf{Location}	& \multicolumn{2}{c}{\textbf{BIC}} \\
\textbf{variable} 	& \textbf{variable}	& \textbf{\textit{OptimalInf.}} & \textbf{\textit{ParticleFilter}} \\
\hline\hline
- & \textit{NoisyMax} 						& 495,063 & 482,358 \\
- & Rational Inattention 					& 524,719 & 486,217 \\
Rational Inattention & \textit{NoisyMax} 		& 445,189 & 439,195 \\
Fixed repetition probability & \textit{NoisyMax} 	& 445,354 & 437,797 \\
Rational Inattention & Rational Inattention 	& 471,995 & 446,187 \\
Fixed repetition probability & Rational Inattention & 472,810 & 446,848 \\
\hline\hline
\end{tabular}
\end{center}
\captionof{table}{\label{tab:ri_bics}
\normalfont \textbf{BICs of the rational-inattention models and the models with fixed repetition probability, combined with the \textit{OptimalInference} and the \textit{ParticleFilter} inference strategies.}}
\end{table}

We implement four variants of this model and compute their BICs (Table \ref{tab:ri_bics}). In a first variant, no cost weighs on the repetition variable ($\psi_1=0$), but a cognitive cost prevents the model subject from choosing the optimal response ($\psi_2 \neq 0$, Table \ref{tab:ri_bics}, second row).
By contrast, in the second variant of the model the repetition variable is subject to a cost ($\psi_1 \neq 0$), while the location variable is not ($\psi_2=0$); the latter follows, however, a \textit{NoisyMax} response-selection strategy (Table \ref{tab:ri_bics}, third row). (If the location variable, instead, were optimal, the model would assign a vanishing probability to most of the subjects' responses, and consequently would yield an infinite BIC.)
In a third variant of the model, there are attentional costs weighing on both the repetition and the location variable ($\psi_1 \neq 0$ and $\psi_2 \neq 0$, Table \ref{tab:ri_bics}, fifth row).
In the fourth variant of the model, the location variable is subject to a cognitive cost ($\psi_2 \neq 0$); the repetition variable, however, is not derived in a rational-inattention approach, but instead is random and governed by a fixed repetition probability (the first variant, above, is a special case of this fourth variant of the model, corresponding to a repetition probability set to zero; Table \ref{tab:ri_bics}, sixth row).
For the sake of comparison, we implement a fifth model that combines a fixed repetition probability with a \textit{NoisyMax} strategy for the location variable (this model does not feature any cognitive cost; Table \ref{tab:ri_bics}, fourth row).

The five models just presented make use of the \textit{OptimalInference} strategy. We implement, in addition, five other models in which the repetition variable and the location variable are chosen as in these five models, but the \textit{OptimalInference} strategy is replaced by the \textit{ParticleFilter} inference strategy (Table \ref{tab:ri_bics}, last column).

\section{Acknowledgements}
This work was supported by the CNRS through UMR8550, the Global Scholar Program at Princeton University, and the Visiting Faculty Program at the Weizmann Institute of Science. A.P.C. was supported by a PhD fellowship of the Fondation Pierre-Gilles de Gennes pour la Recherche. This work was granted access to the HPC resources of MesoPSL financed by the Region Ile de France and the project Equip@Meso (reference ANR-10-EQPX-29-01) of the program Investissements d’Avenir supervised by the Agence Nationale pour la Recherche.


\bibliographystyle{unsrt}
\bibliography{refs}       


\newpage

\appendix

\section{Supplementary tables: statistical tests}

	\begin{tabular}{c|ccccccccccc}
            		\hline \hline
            		$\tilde \tau$ & 0 & 1 & 2 & 3 & 4 & 5 & 6 & 7 & 8 & 9 & 10 \\
            		\hline
W & .307 & .448 & .265 & .335 & .180 & .072* & .016** & .073* & .426 & .160 & .008*** \\
MWU & .269 & .391 & .281 & .212 & .207 & .066* & .034** & .022** & .491 & .568 & .006*** \\
S & .307 & .448 & .268 & .337 & .178 & .070* & .015** & .074* & .426 & .158 & .008*** \\
\hline
N(HI) & 454 & 521 & 613 & 508 & 461 & 436 & 411 & 360 & 318 & 316 & 240 \\
N(HD) & 603 & 616 & 707 & 632 & 637 & 521 & 499 & 421 & 387 & 300 & 231 \\
            		\hline \hline
			\multicolumn{12}{l}{Excluding repetitions} \\
			\hline
W & .353 & .421 & .329 & .352 & .142 & .042** & .008*** & .151 & .391 & .202 & .034** \\
MWU & .357 & .431 & .484 & .170 & .073* & .025** & .007*** & .058* & .343 & .590 & .050** \\
S & .352 & .421 & .332 & .353 & .140 & .041** & .007*** & .153 & .391 & .201 & .035** \\
\hline
N(HI) & 412 & 435 & 495 & 402 & 353 & 333 & 307 & 288 & 237 & 231 & 174 \\
N(HD) & 553 & 529 & 582 & 497 & 493 & 405 & 379 & 318 & 291 & 226 & 186 \\
			\hline \hline
            	\end{tabular}
   		\captionof{table}{\label{tab:lrtau}\textbf{p-values for the one-sided statistical tests of equality of the means of the learning rates in the HI and HD conditions, for each run-length (Fig. \ref{fig:lrs}C).}
		\textnormal{W: Welch's test. MWU: Mann-Whitney's U test. S: Student's test. The N(HI) and N(HD) lines report the number of observations.
The second half of the table reports the same quantities when excluding all occurrences of repetitions (Suppl. Fig. \ref{fig:results_wo_repeats}B).} }

\vspace{0.5in}

		\begin{tabular}{l|l|cc|ccc|cc}
			\hline\hline
			Condition 1 & Condition 2 & $Avg_1$ & $Avg_2$ & W & MWU & S & $N_1$ & $N_2$ 
			\\
			\hline
HI, $\tilde\tau \in [5,6]$ & HI, $\tilde\tau \in [9,10]$ & 0.255 & 0.208 &.0099*** & .0476** & .0119** & 847 & 556		\\
HI, $\tilde\tau \in [5,6]$ & HD, $\tilde\tau \in [5,6]$ & 0.255 & 0.209 &.0055*** & .0089*** & .0052*** & 847 & 1020	\\
HI, $\tilde\tau \in [9,10]$ & HD, $\tilde\tau \in [9,10]$ & 0.208 & 0.264 &.0103** & .0602* & .0100*** & 556 & 531	\\
HD, $\tilde\tau \in [5,6]$ & HD, $\tilde\tau \in [9,10]$ & 0.209 & 0.264 &.0072*** & .0173** & .0050*** & 1020 & 531	\\
			\hline\hline
			\multicolumn{9}{l}{Excluding repetitions} \\
			\hline
HI, $\tilde\tau \in [5,6]$ & HI, $\tilde\tau \in [9,10]$ & 0.338 & 0.285 &.0203** & .0780* & .0233** & 640 & 405		\\
HI, $\tilde\tau \in [5,6]$ & HD, $\tilde\tau \in [5,6]$ & 0.338 & 0.272 &.0017*** & .0009*** & .0016*** & 640 & 784	\\
HI, $\tilde\tau \in [9,10]$ & HD, $\tilde\tau \in [9,10]$ & 0.285 & 0.340 &.0342** & .1721 & .0345** & 405 & 412		\\
HD, $\tilde\tau \in [5,6]$ & HD, $\tilde\tau \in [9,10]$ & 0.272 & 0.340 &.0066*** & .0095*** & .0047*** & 784 & 412	\\
			\hline\hline
		\end{tabular}
		\captionof{table}{\label{tab:lravgs}\textbf{Learning rates averages under various HI/HD and short/long run-length conditions, and p-values for one-sided statistical tests of equality of the means (Fig. \ref{fig:lrs}B).}
		\textnormal{The first two columns indicate the HI/HD and short/long run-length conditions. Learning rates at trials verifying these two conditions have their averages reported in the $Avg_1$ and $Avg_2$ columns. Columns W, MWU, and S provide the p-values for the tests of equality of the means between the two conditions. W: Welch's test. MWU: Mann-Whitney's U test. S: Student's test. $N_1$ and $N_2$ report the number of observations for each condition.
The second half of the table reports the same quantities when excluding all occurrences of repetitions (Suppl. Fig. \ref{fig:results_wo_repeats}A).}}



\begin{center}

	\begin{sidewaystable}[!ht]
            	\begin{tabular}{c|ccccccccccc}
            		\hline \hline
            		$\tilde \tau$ & 0 & 1 & 2 & 3 & 4 & 5 & 6 & 7 & 8 & 9 & 10 \\
            		\hline
                        	p & .0266** & .0331** & .0123** & .0550* & .0048*** & .0000*** & .0387** & .213 & .445 & .0331** & .0000*** \\
			N(HI) & 1255 & 1425 & 1611 & 1440 & 1347 & 1268 & 1111 & 975 & 883 & 795 & 722 \\
			N(HD) & 1591 & 1674 & 1943 & 1804 & 1650 & 1493 & 1333 & 1168 & 983 & 792 & 584 \\
			\hline
			\multicolumn{12}{l}{Excluding repetitions} \\
			\hline
                        	p & .0243** & .0293** & .0556* & .4847 & .0094*** & .0000*** & .4853 & .272 & .953 & .0560* & .0004*** \\
			N(HI) & 1123 & 1147 & 1291 & 1127 & 1003 & 963 & 818 & 748 & 643 & 583 & 533 \\
			N(HD) & 1455 & 1405 & 1559 & 1390 & 1255 & 1115 & 978 & 890 & 723 & 614 & 458 \\
			\hline \hline
            	\end{tabular}
   		\captionof{table}{\label{tab:stdtau}\textbf{p-values for Levene's statistical test of equality of the variances of the corrections in the HI and HD conditions, for each run-length (Fig. \ref{fig:variability}C)}.
		\textnormal{The N(HI) and N(HD) lines report the number of observations. The second part of the table reports the same quantities when excluding all occurrences of repetitions (Suppl. Fig. \ref{fig:results_wo_repeats}).
		}
		}

		\begin{tabular}{c|cccccccccccc}
		\hline\hline
		$\tilde\tau$ & 0 & 1 & 2 & 3 & 4 & 5 & 6 & 7 & 8 & 9 & 10 & 11\\
		p & 0.561 & 0.114 & 0.843 & 0.245 & 0.420 & 0.333 & 0.636 & 1.000 & 0.848 & 0.194 & 0.091* & 0.005*** \\
		N(HI) & 2635& 2211& 1984& 1786& 1620& 1471& 1351& 1215& 1098& 978& 875& 791 \\
		N(HD) & 2647& 2339& 2135& 1943& 1804& 1661& 1534& 1404& 1240& 1111& 946& 781 \\
		\hline\hline
		\end{tabular}
		\captionof{table}{\label{tab:reptau}\textbf{p-values for Fisher's exact test of independence of the repetition propensities in the HI and HD conditions, at each run-length (Fig. \ref{fig:repeat}B).}
		\textnormal{The N(HI) and N(HD) lines report the number of observations.}}
     \end{sidewaystable}
\end{center}


\clearpage
\section{Supplementary figures}


\vspace{0.5in}

{\centering
	\includegraphics[width=0.5\linewidth]{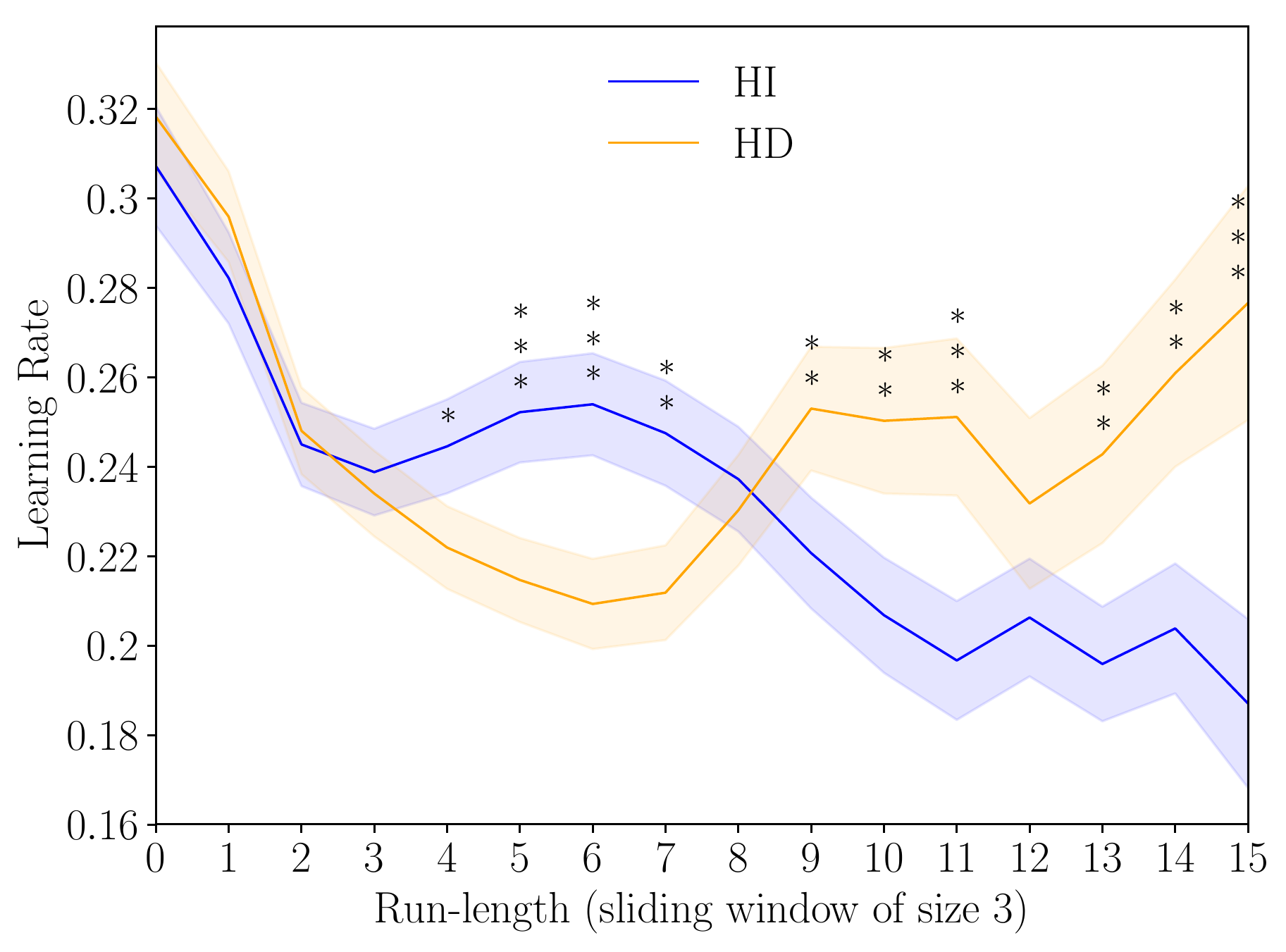}
\par}
	\captionof{figure}{
	\normalfont \textbf{Learning rates at run-lengths greater than 10.}
	In order to curb the fluctuations due to the decreasing amount of data at high run-lengths, we use a sliding window of size 3 over the run-lengths (i.e., we pool together the learning rates corresponding to three consecutive run-lengths). The number reported on the x-axis is the center of this sliding window. In the HD condition, the learning rate, after run-length 10, remains large and appears to increase. In the HI condition, the learning rate keeps decreasing after run-length 10. Although presumably the learning rate in this condition should eventually plateau and remain constant, it does not seem that our subjects reach this stage over the run-lengths for which we have sufficient data to allow for analysis (up to 15).
	}
	\label{fig:lr-vs-rl-longer-rls}

\vspace{0.5in}

	\includegraphics[width=\linewidth]{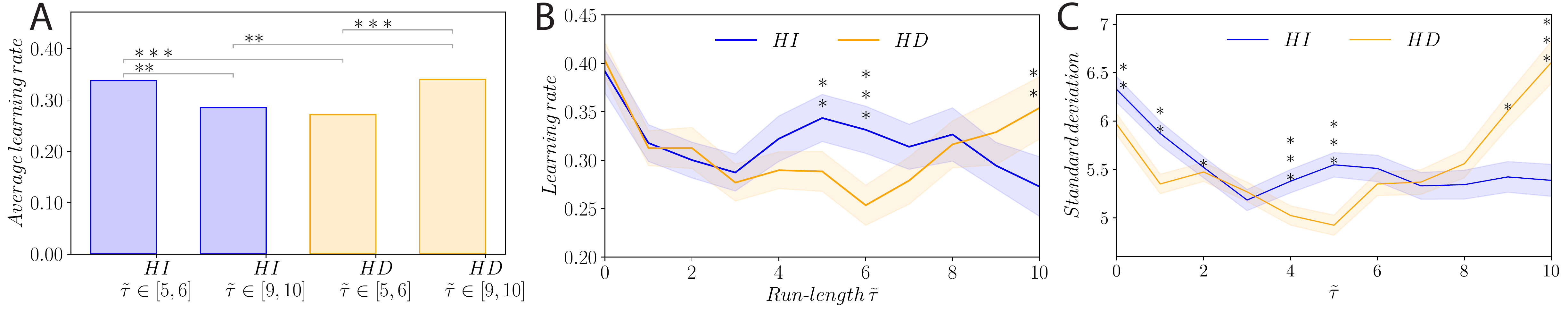}
	\captionof{figure}{\normalfont \textbf{Human learning rates and standard deviations of responses, excluding all occurrences of repetitions.}
	(\textbf{A,B}) as in Figs. \ref{fig:lrs}B,C; (\textbf{C}) as in Fig. \ref{fig:variability}C.
	}
	\label{fig:results_wo_repeats}

\vspace{0.5in}

{\centering
	\includegraphics[width=0.6\linewidth]{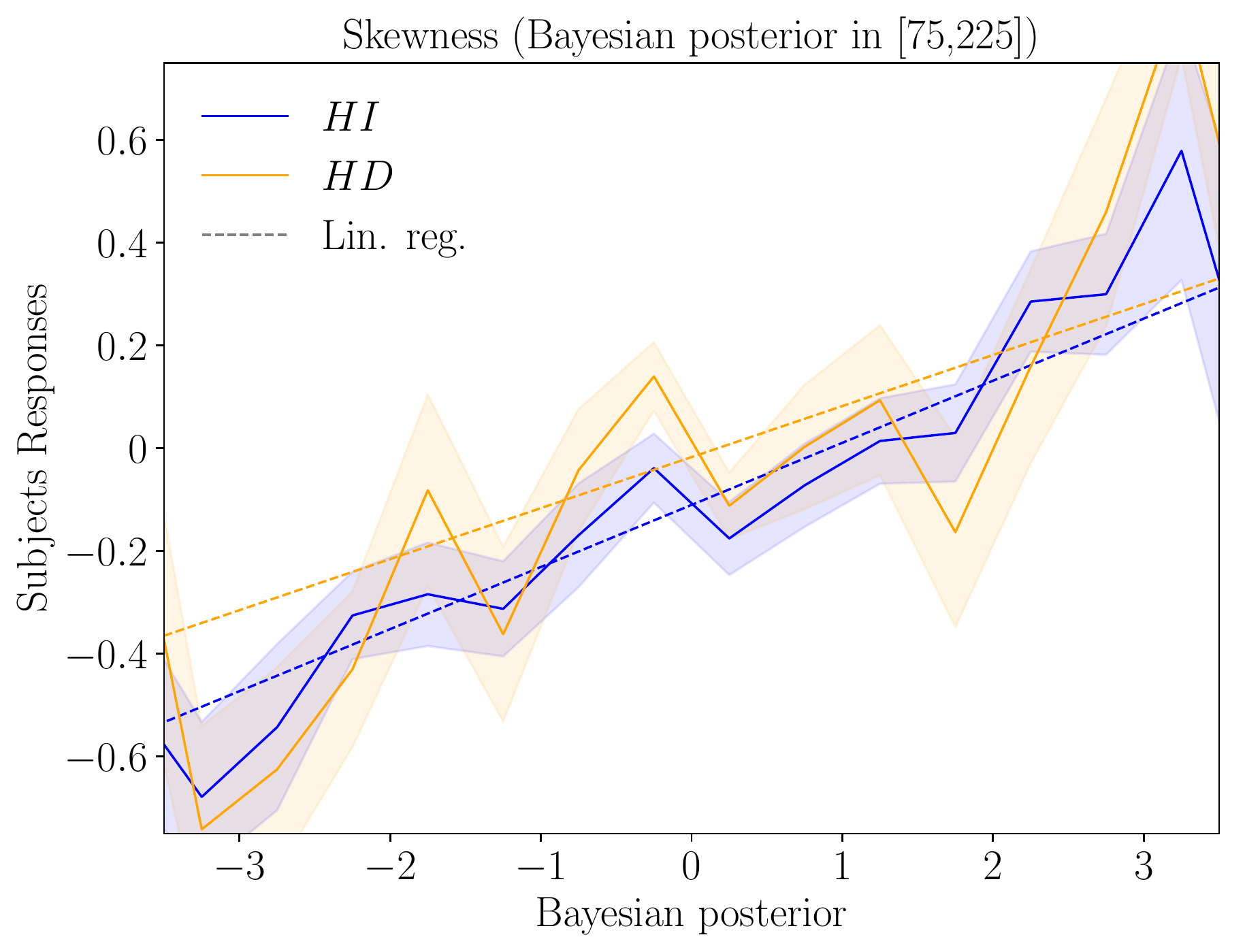}
\par}
	\captionof{figure}{\normalfont \textbf{Skewness in subjects' responses away from the bounds of the response range.}
	Same analysis of the skewness as in Fig. \ref{fig:std_skewness}B, but restricted to the trials in which the support of the Bayesian posterior is entirely contained in the middle interval of width half that of the state space (i.e., [75,225], to be compared to the state space, [0,300]). In both conditions, the correlation between the skewness of the Bayesian posterior and the empirical skewness of the subjects’ responses is positive and significant (HI: Pearson’s r = 0.23, p = 2e-12; HD: r = 0.14, p = 4e-4).
	}
	\label{fig:subj_skew_vs_p_skew_restricted}

\vspace{0.5in}

	\includegraphics[width=\linewidth]{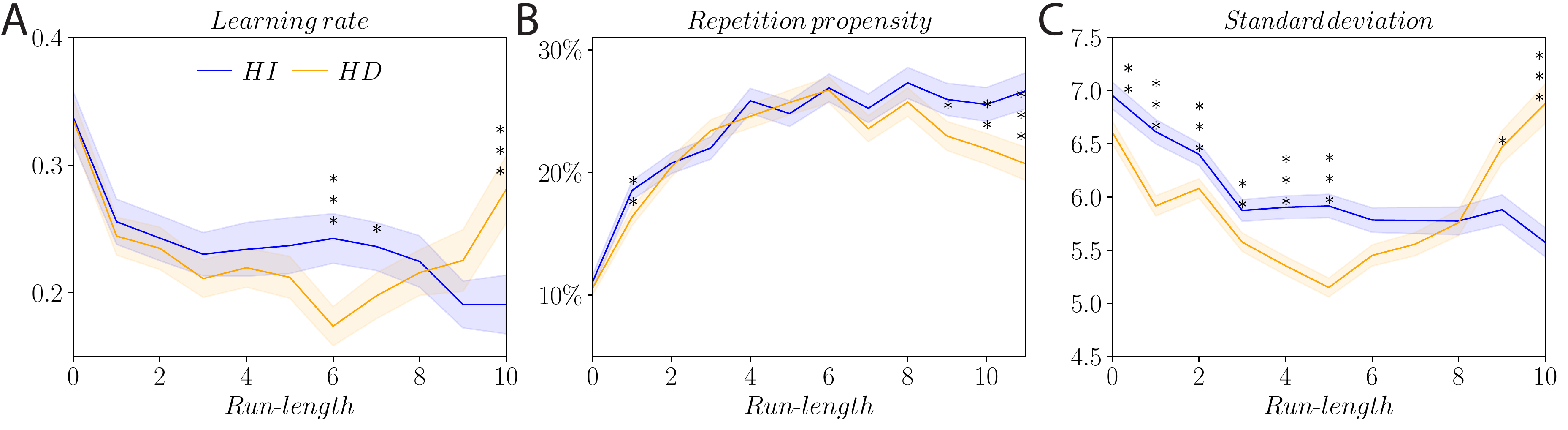}
	\captionof{figure}{\normalfont \textbf{Human learning rates, repetition propensities, and standard deviations of responses, from the complete contingent of subjects including the four subjects excluded from the analysis presented in the main text (see \nameref{sec:Methods}).}
	(\textbf{A}) as in Fig. \ref{fig:lrs}C; (\textbf{B}) as in Fig. \ref{fig:repeat}B; (\textbf{C}) as in Fig. \ref{fig:variability}C.
	}
	\label{fig:metrics-h-4bg}

\vspace{0.5in}

	\includegraphics[width=\linewidth]{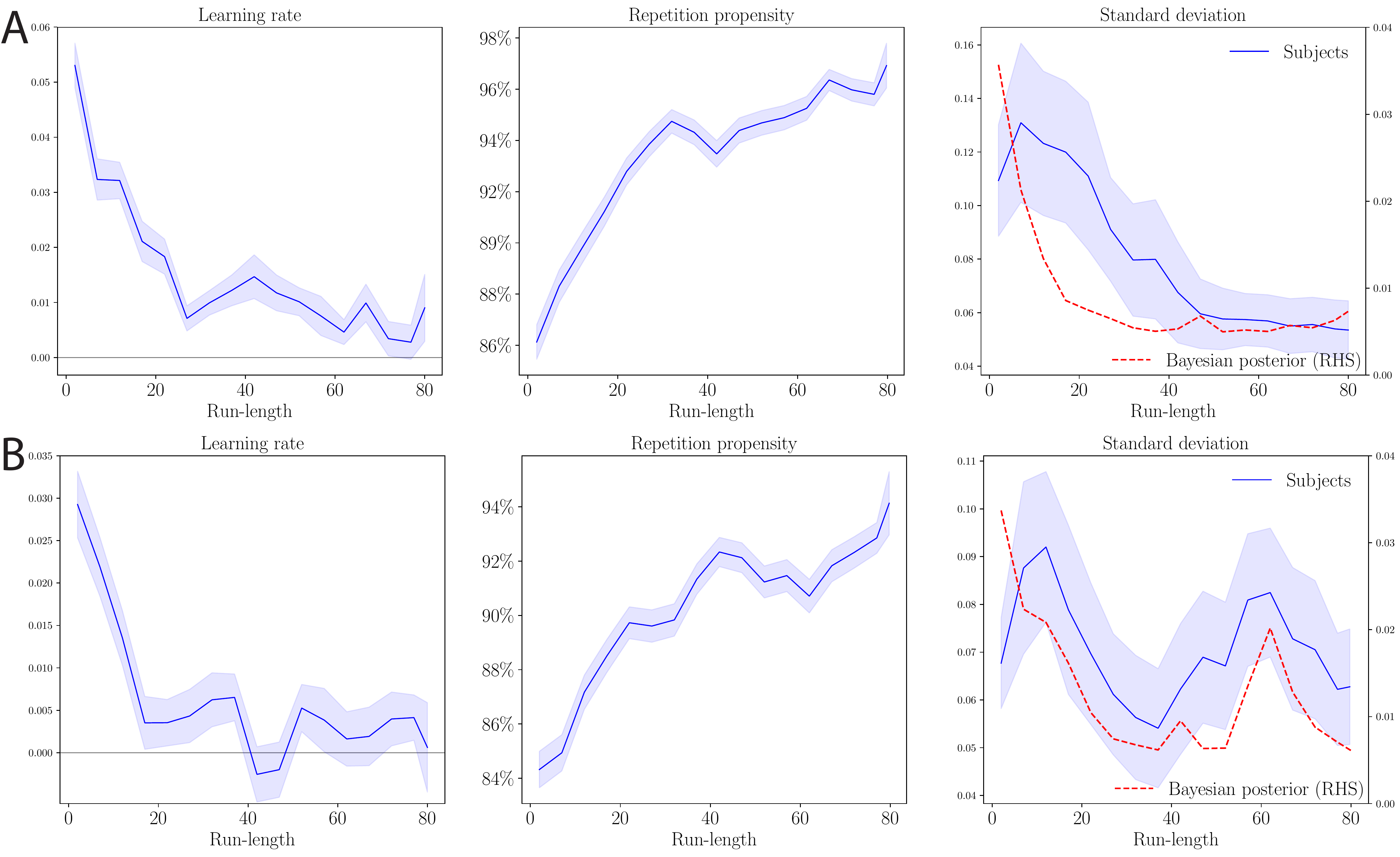}
	\captionof{figure}{\normalfont \textbf{
	Behavior of human subjects in the online inference tasks conducted by \cite{Gallistel2014} (\textbf{A}), and \cite{Khaw2017} (\textbf{B}).}
	\textnormal{
	\textit{Left panels}: Learning rates as a function of the run-lengths, in trials in which the surprise is greater than $.25$.
	\textit{Middle panels}: Repetition propensities.
	\textit{Right panels}: Standard deviations of responses, averaged across subjects.
	In these studies, the stimulus is binary and Bernoulli-distributed. The task is to infer the parameter of the Bernoulli distribution, which is subject to change points with a constant probability of 0.5\% (to be compared to 10\% in the HI condition of our task.) Moreover, some subjects were presented several times with the same sequence of stimuli (in different sessions). It is thus possible to examine the variability of responses within subjects (right panels, across-subject mean of the within-subject standard-deviations). In all panels, in order to mitigate noise, we pool together the responses in windows of five consecutive run-lengths.
	}
	}
	\label{fig:metrics-h-GallistelKSW}


\end{document}